\documentclass[lettersize,journal]{IEEEtran}
\usepackage{amsmath,amsfonts}
\usepackage{algorithmic}
\usepackage{algorithm}
\usepackage{array}
\usepackage[caption=false,font=normalsize,labelfont=sf,textfont=sf]{subfig}
\usepackage{textcomp}
\usepackage{stfloats}
\usepackage{url}
\usepackage{verbatim}
\usepackage{graphicx}
\usepackage{cite}
\usepackage{booktabs}
\usepackage{multirow}
\usepackage{mathrsfs,amssymb}
\usepackage{threeparttable}
\DeclareSubrefFormat{parens}{#1(#2)}

\begin{document}
\title{Joint Optimization of Controller Placement and Switch Assignment in SDN-based LEO Satellite Networks}
\author{Zhiyun Jiang, Wei Li, Menglong Yang}
\markboth{Journal of \LaTeX\ Class Files,~Vol.~14, No.~8, August~2021}%
{Shell \MakeLowercase{\textit{et al.}}: A Sample Article Using IEEEtran.cls for IEEE Journals}
\maketitle
\begin{abstract}
Software-defined networking (SDN) based low earth orbit (LEO) satellite networks leverage the SDN's benefits of the separation of data plane and control plane, control plane programmability, and centralized control to alleviate the problem of inefficient resource management under traditional network architectures.
The most fundamental issue in SDN-based LEO satellite networks is how to place controllers and assign switches. 
Their outcome directly affects the performance of the network.
However, most existing strategies can not sensibly and dynamically adjust the controller location and controller-switch mapping according to the topology variation and traffic undulation of the LEO satellite network meanwhile.
In this paper, based on the dynamic placement dynamic assignment scheme, we first formulate the controller placement and switch assignment (CPSA) problem in the LEO satellite networks, which is an integer nonlinear programming problem. 
Then, a prior population-based genetic algorithm is proposed to solve it.
Some individuals of the final generation of the algorithm for the current time slot are used as the prior population of the next time slot, thus stringing together the algorithms of adjacent time slots for successive optimization.
Finally, we obtain the near-optimal solution for each time slot.
Extensive experiments demonstrate that our algorithm can adapt to the network topology changes and traffic surges, and outperform some existing CPSA strategies in the LEO satellite networks.
\end{abstract}

\begin{IEEEkeywords}
SDN, LEO satellite networks, controller placement, switch assignment
\end{IEEEkeywords}

\section{Introduction}
\IEEEPARstart{L}{o}w earth orbit (LEO) satellite networks have received increasing attention in recent years due to their ability for providing seamless Internet access of the earth's surface regardless of terrain.
However, with the increase of people using satellite networks and the increasing complexity of network functions and protocols, satellite networks under traditional network architectures are experiencing difficulties and inefficiencies in resource management.
Fortunately, the emergence of Software-defined networking (SDN) alleviates these problems.
Compared to traditional network architectures, SDN demonstrates five features: data plane and control plane separation, device simplification, centralized control, networking automation and virtualization, and openness\cite{book1}.
These features of SDN have made it widely used in the management of various network environments, such as Internet of Vehicles\cite{Iov}, UAV network\cite{uav}, Internet of Things\cite{Iot1}\cite{Iot2}, etc.
In the field of satellite networks, SDN is also a potential solution to resource management problems.
OpenSAN\cite{OPENSAN} has pioneered the combination of SDN architecture and satellite networks.
Since then, the concept of software-defined satellite networking (SDSN) has gradually been developed, and its application in the LEO satellite network has also received attention\cite{enhance}.

LEO satellite networks are large-scale networks with numerous nodes and long link distances.
It is necessary to adopt a distributed control plane (i.e., the control plane contains multiple controllers) in order to avoid excessive delay\cite{distr}.
Aiming to meet various demands and improve network performance, multiple controllers need to be placed reasonably in the network to achieve efficient allocation and decision-making of resources and tasks.
Therefore, the placement of controllers and the assignment of switches are critical.
In most current SDSN architectures, the controllers are statically placed on the ground\cite{leo_ground1} or in geostationary-earth-orbit (GEO) satellites. 
The former results in frequent handover of the controller-switch link due to the high-speed motion of LEO satellites, while the latter results in large one-sided propagation delay (more than 100 ms) due to the huge distance between GEO and LEO.
In order to avoid these problems, the strategy of placing the controller on the LEO satellite constellation where the switch is located gradually emerged.

\begin{figure}[h]
  \centering
  \includegraphics[width=3.5in]{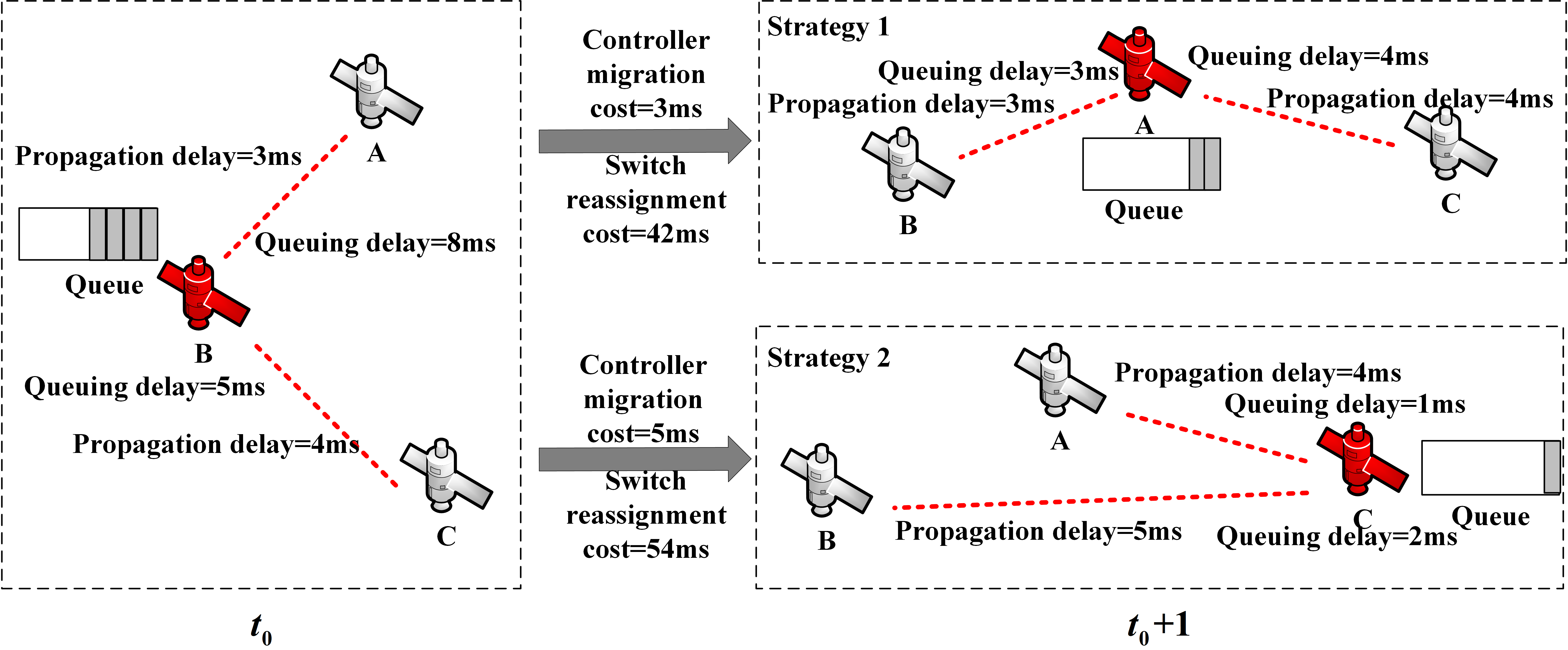}
  \caption{Controller placement and switch assignment strategies of two adjacent time slots.}
  \label{introduction}
\end{figure}

However, there are two issues that are not fully addressed in these existing strategies.
First, they only consider the impact of satellite network topology changes on the performance, ignoring the changes in satellite node traffic. 
The latter significantly impacts some network performances, such as queuing delay and load balance.
Secondly, they ignore the cost of strategy-shifting (i.e., the controller migration cost and the switch reassignment cost), thus severing the connection between strategies for adjacent time slots.
Considering the controller migration cost and switch reassignment cost, adopting an optimal strategy for each time slot in the LEO satellite networks does not mean that these strategies are optimal for a period.
For example, we adopted the controller placement and switch assignment strategy shown in Fig.\ref{introduction} at $t_0$. 
At $t_0+1$, there are two strategies for us to choose: strategy 1 and strategy 2, which activate satellite A and satellite C as controllers, respectively. 
It is obvious that strategy 2 is the optimal strategy for this time slot (12ms  \textless 14ms).
However, taking these two time slots as a whole, it is a better choice to adopt strategy 1 at $t_0+1$ (45ms+14ms \textless59ms +12ms).

Based on the above two points and the fact that solving the problem is NP-hard\cite{np-hard1}\cite{np-hard2}, in this paper, we first build a traffic model to simulate the load on each satellite node and introduce a queuing delay model to estimate the queuing delay of requests on each satellite.
Then, we propose a prior population-based genetic algorithm to place controllers and divide the control domains in terms of the satellites' mobility and traffic fluctuation.
The prior population is exploited to link the strategy-solving processes for two adjacent time slots to facilitate the consideration of the strategy-shifting cost and achieve continuous optimization.
Besides, the prior population can also accelerate the convergence of the algorithm.
An advantage of this continuous optimization is that it avoids predicting the traffic during the whole period, and only needs to know the traffic in the current time slot.
The former is unrealistic in the case of a large time span, while the latter is easy to implement due to the transience of one time slot.
In order to observe the results of the strategy, we develop a visualization tool for controller placement and switch assignment.
In more detail, the main contributions of this paper are summarized as follows:
\begin{itemize}
\item[$\bullet$] In the SDN-based LEO satellite network, we first establish the traffic model of the data plane and the queuing delay model of the controller port.
Then, we model the joint optimization problem of SDN controller placement and switch assignment, and propose the objective function that needs to be minimized, including controller response delay, controller migration cost, switch reassignment cost, controller synchronization cost, and load balance factor.
\item[$\bullet$] We propose a prior population-based genetic algorithm that uses hybrid coding to make a chromosome contain both a controller placement strategy and a switch assignment strategy. 
Besides, some individuals in the final generation population of the previous time slot are used as prior individuals of the current time slot algorithm to achieve continuous optimization for a period.
\item[$\bullet$] We develop a visualization tool for controller placement and switch assignment strategy. 
Then we show the advantages of the algorithm over other strategies in the experimental section. 
Finally, we analyze the impact of the number of controllers and weight settings on the performance. 
\end{itemize}

The rest of this paper is organized as follows.
Section II introduces some current controller placement and switch assignment strategies in terrestrial and satellite networks.
In Section III, we build the calculation model of each part of the cost and formulate the optimization problem.
Subsequently, in Section IV, a prior population-based genetic algorithm is proposed. 
It concatenates the optimization problems of all time slots.
Then the simulation results and analysis are provided in Section V.
Finally, this paper is concluded in Section VI.

\section{Related work}
In this section, we investigate some existing controller placement switch assignment strategies by classification.
\subsection{Controller placement in terrestrial networks}
Heller \emph{et al.}\cite{first_placement} first proposed the controller placement and switch assignment (CPSA) problem in terrestrial networks.
Since then, the CPSA problem has become a hot topic of research.
Various optimization methods were proposed to solve the CPSA problem under respective considerations.

Yao \emph{et al.}\cite{first_traffic} exploited the k-center algorithm first to solve the controller placement problem when considering the load factor of controllers.
Considering the queuing delay of controllers, Chai \emph{et al.}\cite{ground1} applied K-means algorithm, Dijkstra algorithm, and Kuhn-Munkres algorithm to solve the CPSA problem.  
Aiming to minimize inter-controller synchronization cost and sensor-controller communication delay in SDN-enabled wireless sensor networks, Tahmasebi \emph{et al.}\cite{MOP} proposed a multi-objective optimization algorithm based on Cuckoo and successfully obtained a Pareto-front solution set.

Due to the power of deep reinforcement learning (DRL) in solving complex optimization problems, it has also been introduced in several studies to solve CPSA problems.
Wu \emph{et al.}\cite{DRL} proposed a DRL-based approach for controller placement problem to minimize the controller-to-switch latency and balance the load of controllers.
Bouzidi \emph{et al.}\cite{dynamic_clustering} proposed a DRL-based dynamic clustering algorithm, which not only solves the CPSA problem, but also enables dynamic calculation of the optimal number of controllers.
To reduce control latency and packet loss in Internet of vehicles, Yuan \emph{et al.}\cite{vehicles} proposed a multi-agent DRL based dynamic switch-controller assignment algorithm, in which the controllers are statically placed in local devices or remote devices.

\subsection{Controller placement in satellite networks}
Due to the fast-changing nature of satellite network topology and dynamic fluctuations in LEO satellite network nodes, the SDN controller placement strategy in terrestrial networks does not apply to SDSN. 
In the SDSN architecture, the placement strategy of SDN controllers can be divided into two categories.

\emph{1)Static placement schemes:}
The static placement scheme falls into two types of static placement static assignment (SPSA) and static placement dynamic assignment(SPDA).

Hu \emph{et al.}\cite{Soft_leo} proposed a SPSA strategy called SoftLEO, in which a controller is placed in each orbital plane of the LEO satellite network to control the other satellites in that plane.

In SPDA schemes, the most common strategy is to place the controllers on GEO satellites\cite{OPENSAN}\cite{SDSD_research}\cite{SPDA_load} or on geostationary ground stations\cite{SAA} and each controller controls the satellite nodes within its visibility.
In addition, in SDSN with only LEO satellite constellations, some SPDA strategies that place controllers statically on LEO satellite nodes have been proposed.
To reduce the average and maximum controller-to-switch propagation delay, Guo\cite{SPDA2021} proposed a shortest-path based algorithm to obtain the assignment solution of the switches for a known controller placement, and then formulated the controller placement problem as a mixed-integer programming problem to solve it.
In\cite{SPDA2022}, they added to the previous algorithm a dynamic assignment strategy that can balance the load of controllers.

These static placement schemes can avoid incurring controller migration costs. 
However, they either do not take into account topology changes or traffic fluctuations, and thus do not solve the controller placement problem well in realistic situations in SDSN.

\emph{2)Dynamic placement schemes:}
All dynamic placement schemes are dynamic placement dynamic assignment (DPDA) strategies. 
There is no dynamic placement static assignment strategy, based on the fact that if a controller is migrated, then the switches it controlled before the migration will also be reassigned.
As far as we know, in the field of SDSN, there is little work to dynamically place controllers in LEO satellite nodes.

To solve the CPSA problem, Wu \emph{et al.}\cite{BOTH} developed an Accelerate Particle Swarm Optimization to optimize four objective functions, including performance expense, failure tolerance expense, load balance expense, and economic expense.
Papa \emph{et al.}\cite{GLOBECOM1} first studied the CPSA problem in SDSN with only the LEO satellite constellation.
In \cite{TNSM}, the authors proposed a method to quantify the overhead of migration, reassignment, and reconfiguration, then exploited Gurobi framework to minimize the average flow setup time with respect to varying traffic demands.

Some studies have introduced super controllers in the control plane to manage common controllers, forming a hierarchical control plane.
For SDN-based space-terrestrial integrated networks, Zhang \emph{et al.}\cite{DPDA} proposed a controller placement algorithm that enables controllers to be dynamically placed on the ground, LEO satellites, MEO satellites, and GEO satellites. 
Chen \emph{et al.}\cite{INFOCOM} proposed a control relation graph based algorithm to dynamically determine the number of controllers, the location of controllers, and the assignment of switches and then designed a lookahead-based improvement algorithm to reduce network management overhead.

\section{System model}
In this section, we first present our system model, including SDN-based LEO satellite networks architecture, data plane traffic model, expected queuing delay model, and migration cost.
Subsequently, we formulate the CPSA problem based on the system model. 
The notations used in the overall system are summarized in Table I.
\begin{table}[htbp]\scriptsize
\caption{The key notations\label{tab:table1}}
\centering
\begin{tabular}{l|l}
\hline
\hline
\bf Notation&\bf Definition\\\hline
\hline
$N,K,T$&Number of LEO satellites,controllers and time slots\\

$\mathcal{S},\mathcal{T}$&Set of switches and time slots\\

$\mathcal{C},\mathcal{C^{\prime}}$&Set of controllers for the current time slot and the last time slot\\

$y_k^t$& Indicator of satellite $k$ activated as a controller\\

$x_{nk}^t$& Indicator of switch $s_n$ assigned to controller $c_k$ in time slot $t$\\

$\tau^{G}$&Greenwich Mean Time\\

$s_n$&Switch with satellite number $n$\\

$c_k$&Controller with satellite number $k$\\

$\tau^{G\rightarrow L_{r}}$&Local time in region $r$ at Greenwich Mean Time $\tau^{G}$\\

$P_r$&Maximum number of network messages that region $r$ can generate\\

$w_r(\tau^{G})$ &Scaling factor indicating the fluctuation of traffic with time\\

$\eta_1$&Scale factor parameters\\

$P_r(\tau^{G})$&Number of messages that region $r$ sends to controllers at GMT $\tau^{G}$\\

$p_{s_n\rightarrow c}(\tau^{G})$&Number of requests that $s_n$ sends to its controller at GMT $\tau^{G}$\\

$p_{s_n\rightarrow c}^{t_0}(t)$&Number of requests sent by $s_n$ to its controller in time slot $t$, starting at $t_0$\\

$A^{\tau^{G}}_{s_n,r}$&Area covered by $s_n$ over region $r$ in time slot $\tau^{G}$\\

$s_{n\rightarrow l}$& The $l$-th hop satellite on the path from $s_n$ to its controller\\

$\psi^{t}_{nk}$&Response delay for requests sent from $s_n$ to $c_k$ in time slot $t$\\

$\psi^{Q,t}_{nk}$&Queuing delay for requests sent from $s_n$ to $c_k$ in time slot $t$\\

$\psi^{T,t}_{n}$&Transmission delay of $s_n$ in time slot $t$\\

$\psi^{F,t}_{n}$&Forwarding delay of $s_n$ in time slot $t$\\

$\psi^{P,t}_{n,n^{\prime}}$&Propagation delay between $s_n$ and $s_{n^{\prime}}$ in time slot $t$\\

$\psi^{P^{\prime},t}_{n}$&Processing delay of $s_n$ in time slot $t$\\

$t_0$&GMT time at the start of the first time slot\\

$b_k^t$&Backlog length of controller $c_k$ in time slot $t$\\

$\lambda_k$&Processing capacity of controller $c_k$\\

$\rho$&Parameters for fitting the queuing delay\\

$\Delta t$&Duration of one time slot\\

$\Psi^t$&Average response delay of all requests in time slot $t$\\

$cm^t$&Controller migration cost in time slot $t$\\

$sm^t$&Switch migration cost in time slot $t$\\

$cs^t$&Controller synchronization cost in time slot $t$\\

$v_c$&Speed of light\\

$d^t_{c_k,c_j}$&Length of the shortest path link between $c_k$ and $c_j$\\ 

$D,R_s$&Size of $data$ $set$ and transmission speed of migration link\\

$M^t$& Sum of migration cost and synchronization cost\\

$\Delta B^t$&Load balance factor in time slot $t$\\

$B^t_{ave}$&Average load of all controllers in time slot $t$\\

$\boldsymbol{y^t,x^t}$&Solutions to the CPSA problem for time slot $t$\\
\hline
\end{tabular}
\end{table}
\subsection{SDN-Based LEO Satellite Networks Architecture}

\begin{figure}[h]
\centering
\includegraphics[width=3.5in]{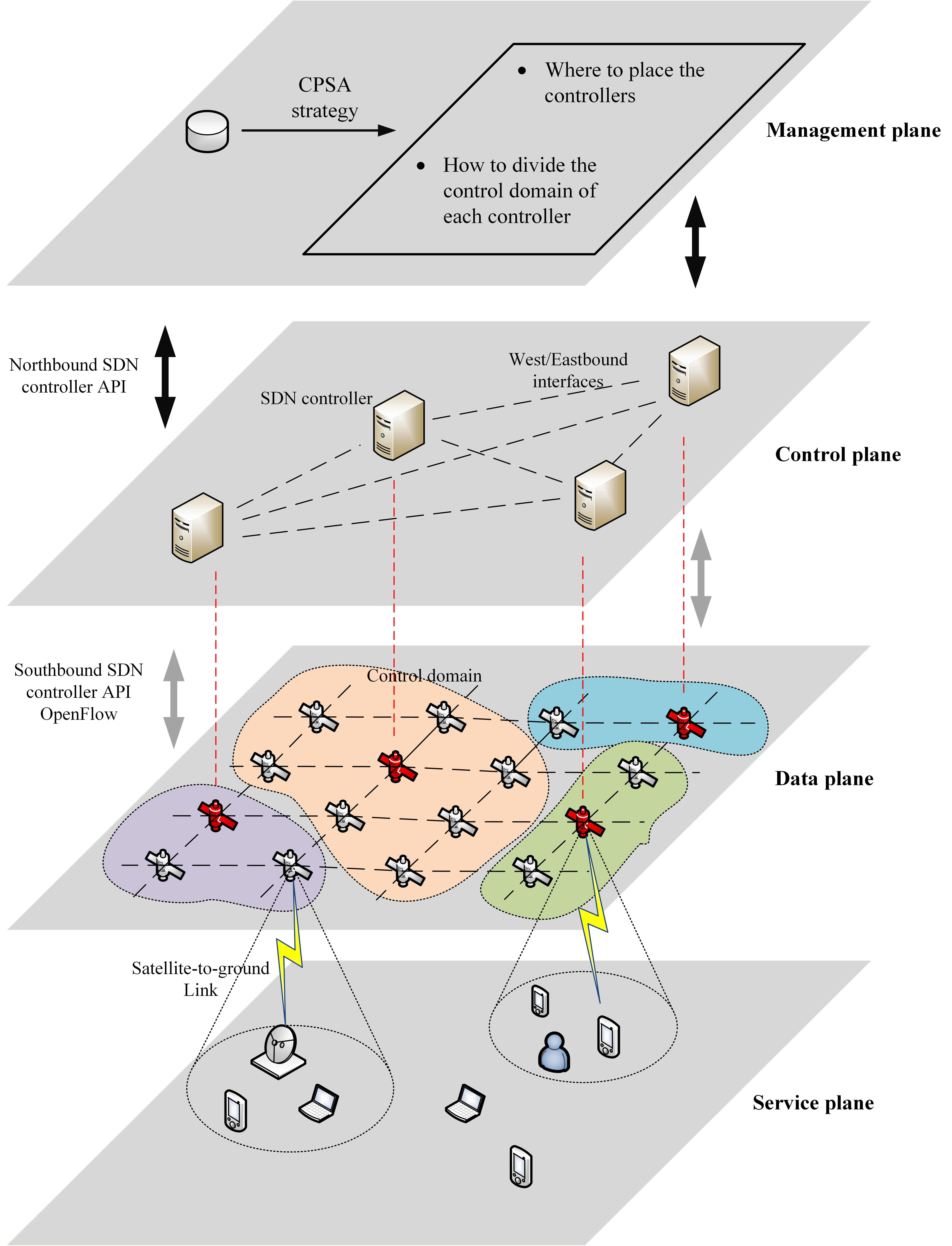}
\caption{SDN-based LEO satellite networks architecture.}
\label{fig_1}
\end{figure}

Fig.\ref{fig_1} illustrates the SDN-based LEO satellite network architecture with four layers.

\emph{1)Service Plane:}
The terminals on the ground that use satellite networks for Internet access and communication constitute the service plane.
The traffic from the terminal devices is uploaded via the satellite-to-ground link to the satellites within their range of visibility, so differences in the distribution of terminals on the ground result in differences in the traffic carried by satellites.
This will be described in detail in Section B.

\emph{2)Data Plane:}
This paper takes $Walker$-$\delta$ LEO satellite constellation into account.
The LEO satellite constellation consists of $N$ satellites, each acting as a switch.
We use $\mathcal{N}=\left\{1,2,\ldots,N\right\}$ to denote the set of LEO satellites and $\mathcal{S}=\left\{s_1,s_2,\ldots,s_N\right\}$ to denote the set of SDN switches.
The switches use the Southbound interface and OpenFlow to obtain flow tables from the controllers and use them to determine the forwarding action for messages coming from the service plane.

\emph{3)Control Plane:}
Since SDN is layered logically, each satellite can act as either a switch or a controller.
We deploy a total of $K$ controllers in the control plane and denote the set of controllers by $\mathcal{C}$. 
If the satellite indexed $k$ is activated as a controller, then this controller is denoted as $c_k$, with $c_k \in \mathcal{C}$.
Obviously, the elements in the set $\mathcal{C}$ change over time under the dynamic controller placement scheme, but the length of $\mathcal{C}$ is constant (i.e., $\left|\mathcal{C}\right|=K$).

Each controller manages the switches within a control domain which is a subset of the network, providing these switches with functions such as route calculation and tracking statistics.
In addition, messages are exchanged between controllers via West/Eastbound interfaces so that each controller has a global view of the network.

\emph{4)Management Plane:}
It collects network measurement data from the Network Measurement module in the control plane via the Northbound interface\cite{dynamic_clustering}, and these data support the CPSA algorithm. 
Then, the results obtained by the algorithm will be fed back to the control plane through the Northbound interface to determine the placement of controllers and the division of the control domain of each controller.

To address the CPSA problem of LEO satellite networks over times, we divide a period of time evenly into $T$ slots and denote them by $\mathcal{T}=\left\{t=1,\ldots,T\right\}$.
We assume that the topology of the satellite network and the traffic maintain stable within a time slot.
At the initial stage of each time slot, the locations of the controllers and the control domains will be updated with the changes of traffic and network topology.
In this paper, we assume that the communication between the controllers and the switches is executed in an $in$-$band$ manner\cite{in-band}.
We define $y_k^{t} \in \left\{0,1\right\}$ to indicate whether the satellite indexed $k$ is activated as a controller in time slot $t$ ($y_k^{t}=1$) and $x_{nk}^t\in \left\{0,1\right\}$ to indicate whether switch $s_n$ is assigned to controller $c_k$ in time slot $t$ ($x_{nk}^{t}=1$).

\subsection{Data Plane Traffic Model}
\begin{figure}[h]
\centering
\includegraphics[width=3.5in]{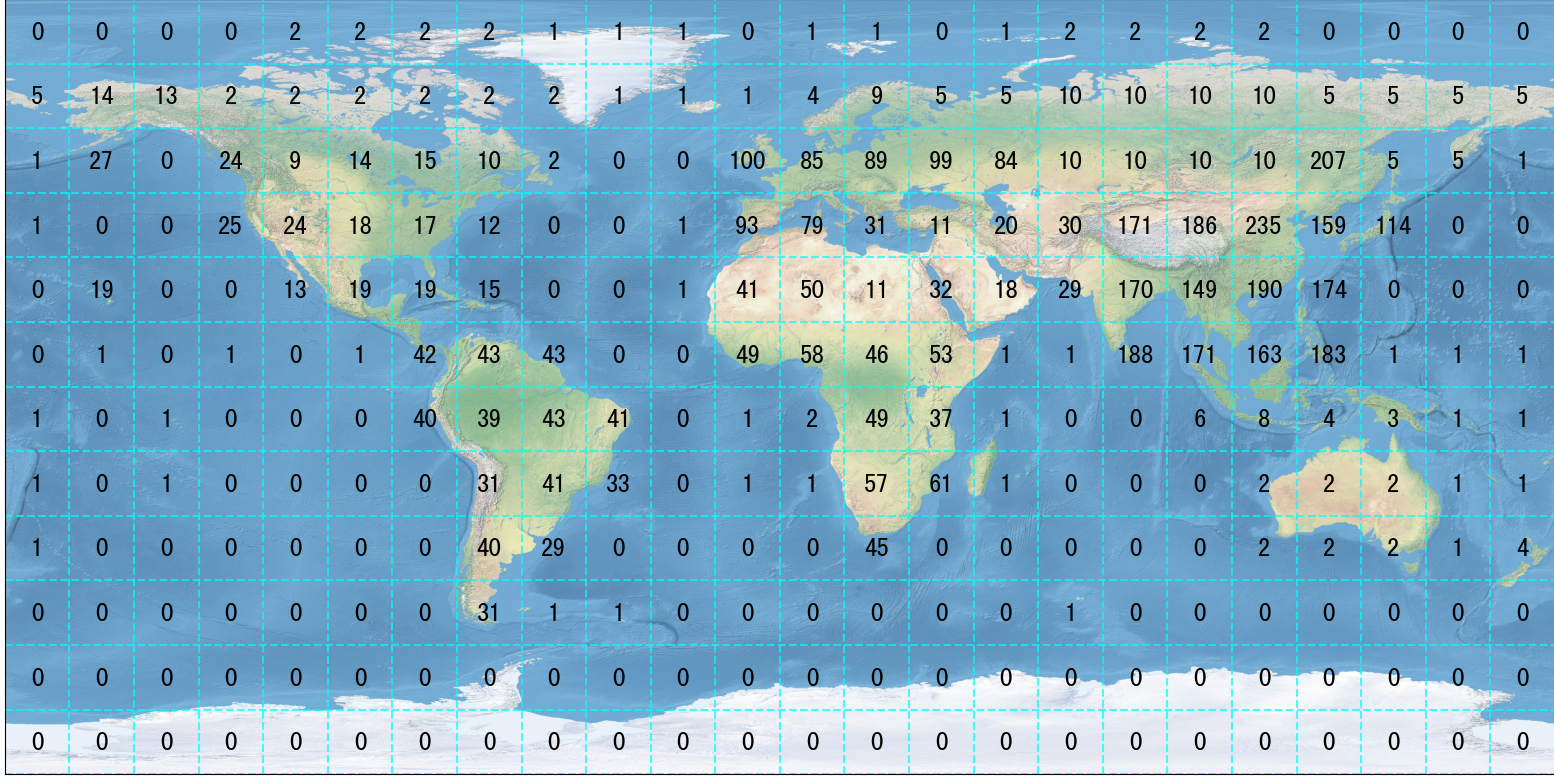}
\caption{Maximum number of Internet users per region, divided by $10^6$.}
\label{fig_2}
\end{figure}

For the spatial dimension, based on\cite{traffic_model}, the earth's surface is divided by $15^{\circ}\times 15^{\circ}$ to get 288 regions. 
Then, according to the distribution of Internet users in the world in 2022\cite{distribution}, the number of Internet users in each region is obtained, as shown in Fig.\ref{fig_2}.

The number of people using the network in a region change over time, which can further lead to fluctuations in traffic on the satellites.
The number of network messages transmitted region $r$ to the SDN controllers at GMT $\tau^{G}$ is:
\begin{equation}\label{}
p_r(\tau^{G})=P_r w_r(\tau^{G}) \eta_1
\end{equation}
where $P_r$ is the maximum number of network messages transmitted from the region $r$ to the LEO satellite network, which is related to the maximum number of Internet users in the region $r$.
This paper assumes that every 100 Internet users send one network message to the LEO satellite network in one time slot. 
The messages sent by the switch to its controller are called requests.
Their quantify is related to the number of $Packet$-$In$ messages, which accounts for only a fraction of all network messages of the switch.
Hence, a parameter $\eta_1$ is added to indicate the quantitative correspondence between the network messages and the requests.
$w_r(\tau^{G})$ is a scaling factor indicating the variation of the Internet users in the region $r$ with time, and it is given by:
\begin{equation}\label{}
w_r(\tau^{G})=\begin{cases}
0& \text{$0\leq\tau^{G\rightarrow L_{r}}<6$}\\
0.25(\tau^{G\rightarrow L_{r}})-6& \text{$6\leq\tau^{G\rightarrow L_{r}}<10$} \\
1& \text{$10\leq\tau^{G\rightarrow L_{r}}<22$} \\
1-0.25(\tau^{G\rightarrow L_{r}}-22)& \text{$22\leq\tau^{G\rightarrow L_{r}}<24$}
\end{cases}
\end{equation}
where $\tau^{G\rightarrow L_{r}}$ is the local time of region $r$ converted by $\tau^{G}$ according to the time zone.
 
Since an area may be covered by multiple satellites in the LEO satellite network simultaneously, network messages from that area are transmitted to multiple satellites. 
In this paper, we use a strategy based on the coverage area ratio to determine the number of messages each satellite receives from the region.
Therefore, the total number of requests sent by switch $s_n$ to its controller during a time slot starting at $\tau^{G}$ is:
\begin{equation}\label{}
p_{s_n\rightarrow c}(\tau^{G})=\sum\nolimits_{r=1}^{288}\left[p_r(\tau^{G})\frac{A^{\tau^{G}}_{s_n,r}}{\sum\nolimits_{n^{\prime}=1}^{N}A^{\tau^{G}}_{s_{n^{\prime}},r}} \right]
\end{equation}
where $A^{\tau^{G}}_{s_n,r}$ is the area covered by satellite $s_n$ over region $r$.
If the GMT time at the beginning of the scenario (i.e., the start moment of the first time slot) is $t_0$, we define the total number of requests sent by the switch $s_n$ to its controller in the $t$-th time slot as:
\begin{equation}\label{}
p_{s_n\rightarrow c}^{t_0}(t)=p_{s_n\rightarrow c}(t_0+\Delta t(t-1))
\end{equation}
where $\Delta t$ is the duration of one time slot.

\subsection{Expected Response Delay Model}
\begin{figure}[h]
\centering
\includegraphics[width=3.5in]{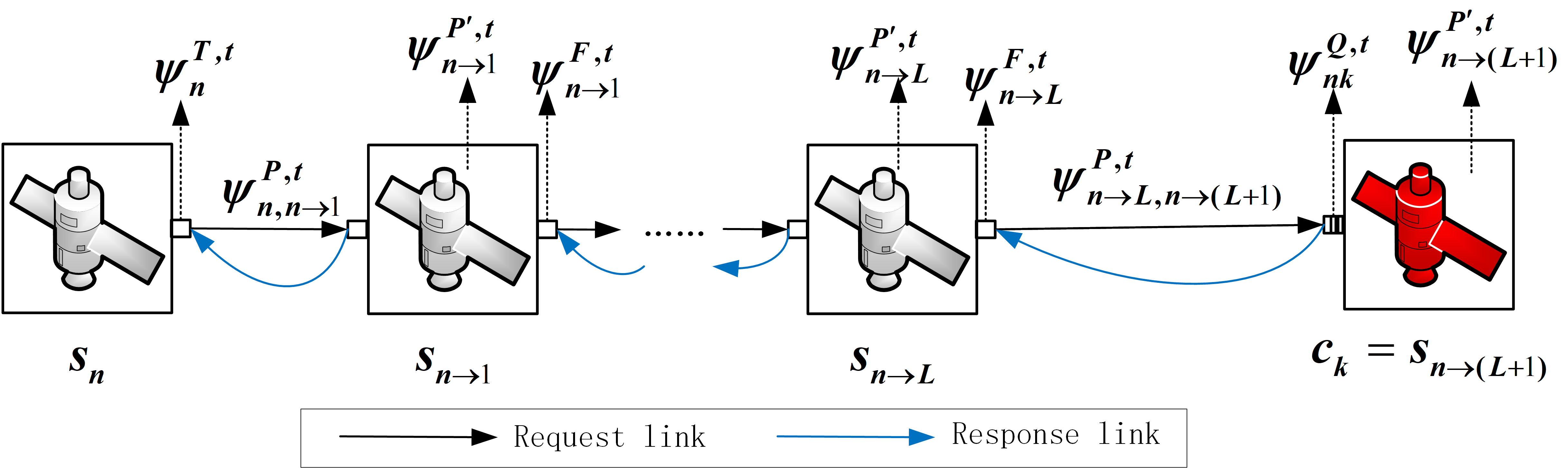}
\caption{An example of switch $s_n$ sending a request to controller $c_k$ and receiving a response.}
\label{fig_3}
\end{figure}

The whole process, from transmitting a request to receiving the corresponding response in time slot $t$, is shown in Fig.\ref{fig_3}.
We assume the entire process completes within one time slot, and the request and response processes use the same routing path.
The time consumed by this process is the response time of a request, including transmission delay, propagation delay, forwarding delay, queuing delay, and processing delay.

In the figure, switch $s_n$ sends a request to its controller $c_k$ at time slot $t$. 
The hop number of the routing path for this request is $L$.
$s_{n\rightarrow l}$ denotes the $l$-th hop switch on the path from $s_n$ to $c_k$.
The $(L+1)$-th hop switch $s_{n\rightarrow(L+1)}$ is the controller $c_k$ of $s_n$.
Then the expected response delay is:
\begin{equation}\label{}
\begin{aligned}
\psi^{t}_{nk}=&2\left[\sum\nolimits_{l=0}^L\psi^{P,t}_{n\rightarrow l,n\rightarrow (l+1)}+\sum\nolimits^{L+1}_{l=1}\psi^{P^{\prime},t}_{n\rightarrow l}+\sum\nolimits^{L}_{l=1}\psi^{F,t}_{n\rightarrow l}\right] \\ &+\psi^{T,t}_n+\psi^{Q,t}_{nk}\\
\end{aligned}
\end{equation}
where $\psi^{T,t}_n$ is the transmission delay of $s_n$ at time slot $t$.
$\psi^{P,t}_{n\rightarrow l,n\rightarrow (l+1)}$ is the propagation delay between two adjacent hopping switches, which is related to the distance between them. 
$\psi^{P^{\prime},t}_{n\rightarrow l}$ and $\psi^{F,t}_{n\rightarrow l}$ are the processing delay and forwarding delay of $s_{n\rightarrow l}$, respectively.
$\psi^{Q,t}_{nk}$ is the expected queuing delay for requests sent from switch $s_n$ to controller $c_k$, which is calculated by the expected queuing delay model\cite{INFOCOM}:
\begin{equation}\label{}
\begin{aligned}
\psi^{Q,t}_{nk}=&\rho(\sum\nolimits_{n^{\prime}=1}^N x^t_{n^{\prime}k})^2+\left[\frac{b^t_k}{\lambda_k}-\psi^{T,t}_n-\sum\nolimits_{l=0}^L\psi^{P,t}_{n\rightarrow l,n\rightarrow (l+1)}\right.\\ &\left.-\sum\nolimits^{L}_{l=1}(\psi^{P^{\prime},t}_{n\rightarrow l}+\psi^{F,t}_{n\rightarrow l}) \right]^{+}\\
\end{aligned}
\end{equation}

\begin{equation}
\begin{aligned}
    D^{label}_{i,l}=&\left[ \sum\limits_{t \in \mathcal{Y}_i, t=\mathbb{P}_1}\!  \! \! \! \! \! \!  A_l^i(q_{\hat{r_i}},q_t),\!\!\!\! \sum\limits_{t \in \mathcal{Y}_i, t=\mathbb{P}_2}\! \! \! \! \!  \! \!  A_l^i(q_{\hat{r_i}},q_t), \sum\limits_{t \in \mathcal{Y}_i, t=\mathbb{P}_2}\! \!  \! \! \! \! \! A_l^i(q_{\hat{r_i}},q_t) \right]
\end{aligned}
\end{equation}

where $\rho$ is a parameter that fits the queuing delay obtained by $ping$ in the joint SDN architecture of ONOS\cite{ONOS} and Mininet.
$\lambda_k$ represents the processing capacity of controller $c_k$, which is the number of requests that $c_k$ can process per second.
$b^t_k$ is the backlog length of controller $c_k$ at time slot $t$ (i.e., the number of requests that the controller has not completed processing by the end of the previous time slot), which is given by:
\begin{equation}\label{backlog}
b^t_k=\left[\sum\nolimits^N_{n=1}p_{s_n\rightarrow c}^{t_0}(t-1) x^{t-1}_{nk}y^{t-1}_k+b^{t-1}_k-\lambda_k\Delta t\right]^{+}
\end{equation}
We set $b^t_k=0$,$\forall c_k$ in the first time slot.
The average response delay of all requests in time slot $t$ is given by
\begin{equation}\label{}
\Psi^t=\frac{\sum\nolimits_{c_k \in \mathcal{C}}\sum\nolimits_{n=1}^N\psi^{t}_{nk}p_{s_n\rightarrow c}^{t_0}(t)x^t_{nk}}{\sum\nolimits_{n=1}^N p_{s_n\rightarrow c}^{t_0}(t)}
\end{equation}

\subsection{Migration and Synchronization}
As the topology of the LEO satellite network and the traffic from the ground to satellites change with time, in order to balance the controllers' load and ensure the response delay is tolerable, three phases of controller migration, switch reassignment, and controller synchronization occur sequentially at the beginning of each time slot. 
The following types of costs are incurred accordingly.

\emph{1)Controller migration cost:}
Controller migration means that the satellites activated as controllers may change over time due to a dynamic placement scheme.
In this phase, the satellite newly activated as a controller in a certain time slot acquires $data$ $store$ from any of the controller satellites in the previous time slot, thus grasping the state information of the whole network\cite{TNSM}.
Here we adopt the shortest distance strategy (i.e., the controller of previous time slot closest to the newly activated controller is selected for data acquisition).
The controller migration cost for the entire network at time slot $t$ is:
\begin{equation}\label{}
cm^t=\sum\nolimits_{k=1}^N\left[(y^t_k\oplus y^{t-1}_k)y^t_k(\frac{\underset{\forall c^{\prime} \in \mathcal{C^{\prime}}}{\min}d^t_{c_k,c^{\prime}}}{v_c}+\frac{D}{R_s})\right]
\end{equation}
where $\mathcal{C^{\prime}}$ is the controller set of $t-1$ time slot.
$d^t_{c_k,c^{\prime}}$ is the length of the shortest path link between $c_k$ and $c^{\prime}$ at time slot $t$.
$v_c$ is the speed of light.
$D$ and $R_s$ are the size of the $data$ $set$ and the transmission speed of the migration link, respectively.

\emph{2)Switch reassignment cost:}
Switch reassignment occurs at the end of the controller migration phase.
During this phase, the migrated switch interacts with the newly selected controller to enter the control domain of that controller.
The protocol for migrating a switch is described in detail in\cite{dynamic_migration}.
Six messages containing $hello$ and $handshake$ need to be exchanged between the switch and the controller to which it is reassigned\cite{TNSM}.
We assume that the links used for this process are based on the shortest-path strategy.
The switch reassignment cost of the LEO satellite networks in time slot $t$ is: 
\begin{equation}\label{}
sm^t=\sum\nolimits_{n=1}^N\sum\nolimits_{\forall c_k \in \mathcal{C}} \left[6\frac{d^t_{s_n,c_k}}{v_c}(x^t_{nk}\oplus x^{t-1}_{nk})x^t_{nk}\right]
\end{equation}

\emph{3)Synchronization cost:}
During the controller synchronization phase, data needs to be exchanged between the controllers so that each controller has information about the status of the entire network, which creates inter-controller delay cost\cite{cs}.
We use ONOS's $leaderless$ mode\cite{leaderless} for controller synchronization, which means that all controllers are fully distributed, and each controller has to send its own information to all other controllers.
We also use the shortest-path strategy to determine the links for exchanging messages between controllers, and the synchronization cost of the LEO satellite networks in time slot $t$ is:  
\begin{equation}\label{}
cs^t=\sum\nolimits^N_{k=1}\sum\nolimits^N_{j=1,j\neq k}y^t_k y^t_j\frac{d^t_{c_k,c_j}}{v_c}
\end{equation}

Based on the migration and synchronization cost model above, the total cost generated at the beginning of the t time slot is
\begin{equation}\label{}
M^t=cm^t+sm^t+cs^t
\end{equation}

\subsection{Load Balance Factor:}
Balancing the load of controllers is desirable to fully utilize the processing power of controllers and avoid wasting resources.
We utilize the mean squared deviation as $load$ $balance$ $factor$ to measure the load balancing of controllers in the whole network.
It is calculated by the equation as follows:
\begin{equation}\label{}
\Delta B^t=\sqrt{\frac{\sum\nolimits_{c_k \in \mathcal{C}}\left[\sum\nolimits^N_{n=1}p_{s_n\rightarrow c}^{t_0}(t)x^t_{nk}-B_{ave}^t\right]^2}{\left|\mathcal{C}\right|}}
\end{equation}
where $B^t_{ave}$ is the average number of requests for all controllers. 
It is given by:
\begin{equation}\label{}
B_{ave}^t=\frac{\sum\nolimits_{c_k \in \mathcal{C}}\sum\nolimits_{n=1}^N p_{s_n\rightarrow c}^{t_0}(t)x^t_{nk}}{\left|\mathcal{C}\right|}
\end{equation}

In theory, the small value of $\Delta B^t$ means the preferable balanced load.
Therefore, we take minimizing $\Delta B^t$ as an objective.
\subsection{Problem Formulation}

Given the above cost model, we formulate the cost function in the Formula\eqref{problem} and then minimize it by jointly optimizing the controller placement and switch assignment.
The CPSA problem is formulated as follows:

\begin{equation}\label{problem}
\underset{\boldsymbol{y^t},\boldsymbol{x^t}}{\min} \qquad w_1\Delta B^t+w_2\Psi^t +w_3M^t
\end{equation}

\centerline{$s.t.$  $(1)-(14)$}

\begin{equation}\label{c1}
y_{k}^t \in \left\{0,1\right\}, \ \forall k\in\mathcal{N},t\in\mathcal{T}
\end{equation}
\begin{equation}\label{c2}
x_{nk}^t \in \left\{0,1\right\}, \ \forall n\in\mathcal{N},k\in\mathcal{N},t\in\mathcal{T}
\end{equation}
\begin{equation}\label{c3}
x_{nk}^t\leq y_k^t ,\ \forall  n\in\mathcal{N},k\in\mathcal{N},t\in\mathcal{T}
\end{equation}
\begin{equation}\label{c4}
\sum\nolimits_{k=1}^N y_k^t = K ,\ \forall t\in\mathcal{T}
\end{equation}
\begin{equation}\label{c5}
\sum\nolimits_{k=1}^N x_{nk}^t=1         ,\ \forall n\in\mathcal{N},t\in\mathcal{T}
\end{equation}
where $\boldsymbol{y^t}=\left\{y^t_k|1\leq k\leq N\right\}$ represents the controller placement and $\boldsymbol{x^t}=\left\{x^t_{kn}|1\leq k\leq N,1\leq n\leq N\right\}$ represents the assignment indicator between switches and controllers.
Constraint\eqref{c1} forces $y_{k}^t$ to be 0 or 1 to indicate whether satellite $k$ is activated as a controller.
Constraint\eqref{c2} shows that the indicator of switch assignment between switch $s_n$ and controller $c_k$ in time slot $t$, $x^t_{kn}$, equals 0 or 1.
Constraint\eqref{c3} ensures that each switch can only be managed by activated controllers. 
Constraint\eqref{c4} forces the number of controllers to be $K$ at any time slot.
Constraint\eqref{c5} means that a switch can only be assigned to one controller at a time slot.
$w_1$,$w_2$, and $w_3$ are weighting factors representing the importance of these three objectives.

According to the above definition, the formulated problem in\eqref{problem} has three features: nonlinear optimization, integer optimization, and the coupling between controller placement and switch assignment.
These features make conventional integer programming (IP) methods inapplicable to our continuous optimization scenarios where optimization is performed at every time slot.
Here, in order to address this issue, we design an optimization algorithm for continuous optimization scenarios based on genetic algorithm.

\section{Proposed prior population-based genetic algorithm}

\begin{figure*}[h]
    \centering
    \includegraphics[width=7in]{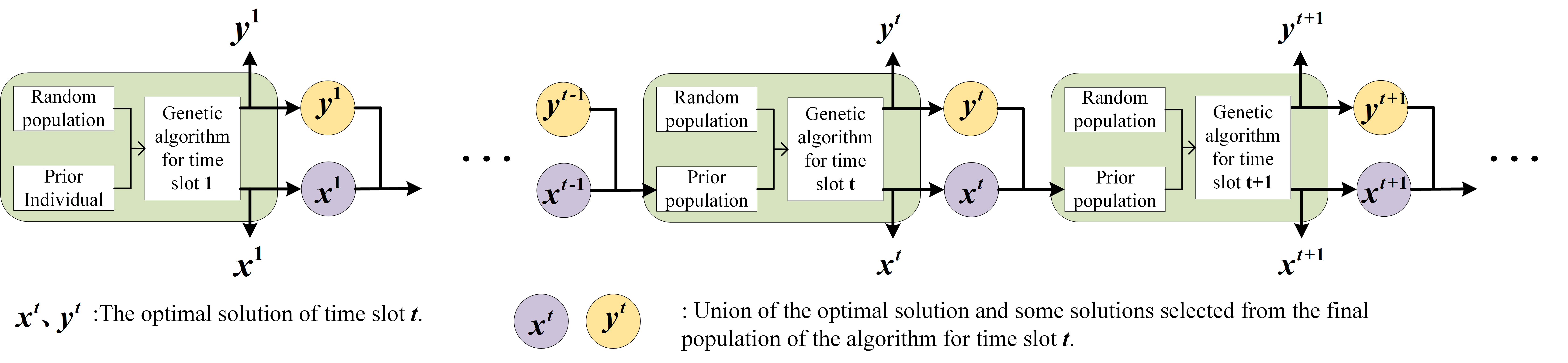}
    \caption{Continuous optimization process of the proposed algorithm.}
    \label{al_vis}
\end{figure*}

According to the definition of the cost function, the solution of the current time slot has an impact on the solution of the following time slots. 
Since the topology of the satellite network and traffic do not change dramatically between two adjacent time slots, some outstanding solutions or the optimal solution for the current time slot are also likely competent for the next time slot.
Based on the above point, we use the optimal solution and some individuals from the final population of the previous time slot as the initial population of the genetic algorithm for the current time slot, and let them speed up the algorithm's convergence. 
In addition, to expand the genetic algorithm's search space and avoid falling into the local optimum, we still keep a part of the initial population of the genetic algorithm for the current time slot as the random population.
The proposed prior population-based genetic algorithm for continuous optimization is visualized in Fig.\ref{al_vis}.
A prior population connects the algorithms of every two adjacent time slots. 
This strategy-solving process has two other benefits. 
Firstly, it does not need to predict the traffic in the entire time period, but only needs to know the current time slot traffic. 
The latter is often easy to implement. 
Secondly, the weights (i.e., $w_1$, $w_2$ and $w_3$) of each objective can also be adjusted over time according to the needs of each time slot.

\subsection{Generation of the Prior Individual for the First Time Slot.}
According to the algorithm optimization process shown above, it is noted that for the first time slot, there is no prior population passed from the previous time slot. 
Based on this particularity, in this paper, we propose a clustering-based population generation algorithm to get the prior individual of the algorithm for the initial time slot, as shown in Alg.\ref{t=0}.
The cluster center is the controller, and each switch is controlled by the controller represented by the cluster center closest to it.

In detail, the algorithm takes the number of controllers $K$, the satellite node distance matrix $D_0$ and the satellite topology $G_0$ as inputs.
In the initial phase, $K$ vertices are randomly selected as the centers of the clusters. 
For the iteration phase, each other vertex is assigned to the corresponding cluster according to the shortest distance principle. 
Then, the centers of the clusters are recalculated in each cluster to update $C$. 
The iteration ends when the updated $C$ is the same as the one before the update (i.e.,$C_0$) or the number of iterations reaches the upper limit.
The final clustering centers and the membership of each cluster are encoded as a chromosome and then output by the algorithm.

Subsequently, the genetic algorithm (Alg.\ref{t>0}, which is described in detail in Section B) for the first time slot iterates with the chromosome output by Alg.\ref{t=0} as the prior individual.
Since the backlog length of each controller is empty at the initial time slot, the queuing delay is relatively small and the response delay of the controller approximates the propagation delay. 
Therefore, the performance of the prior individual obtained by Alg.\ref{t=0} is relatively good compared to the random individuals.
The genetic algorithm based on this individual can accelerate convergence.

\begin{algorithm}[htbp]
\caption{The clustering-based prior individual generation algorithm}\label{t=0}
\hspace*{0.02in} \raggedright {\bf Input:} $K,D_0,G_0(V_0,E_0)$\\
\hspace*{0.02in} \raggedright {\bf Output:} $P_{l}^0$\\
\begin{algorithmic}[1]
    \STATE $C_0=\emptyset,k=0$
    \STATE $C \gets$Randomly select $K$ vertices from $V_0$ as clustering centers.
    \WHILE {$C\neq C_0$ or $k<m_k$}
      \STATE $C_0=C$
      \FOR {$v \in V_0$}
        \STATE $c_m=argmin D_0(v,c) ,\forall c \in C$
        \STATE Add $v$ to $cluster_{c_m}$.
      \ENDFOR
      \STATE Update $C$.  
      \STATE $k=k+1$
    \ENDWHILE
    \STATE $\boldsymbol{x^1},\boldsymbol{y^1}$ are obtained from $C$ and $cluster_c ,\forall c \in C$
    \STATE $P_l^0=encode(\boldsymbol{x^1},\boldsymbol{y^1})$
    \RETURN $P_l^0$
\end{algorithmic}
\label{alg2}
\end{algorithm}

\subsection{The Prior Population-based Algorithm Description}
We describe the genetic algorithm for each time slot in detail in Alg.\ref{t>0}.
We use a $1\times K$ matrix $\boldsymbol{b^t}=\left\{b_k^t \right\}$, and a $1\times N$ matrix $\boldsymbol{p^t}=\left\{p_{s_n\rightarrow c}^{t_0}(t)\right\}$ to represent the backlog length of controllers and the number of requests to be sent by each satellite, respectively.
In addition, the function $f(.)$ is used to calculate the objective function value.

In the initial phase (lines 1-10), we first initialize $\boldsymbol{b^t}$ or calculate it based on $\boldsymbol{p^{t-1}}$ and $\boldsymbol{b^{t-1}}$.
Subsequently, we generate the data $\boldsymbol{p^t}$ based on the Two-Line Orbital Element (TLE) and the established traffic model. 
Finally, based on the previous control placement optimal solution $\boldsymbol{y^{t-1}}$, the switch assignment optimal solution $\boldsymbol{x^{t-1}}$, and the subset of the final generation population $P^{t-1}_l$, the initial population $P_0^t$ of the genetic algorithm is generated, and the fitness of each individual is calculated.
$P_0^t$  is the result of the union of $P_r^t$ and $P_p^t$, and it is necessary to ensure that the number of individuals in $P_0^t$ is equal to the population size $N_p$.

In the iteration phase of the algorithm (lines 11-20), we first select the optimal individual based on fitness to ensure that it can appear in the population of the next generation.
According to the selection operator, the individuals that need to be crossed over and mutated are selected from $P_g^t$, and the set of selected individuals is $P^t_{g,s}$.
Unselected individuals do not participate in crossover and mutation.
After crossover and mutation, the individuals in $P_{g,s}^t$ produce $N_p-1$ offspring individuals in the collection. 
These individuals together with $I_{g,b}^t$ constitute the next generation population $P^t_{g+1}$.
Since the optimal individual of each generation is retained in the population of the next generation, the optimal individual of the last generation is the optimal individual of the whole iteration.

In order to reduce the number of iterations, we adopt the strategy of adaptively adjusting the number of iterations in the algorithm (lines 17-19).
We compare the objective function value $f(I^t_{g,b},\boldsymbol{p^t},\boldsymbol{b^t})$ of the optimal individual of the current generation with the global optimal objective function value. 
When the difference between the two is less than the threshold value $\delta$, we consider that the evolution of the current generation has fallen into evolutionary trap. 
The whole iteration process does not end until the number of generations reaches the upper limit $m_g$ or the number of evolutionary traps exceeds the maximum allowed times $m_c$.

After the whole iteration, we randomly select some individuals from the final generation population to prepare for the algorithm of next time slot. 
Finally, the optimal individual is decoded to obtain the solution of the current time slot.

\begin{algorithm}[htbp]
    \caption{Pseudocode prior population based genetic algorithm}\label{t>0}
    \hspace*{0.02in} \raggedright {\bf Input:} $\boldsymbol{x^{t-1}},\boldsymbol{y^{t-1}},\boldsymbol{b^{t-1}},P_{l}^{t-1},\boldsymbol{p^{t-1}}$\\
    \hspace*{0.02in} \raggedright {\bf Output:} $\boldsymbol{x^t}$,$\boldsymbol{y^t}$,$\boldsymbol{b^t}$,$P_{l}^t$,$\boldsymbol{p^t}$\\
    \begin{algorithmic}[1]
    \IF {$t=1$}
      \STATE Initialize all elements in $\boldsymbol{b^t} $ to 0. 
    \ELSE  
      \STATE $\boldsymbol{b^t} \gets$ based on Eq.(\ref{backlog}), $\boldsymbol{b^{t-1}}$ and $\boldsymbol{p^{t-1}}$.
    \ENDIF
    \STATE $\boldsymbol{p^t} \gets$ based on TLE and the established traffic model.
    \STATE $P_{p}^t \gets P_{l}^{t-1}\cup encode(\boldsymbol{x^{t-1}},\boldsymbol{y^{t-1}})$
    \STATE Generate a random population $P_{r}^t$ satisfying $\left|P_{r}^t\right|=N_p-\left|P_{p}^t\right|$.
    \STATE $P^t_0 \gets P_{r}^t\cup P_{p}^t$
    \STATE $M_{f} \gets$ calculate the fitness of $P^t_0$
    \WHILE {$True$}
      \STATE $I_{g,b}^t \gets P^t_g[argmax(M_{f})]$
      \STATE $P_{g,s}^t,\left|P_{g,s}^t\right|=N_p-1 \gets$ the selection operation on $P^t_g$.
      \STATE Perform mutation and crossover operation on $P_{g,s}^t$
      \STATE $P^t_{g+1} \gets P_{g,s}^t\cup I^t_{g,b}$
      \STATE $M_{f} \gets$ calculate the fitness of $P^t_{g+1}$
      \IF{the difference between $f(I_{g,b}^t,\boldsymbol{p^t},\boldsymbol{b^t})$ and the global optimal value is less than $\delta$ in consecutive $m_c$ generations \textbf{or} the number of iterations reaches $m_g$}
        \STATE Break.
      \ENDIF

    \ENDWHILE
    \STATE Randomly select some individuals from $P^t_g$ as $P_{l}^t$
    \STATE $\boldsymbol{x^t},\boldsymbol{y^t} \gets$ decode $I_{g,b}^t$
    \RETURN $\boldsymbol{x^t},\boldsymbol{y^t},\boldsymbol{b^{t}},P_{l}^t,\boldsymbol{p^{t}}$
    \end{algorithmic}
    \label{alg1}
  \end{algorithm}

\subsection{Algorithm Elements}
\begin{figure}[h]
    \centering
    \includegraphics[width=3.5in]{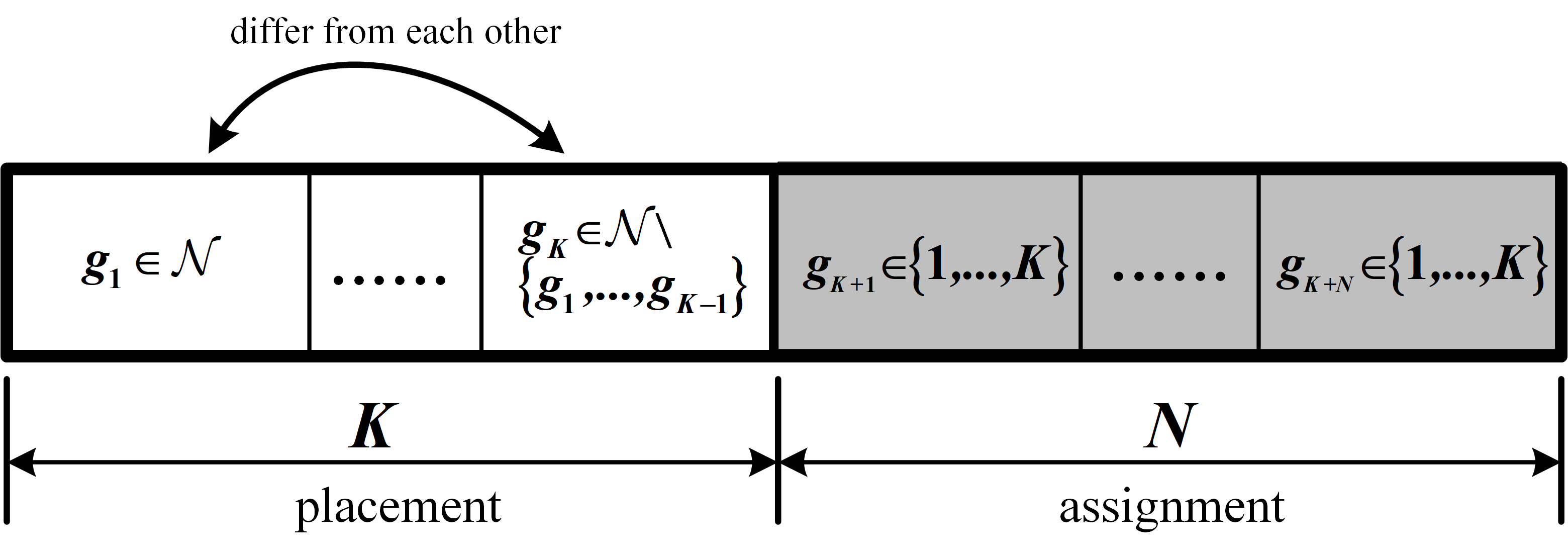}
    \caption{One chromosome in the genetic algorithm with the placement and assignment included.}
    \label{chorm}
\end{figure}

\emph{1)Chromosome:}
The chromosome is the representation of the solution after coding.
Here, we divide a chromosome into two parts, representing the controller placement strategy and switch assignment strategy, as shown in Fig.\ref{chorm}.
$g_i$ represents the $i$-th gene on the chromosome.

For the controller placement strategy part, it consists of $K$ genes. 
The value of a gene is the index of the satellite where a controller is located.
Thus, it takes values from 1 to $N$. 
Due to the constraint that controller locations are mutually exclusive, the genes' values should differ.
Based on the above, $g_k \in \mathcal{N}\backslash\left\{g_1,\ldots,g_{k-1} \right\}$, $\forall k \in \left\{1,\ldots,K\right\}$.
Here we use $permutation$ $coding$ as the form of coding for this part.
Subsequently, the second part has $N$ genes, indicating the assignment of the $N$ switch satellites.
When $g_{n+K}=k$, the $s_n$ selects the satellite indexed with $g_k$ of this chromosome as its controller.
This part adopts $real$ $number$ $encoding$ as the encoding form.

In the decoding stage, the chromosome's first part does not need to be decoded.
The second part is decoded according to the values of the genes in the first part of this chromosome.

\emph{2)Selection:}
The purpose of the selection operation is to select superior individuals from the population for mutation and crossover, and these individuals are called parents. 
The degree of excellence of individuals is measured by fitness. 
There are many selection strategies, such as tournament methods, roulette methods, truncation methods, Monte Carlo methods, etc\cite{selection_compare}. 
In this paper, we adopt the tournament method, which is less burdensome in computation and maintains the diversity of solutions compared to other selection procedures\cite{selection}.

\emph{3)Crossover and mutation:}
The purpose of both crossover and mutation is to produce new individuals called offspring. 
The former is the main operation that continuously opens up the solution search space and improves the global search capability of the algorithm. 
Since the hybrid coding form of the chromosome, we adopt different crossover operators and mutation operators for different parts.
For the first part of the chromosome (i.e., the controller placement part), we adopt the partially-matched crossover method, in which the conflict detection mechanism can ensure that the individual generated after crossover satisfy the constraint that the controllers' positions are not identical to each other.
The second part (i.e., the switch assignment part) adopts the two-point crossover method, which can efficiently and effectively generate offspring individuals representing the new switch assignment strategies.
The mutation operation is then used to avoid falling into a local optimum solution during the optimization process.
In this paper, we take the reverse mutation method for the first part of the chromosome to ensure that the controllers' positions differ.
Moreover, the breeder GA mutation method\cite{mutation} is used for the second part of chromosomes.

\subsection{Algorithm Complexity Analysis}
Due to the adaptive generation count mechanism, we cannot determine the number of generations. 
Here, We analyze the complexity bound.
The initialization population comes partly from the algorithm's input and partly from random generation. 
So, the complexity of the initialization population phase is $O(\left|P_{r}^t\right|(K+N))$.
The complexity of generating the population for every generation is $O(N_p(K+N))$.
The optimal individual is selected from the population at each iteration, and the complexity of this part is $O(N_p)$.
Then, the complexity of population generation, selection operation, mutation and crossover operation for maximum $m_g$ generations is given by $O(\left|P_{r}^t\right|(K+N))+O(N_p m_g (K+N))+O(m_g N_p)+O(m_g(N_p-1)N_p)+O(m_g(N_p-1))$, which is dominated by $O(N_p m_g (K+N))$.
The complexity of fitness calculation for $m_g$ generations of populations is $O(N_p m_g(K+N))$.
The complexity of selecting some individuals from the population of the last generation is $O(\left|P_{l}^t\right|N_p)$.
Hence, the overall complexity bound of prior population-based genetic algorithm can be approximated by $O(N_p m_g(K+N))$.

For the first time slot, its prior population is generated by Alg.\ref{t=0} with a complexity bound of $O(K m_k N)$.
So, the complexity bound for solving the strategies for all time slots in the entire period is $O(K m_k N)+O(T N_p m_g(K+N))\approx O(T N_p m_g(K+N))$.

\section{Simulation and result}
\subsection{Scene and Parameter Settings}
\begin{table*}[htbp]
    \caption{Major simulation parameters\label{t2}}
    \centering
    \begin{threeparttable}[b]
    \begin{tabular}{|ll||ll||ll|}
    \hline
    Parameter&Value&Parameter&Value&Parameter&Value\\\hline\hline
    Constellation type &Walker &$R_s$&1Gbps&Population size&200\\\hline
    Number of orbits &8 &$\Delta t$&60s&$m_g,m_c$&500,300\\\hline
    Number of satellites &72& $\lambda_k$&4000 requests/s\cite{INFOCOM}&$\delta$&$10^{-9}$\\\hline
    Inclination & 53deg &$\rho$&0.09\cite{INFOCOM}&Crossover probability&0.9\tnote{1},0.7\tnote{2}\\\hline
    Satellite half angle of view &35.5deg& $\eta_1$& 0.05&Mutation probability\tnote{1}&0.3\\\hline
    Orbit height&780km& $D$& 100MB&Mutation parameter\tnote{2}&$P_{shrink}=0.5,Gradient=20$\\\hline
    \end{tabular}
    \begin{tablenotes}
        \item [1] This parameter belongs to the chromosome of the controller placement section.
        \item [2] This parameter belongs to the chromosome of the switch assignment section.
    \end{tablenotes}
    \end{threeparttable}
    \end{table*}
Based on the TLE of the satellite constellation and CZML file, we build the satellite network scenario with Ceisum\cite{cesium}. 
The main parameters used in the simulation are listed in Table\ref{t2}.
In detail, 72 LEO satellites are uniformly distributed in 8 orbits at a height of 780km and inclination of 53deg.
The inter-satellite link is established in the form of +Grid\cite{hypatia}.
We take 1/1/2022 00:00:00 GMT as the starting time of the simulation and the simulation lasts for one day.
Since one time slot lasts 60 seconds, $\mathcal{T}$ contains 1440 time slots.
In the table, $P_{shrink}$ is the compression rate, and $Gradient$ is the number of gradient divisions, both of which are used to control the mutation distance for the breeder GA mutation method.

\begin{figure*}[h]
    \centering
    \includegraphics[width=7in]{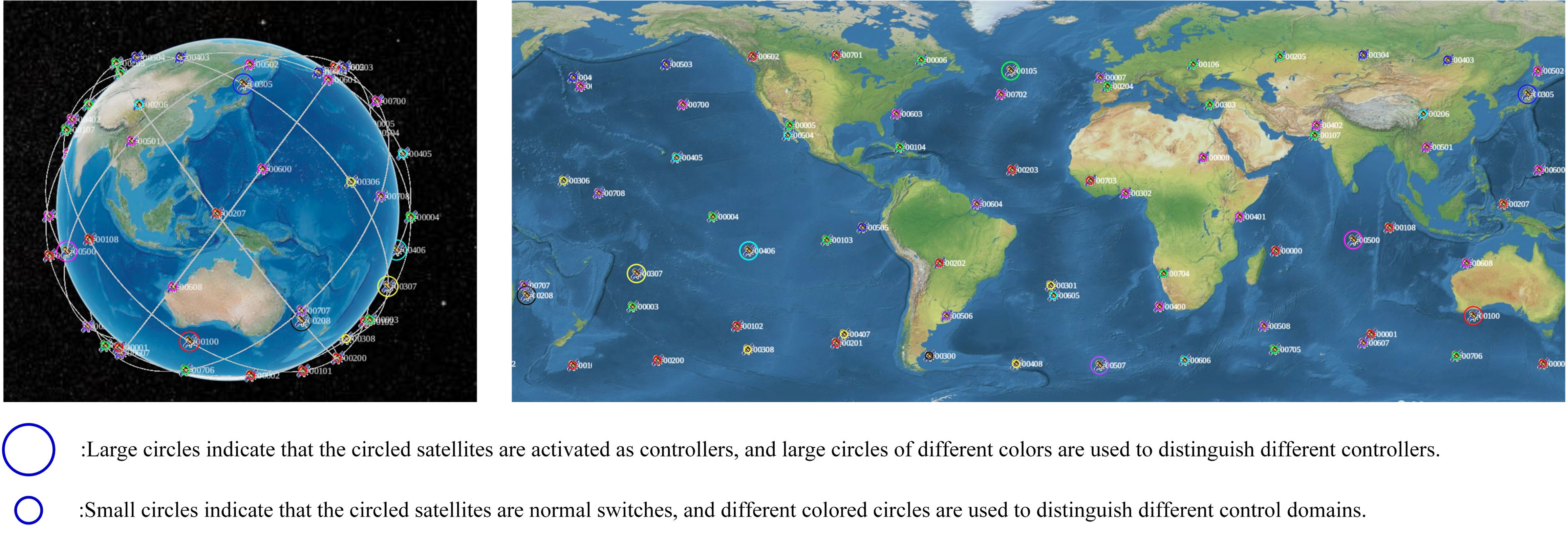}
    \caption{Visualization of satellite constellation and controller deployment switch assignment policy results.}
    \label{SDN_vis}
\end{figure*}

\begin{figure}[h]
    \centering
    \includegraphics[width=2.5in]{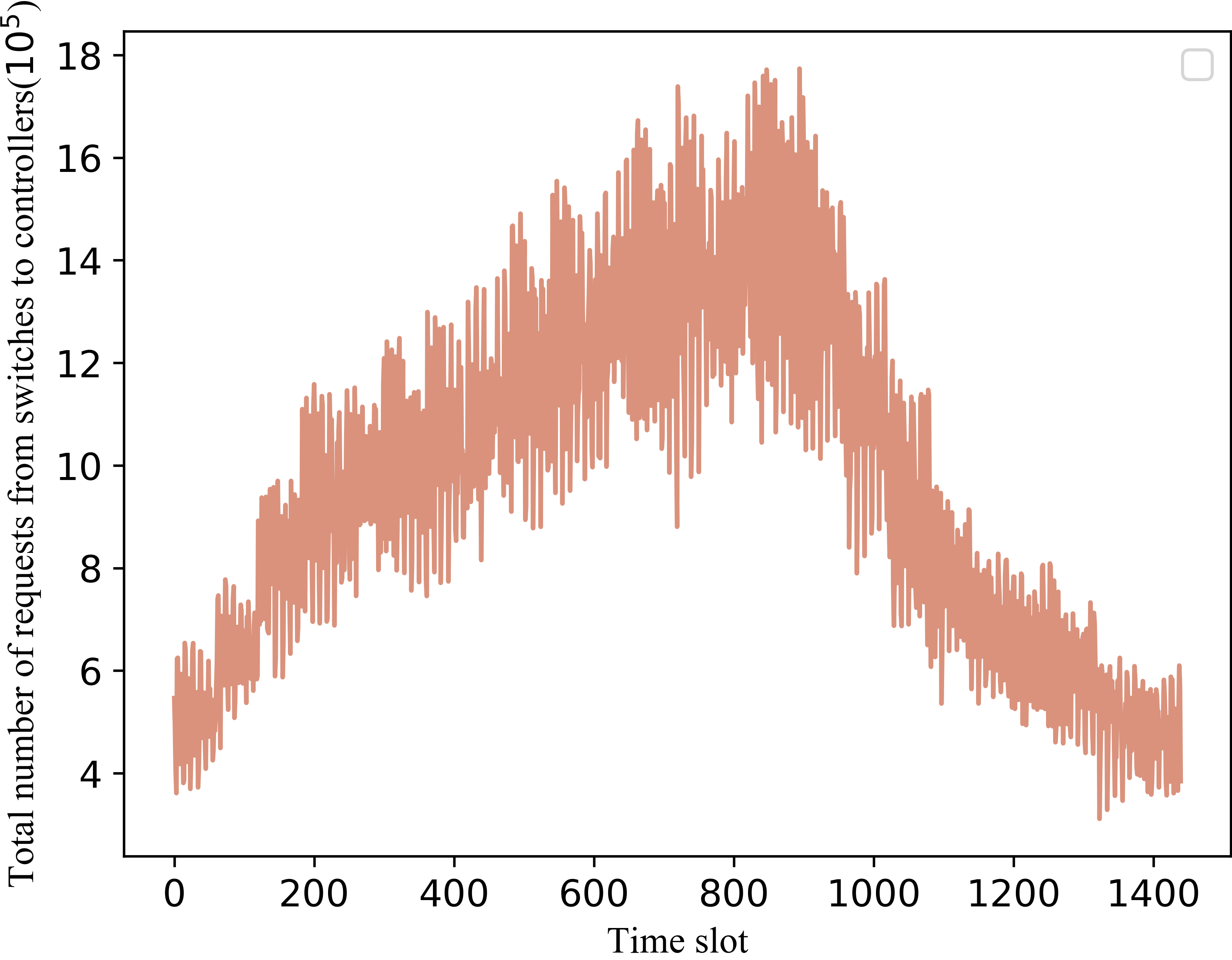}
    \caption{The change of the total number of requests from switches to controllers over time in the traffic model.}
    \label{req_num}
\end{figure}

In addition, we have developed a visualization tool for controller placement and switch assignment in the satellite constellation. 
The visualization of our LEO satellite constellations is shown in Fig.\ref{SDN_vis}.
To facilitate our subsequent analysis of the results, we show the change of the total number of requests per time slot over time in Fig.\ref{req_num}.

\begin{figure}[h]
    \centering
    \subfloat[]{\includegraphics[width=1.7in]{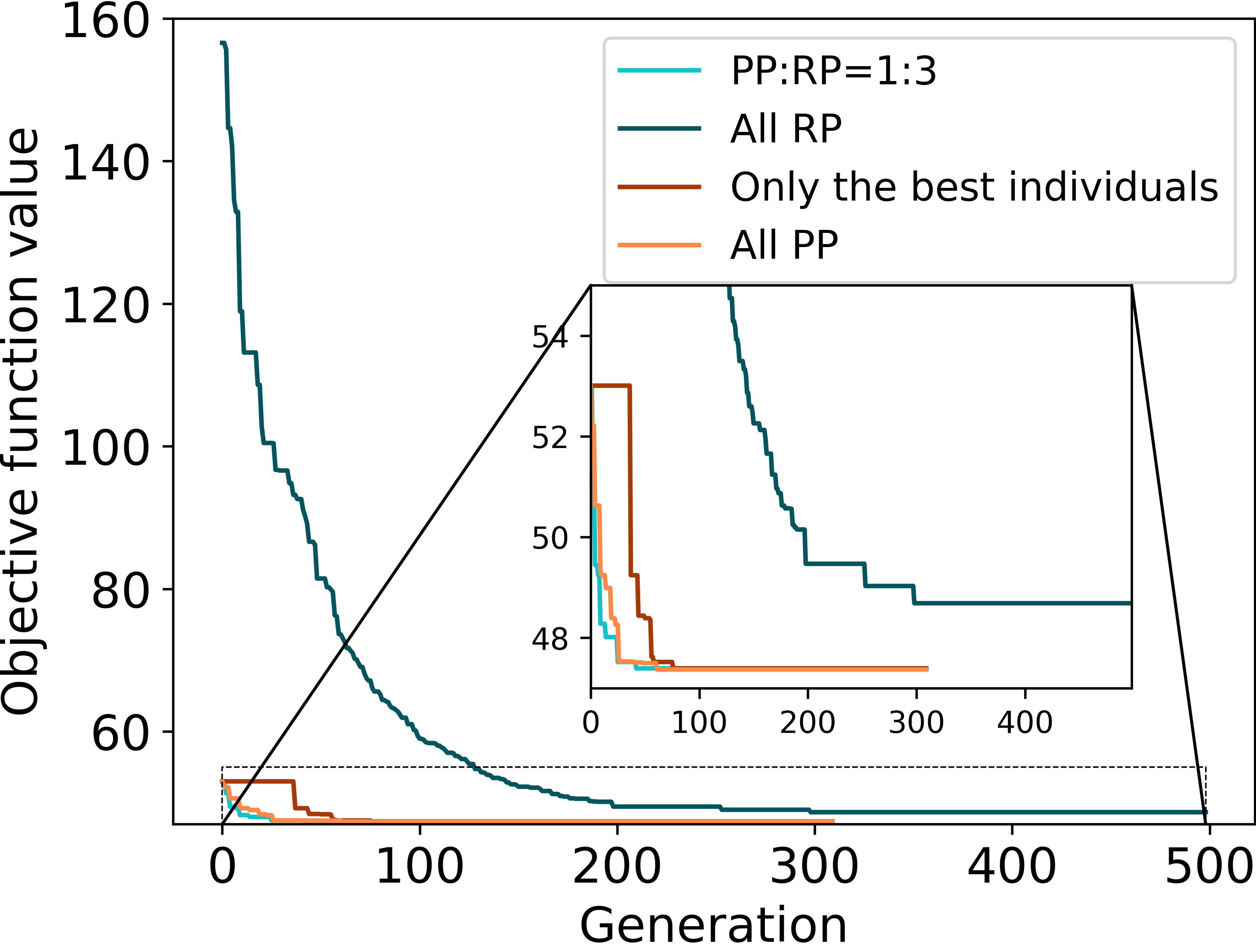}%
    \label{t!=1}}
    \hfil
    \subfloat[]{\includegraphics[width=1.7in]{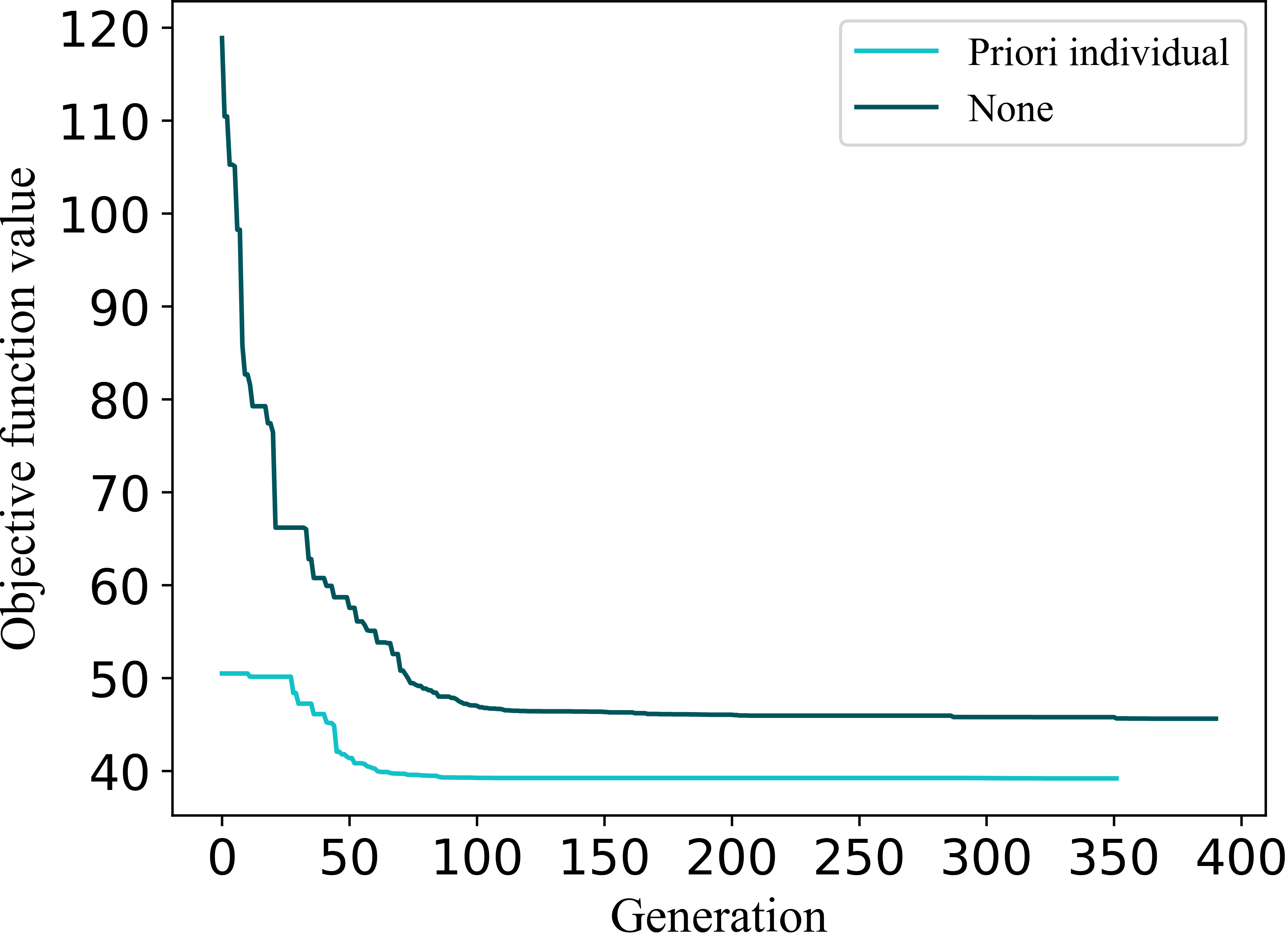}%
    \label{t=1}}
    \caption{Effect of a prior population on iteration.(a)Algorithm of the non-first time slot.(b)Algorithm of the first time slot.}
    \label{pop}
    \end{figure}

Fig.\ref{pop}\subref{t!=1} plots the iterations of our algorithm at a given time slot for different ratios of prior population (PP) size to random population (RP) size.
It can be seen that the algorithm of this time slot converges to a new optimal value in about 20 generations of evolution when there is a PP in the initial population.
In comparison, with only RP, the convergence takes more than 200 generations, which fully demonstrates the rapidity of our algorithm.
The figure also shows that the convergence result of the algorithm with PP is better than that without PP.
In order to ensure the diversity of the population while using the prior population, we set the ratio of PP to RP as 1:3 in this paper (i.e., PP and RP include 50 individuals and 150 individuals, respectively).
As for the first time slot, we can observe from Fig.\ref{pop}\subref{t=1} that the solution of the clustering algorithm as a priori individual of the strategy solution algorithm speeds up the convergence and the quality of the solution obtained is better than the solution obtained without a prior individual.

\subsection{Performance with Different Strategies.}
In this section, we compare the performance of the proposed prior population-based genetic algorithm with the following strategies.
The weights of our algorithm are set to $w_1=0.001,w_2=1,w_3=0.002$.
All of these strategies place the controllers on LEO satellites.
\begin{itemize}
\item[$\bullet$]\textbf{Minimum average flow step time (MAFST)\cite{TNSM}.} This strategy minimizes the average flow setup time in the LEO satellite network with a determined number of controllers, which is calculated from the propagation delay between controllers and switches. 
The optimization problem is built as a mixed integer programming problem, and the final controller placement and switch assignment results are obtained using the Gurobi optimizer.
\item[$\bullet$]\textbf{Soft LEO\cite{Soft_leo}.} In this strategy, one satellite within each orbital plane is selected as the controller to manage the other satellites within this orbital plane. 
The satellites acting as controllers have the same in-orbit indexes, and their role as controllers does not change over time.
\item[$\bullet$]\textbf{Modified density peaks clustering (MDPC)\cite{MDPC}.} The clustering centers of the LEO satellite network topology graph are obtained as controllers by running the MDPC algorithm separately at each timestamp over a period of time, and each switch is assigned to the one closest to it.
\item[$\bullet$]\textbf{Static placement and dynamic assignmemt (SPDA)\cite{SPDA2021}.} The strategy takes minimizing the average propagation delay and the maximum propagation delay of all switch-controller pairs over a time period as the optimization objectives and exploits Gurobi optimizer and the proposed shortest-path based dynamic assignment algorithm to obtain the results.
\end{itemize}

\begin{figure}[h]
    \centering
    \subfloat[]{\includegraphics[width=1.7in]{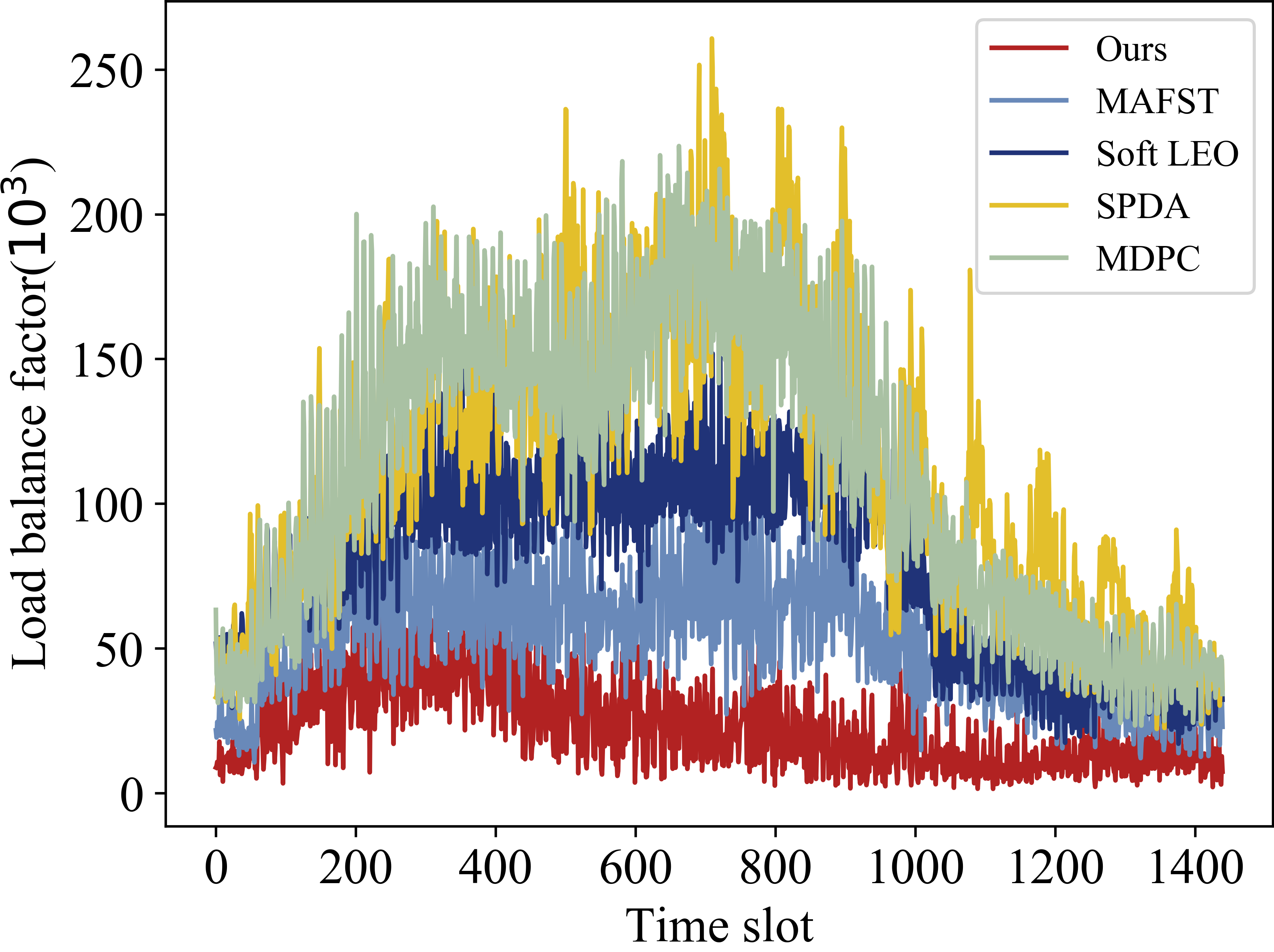}%
    \label{fig_1_case}}
    \hfil
    \subfloat[]{\includegraphics[width=1.7in]{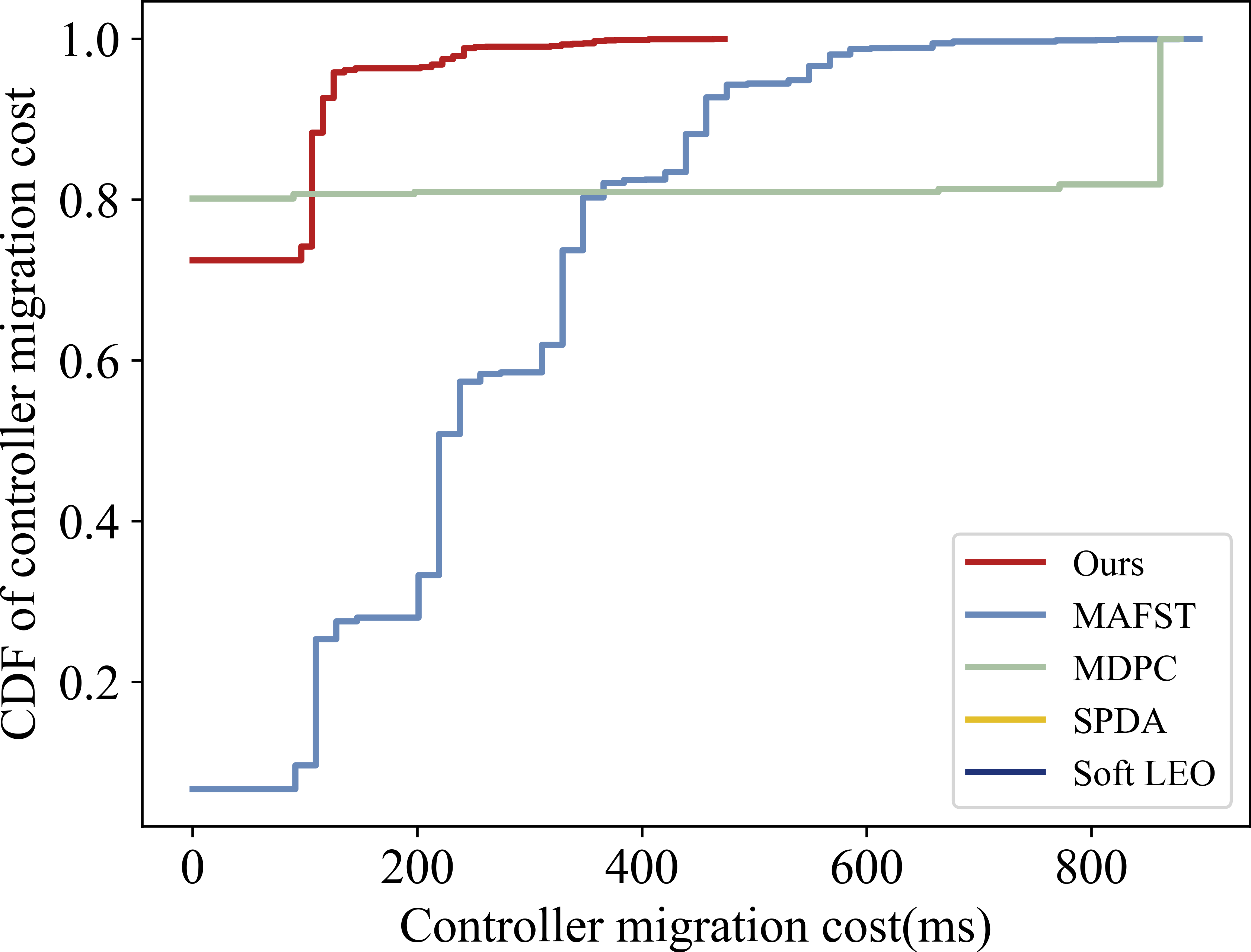}%
    \label{fig_2_case}}
        \hfil
    \subfloat[]{\includegraphics[width=1.7in]{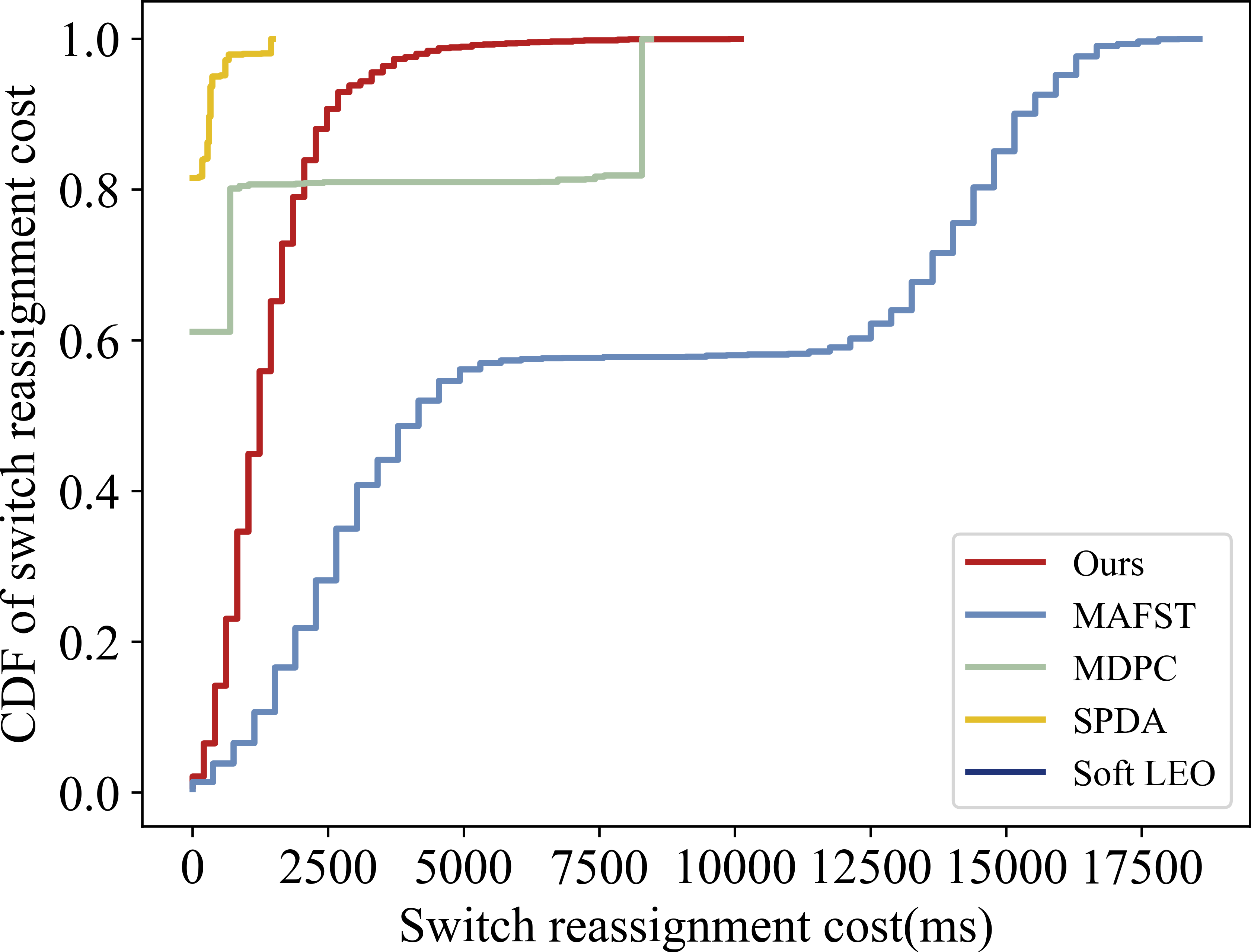}%
    \label{fig_3_case}}
    \hfil
    \subfloat[]{\includegraphics[width=1.7in]{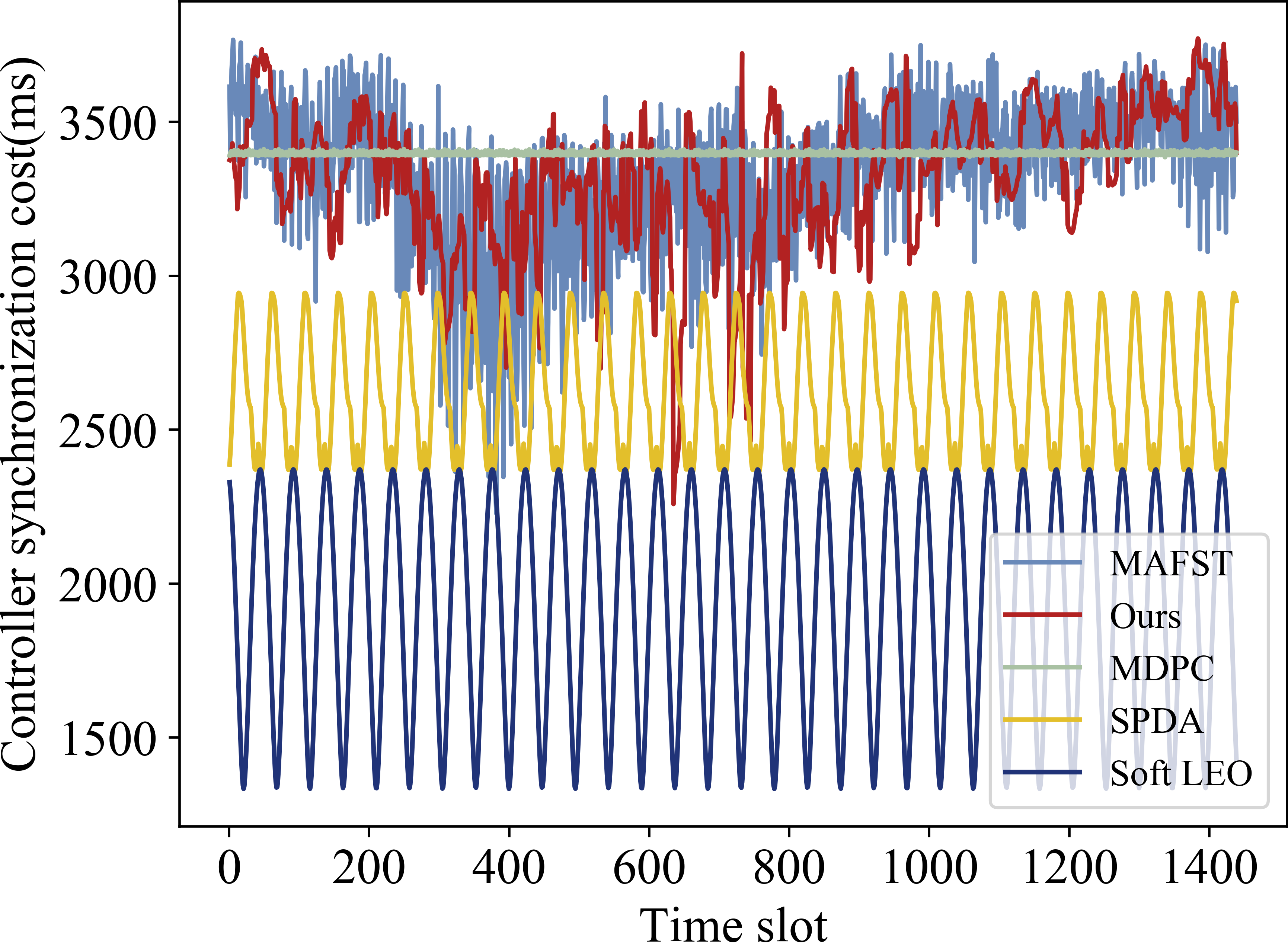}%
    \label{fig_4_case}}
    \hfil
    \subfloat[]{\includegraphics[width=3in]{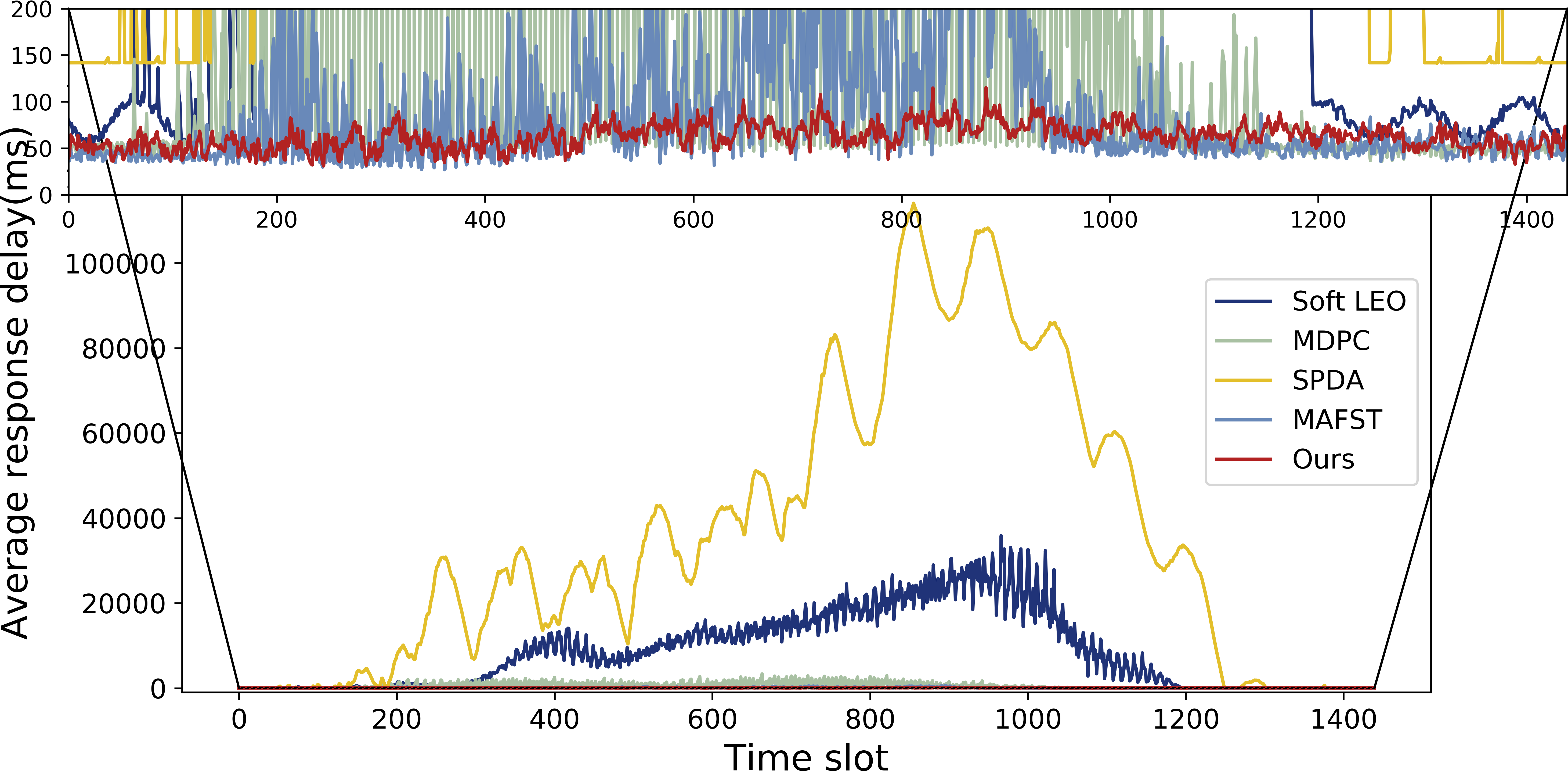}%
    \label{fig_5_case}}
    \caption{Performance and cost under different strategies. (a)Load balance factor. (b)Controller migration cost. (c)Switch reassignment cost. (d)Controller synchronization cost. (e)Average controller response delay.}
    \label{strategies}
    \end{figure}

Since the Soft LEO strategy can only set 8 controllers in this paper's LEO satellite constellation scenario, we set the number of controllers for all algorithms to 8 for fairness.
From Fig.\ref{strategies}\subref{fig_1_case}, we can observe that our algorithm is the best in load balancing.
In addition, combined with Fig.\ref{req_num}, it is clear that the increase in the total number of requests leads to the bigger load balancing factor.
Fig.\ref{strategies}\subref{fig_2_case} shows the cumulative distribution function of controller migration cost under different strategies.
Since Soft LEO and SPDA belong to static placement strategies and do not migrate controllers, their controller migration costs are always 0.
Among the other three dynamic placement strategies, the controller migration cost of our algorithm is the smallest.
Subsequently, Fig.\ref{strategies}\subref{fig_3_case} shows the cumulative distribution function of switch reassignment cost under different strategies.
Soft LEO is a static assignment strategy. 
Thus, its switch reassignment cost is always 0.
We can see from the other four dynamic assignment strategies that SPDA outperforms the other three DPDA mechanism algorithms in switch reassignment.
This is due to the fact that no controller migration occurs in SPDA. 
In contrast, once a controller is migrated under the other three strategies, the switches originally managed by that controller have to be reassigned.
It is evident from the figure that our proposed algorithm can generate less reassignment cost than MAFST.
Although compared with MDPC, our algorithm has no obvious advantage from the figure, after calculation, the total reassignment cost of all time slots of our algorithm is $2.18 \times 10^6$ms, which is better than $2.54\times 10^6$ms of MDPC.
Fig.\ref{strategies}\subref{fig_4_case} depicts the variation of controller synchronization cost with time.
As we can see, the controller synchronization costs of Soft LEO and SPDA change periodically due to the static placement.
While in the Soft LEO strategy, there is a direct inter-satellite link (ISL) between two controllers in adjacent orbits, and any two controllers only exchange information through one or more ISLs between orbits, so the synchronization cost of Soft LEO is the minimum.
The remaining three dynamic placement strategies do not differ much, and our algorithm yields a total synchronization cost of $4.78\times 10^6$ms, while MDPC and MAFST generate $4.89\times 10^6$ms and $4.81\times 10^6$ms synchronization costs, respectively.
Finally, Fig.\ref{strategies}\subref{fig_5_case} shows the variation of the average response delay with time.
Our algorithm can get the most petite average response delay in the most of the time. 
In addition, the average response delay obtained by our algorithm is between 30ms-115ms over the whole period, which has the best stability compared to other strategies.
MAFST is second only to our algorithm. 
It can also achieve good results, but it does not perform well in terms of delay jitter.
The other three strategies cause excessive queuing delay at the controllers' interfaces when the number of requests surges, resulting in unacceptable response delay. 

In summary, our proposed method achieves the optimal load balancing among satellites, and has the least controller migration and switch reassignment cost in dynamic strategies.
Besides, it also has satisfied performance in the controller synchronization cost and the average response delay compared with other methods.
Therefore, our proposed algorithm has flexibility to adapt to the dynamic LEO satellite network applications.
The obtained strategy is exploited as the underlying technology to support the research of upper level technologies in SDN-based LEO satellite networks (e.g., routing and mobility management).

\subsection{Performance with Different Number of Controllers.}
\begin{figure}[h]
    \centering
    \subfloat[]{\includegraphics[width=1.7in]{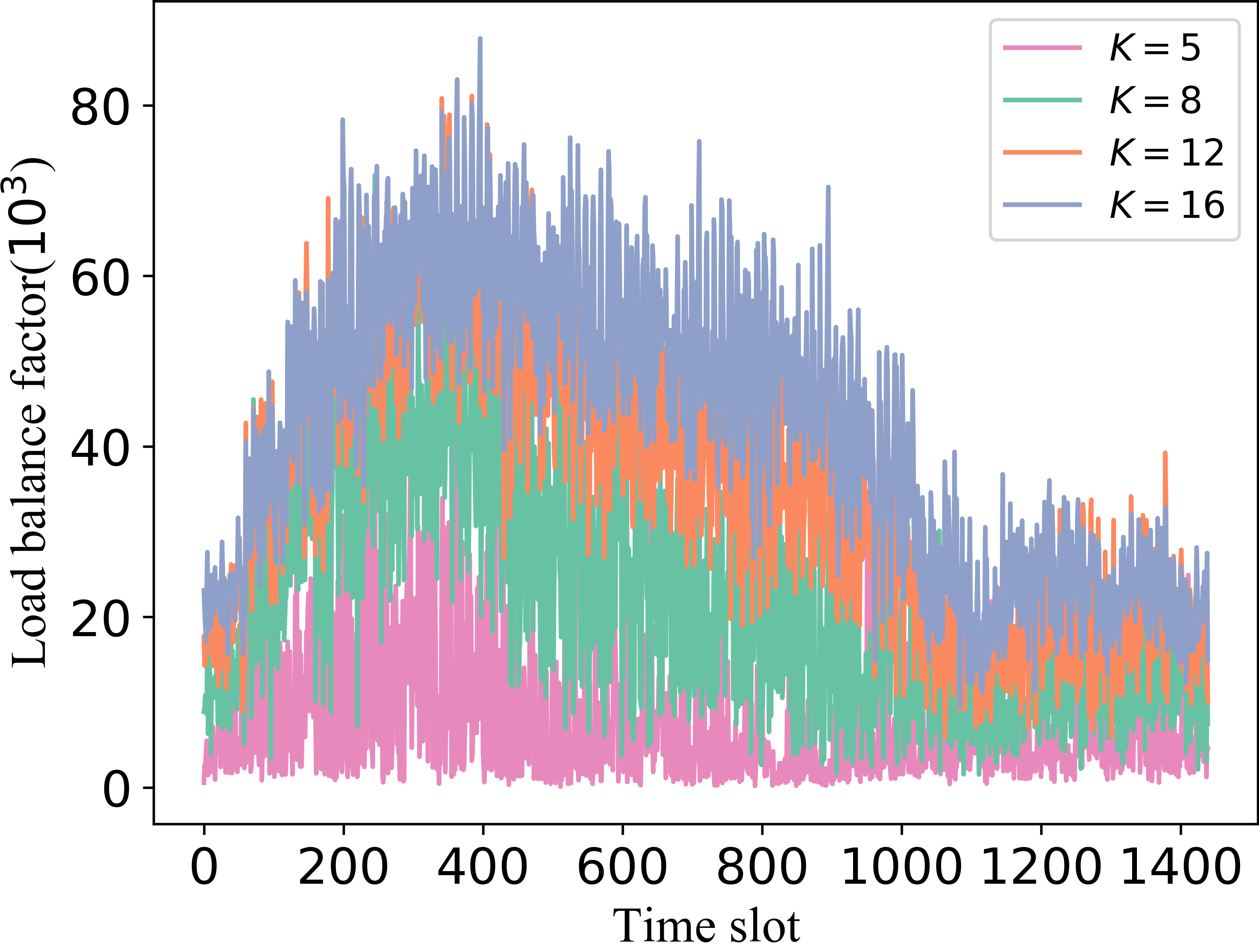}%
    \label{fig_first_case}}
    \hfil
    \subfloat[]{\includegraphics[width=1.7in]{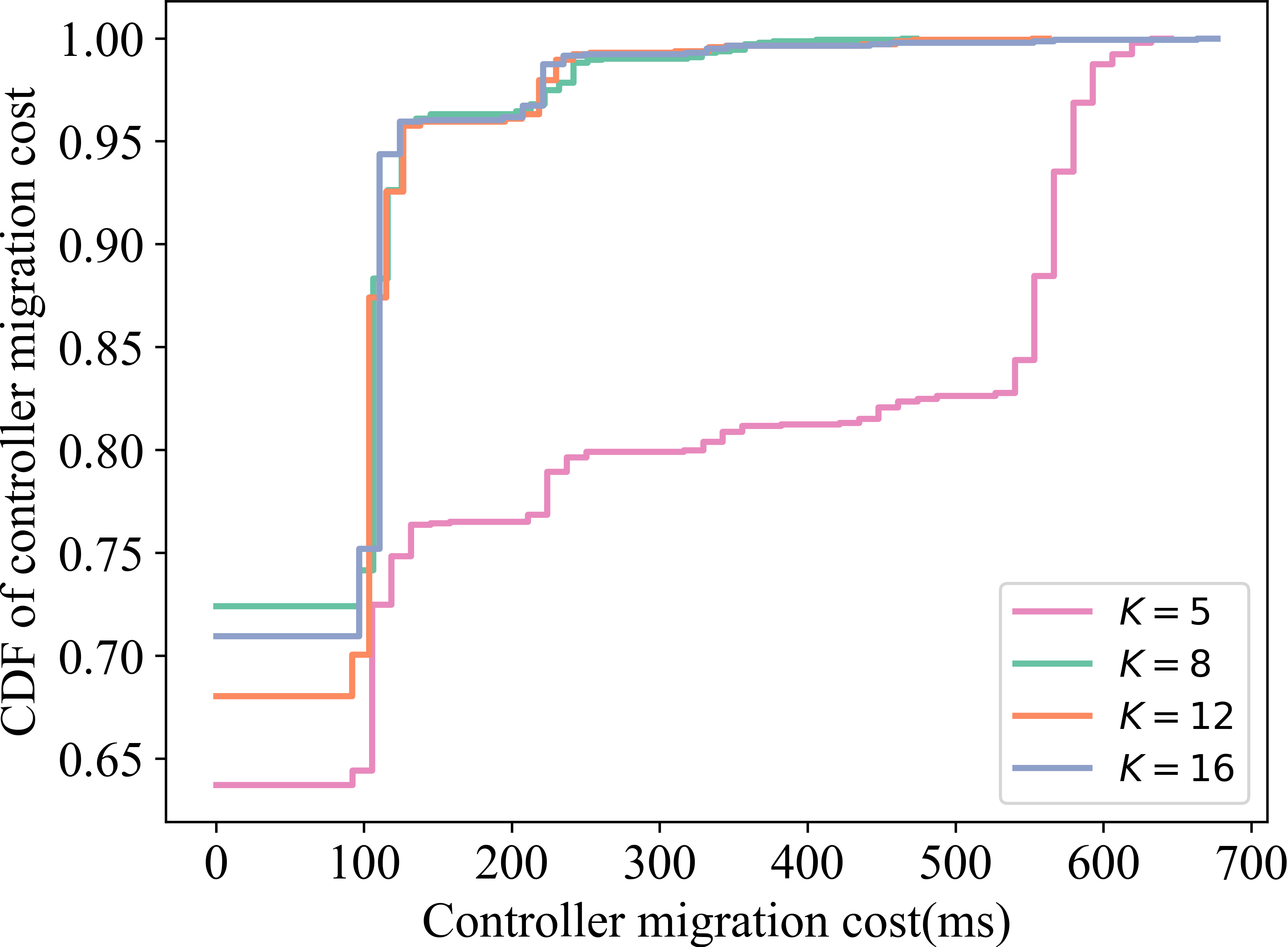}%
    \label{fig_second_case}}
        \hfil
    \subfloat[]{\includegraphics[width=1.7in]{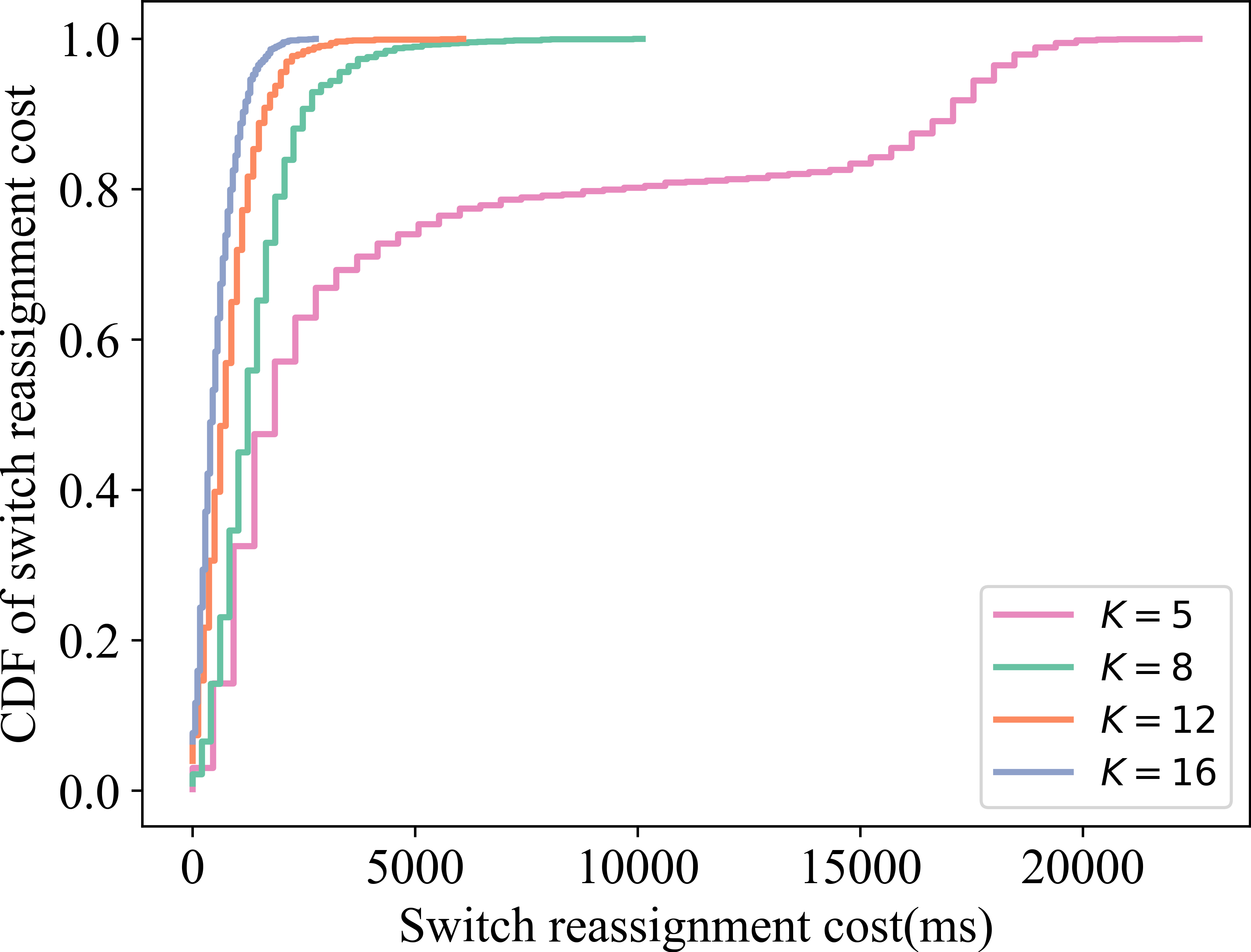}%
    \label{fig_third_case}}
    \hfil
    \subfloat[]{\includegraphics[width=1.7in]{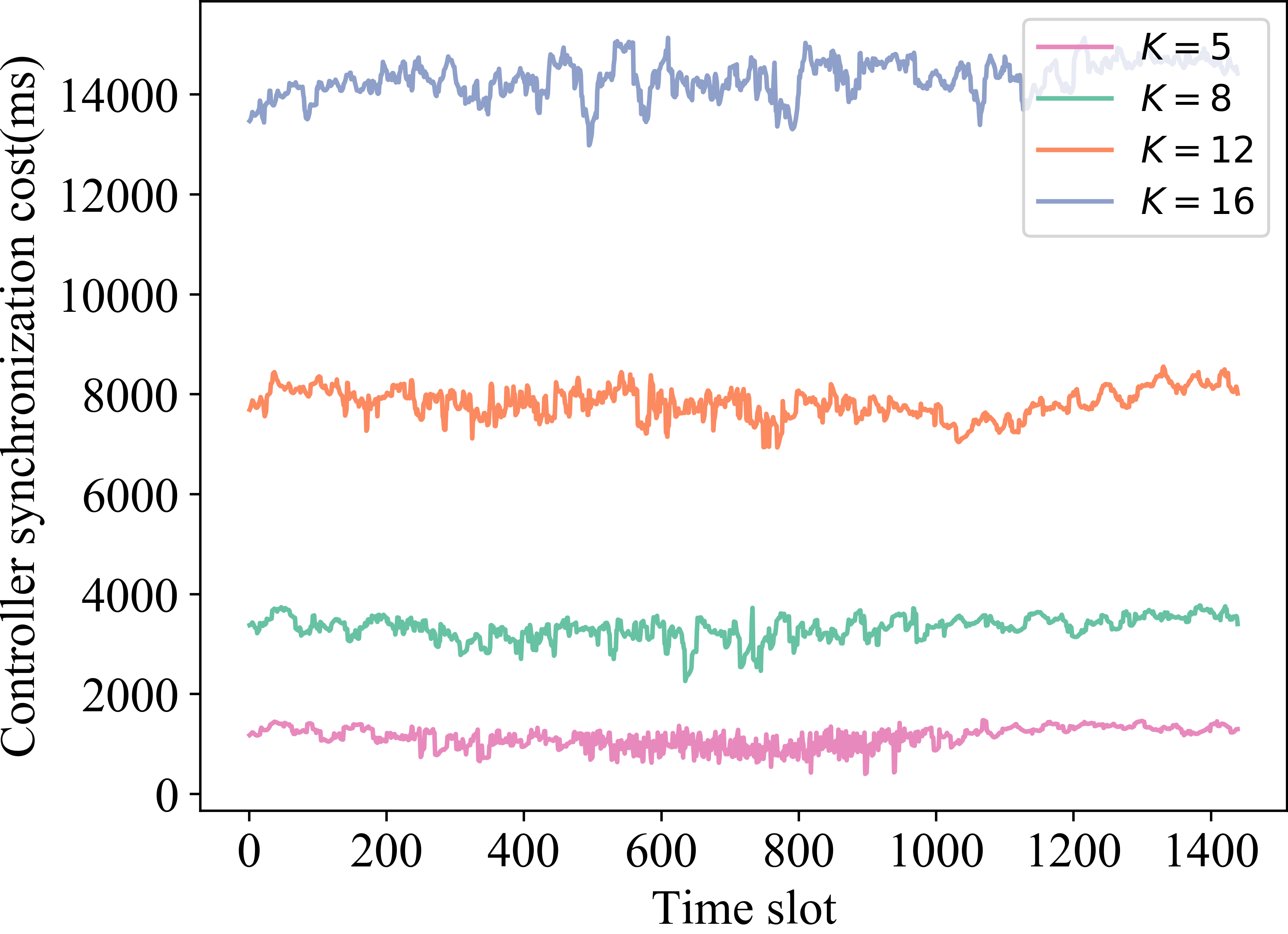}%
    \label{fig_forth_case}}
    \hfil
    \subfloat[]{\includegraphics[width=2in]{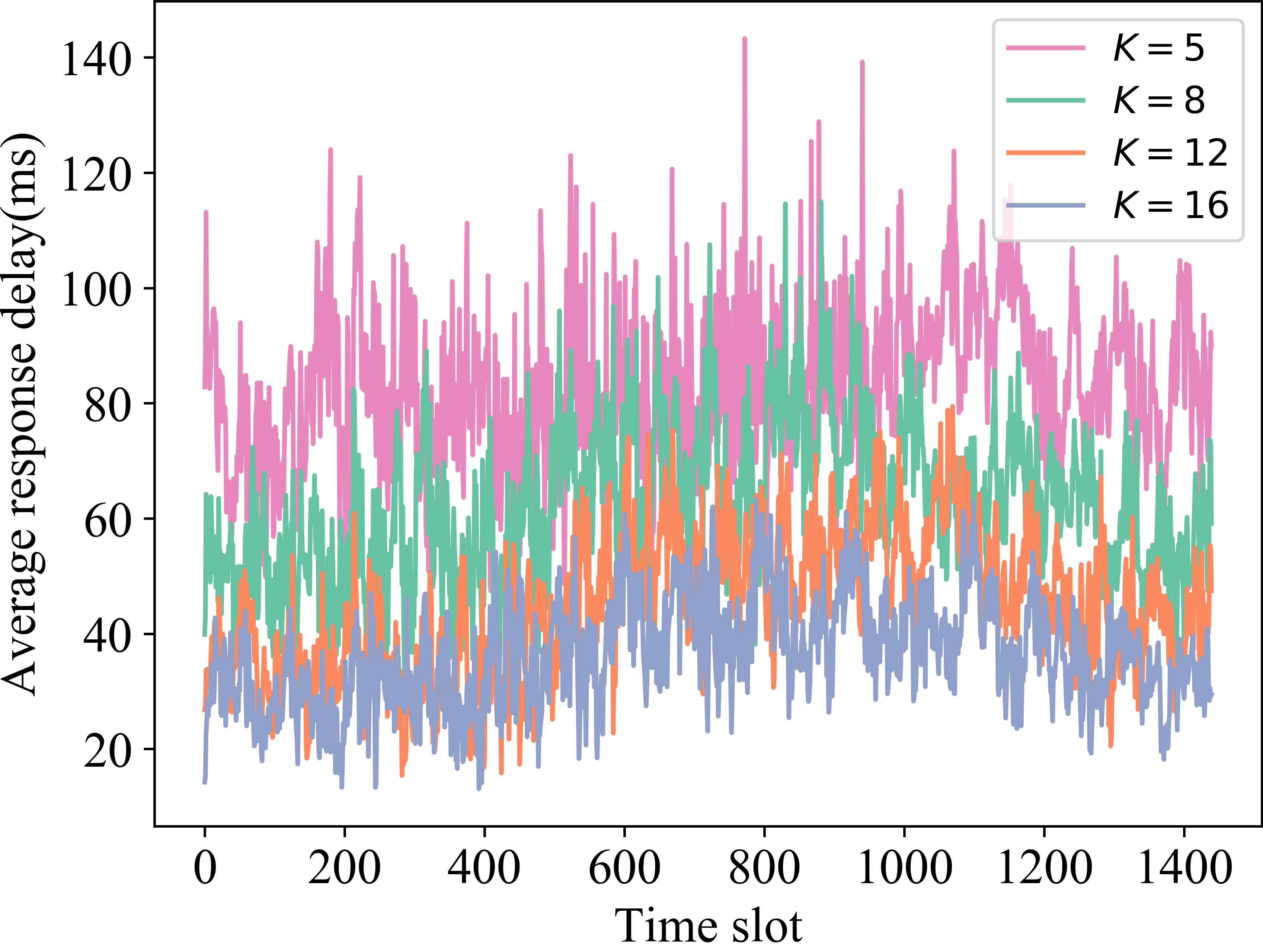}%
    \label{fig_fifth_case}}
    \caption{Performance with different number of controllers. (a)Load balance factor. (b)Controller migration cost. (c)Switch reassignment cost. (d)Controller synchronization cost. (e)Average controller response delay.}
    \label{num_c}
    \end{figure}

In this section, we aim to justify the performance under different numbers of controllers.
Fig.\ref{num_c} shows the results of four experiments with different numbers of controllers,
The objective function weights of these experiments are $w_1=0.001,w_2=1,w_3=0.002$.
From Fig.\ref{num_c}\subref{fig_first_case}, we can observe that with the increase of the number of controllers, the load balance factor becomes larger, which means that the increase of controllers makes it difficult to balance the load of the controllers.
Fig.\ref{num_c}\subref{fig_second_case} and Fig.\ref{num_c}\subref{fig_third_case} are the cumulative distribution functions of controller migration cost and switch reassignment cost, respectively.
As for the controller migration cost, we find that an enormous controller migration cost is incurred when the number of controllers is 5. 
It is due to the fact that these controllers' processing capacity is insufficient to process all the requests at some time slots, and controller migration needs to occur frequently to avoid excessive queuing delay.
In contrast, when the number of controllers is large enough to have sufficient processing power (i.e., all requests for each time slot can be processed at the current time slot), the controller migration cost incurred at different controller numbers does not differ significantly.
In Fig.\ref{num_c}\subref{fig_third_case}, it can be observed that the occurrence of switch reassignment decreases with the increase of controllers.
Fig.\ref{num_c}\subref{fig_forth_case} shows the variation of controller synchronization delay with time under different numbers of controllers.
It is clear that the increase of controllers heightens the controller synchronization cost, which conforms to the definition and calculation model of controller synchronization cost.
As for average response delay, we can observe from Fig.\ref{num_c}\subref{fig_fifth_case} that it decreases with adding the number of controllers. 
\subsection{Performance with Different Weight Settings.}
In this section, we focus on the impact of each weight on performance. 
In each experiment, the number of controllers is 8.
The cost of controller synchronization for all experiments is analyzed uniformly in the last part.

\begin{figure}[h]
  \centering
  \subfloat[]{\includegraphics[width=1.7in]{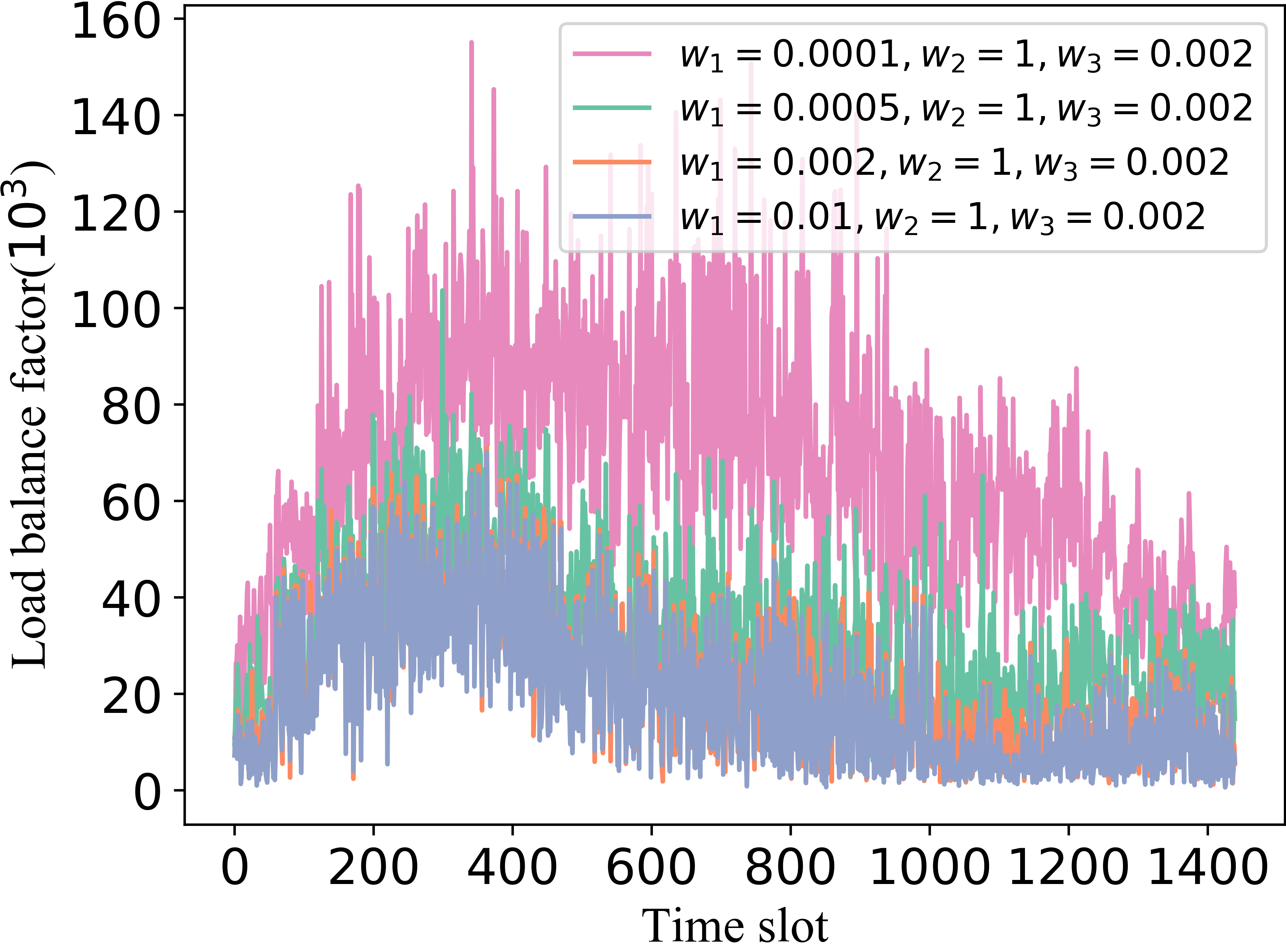}%
  \label{1}}
  \hfil
  \subfloat[]{\includegraphics[width=1.7in]{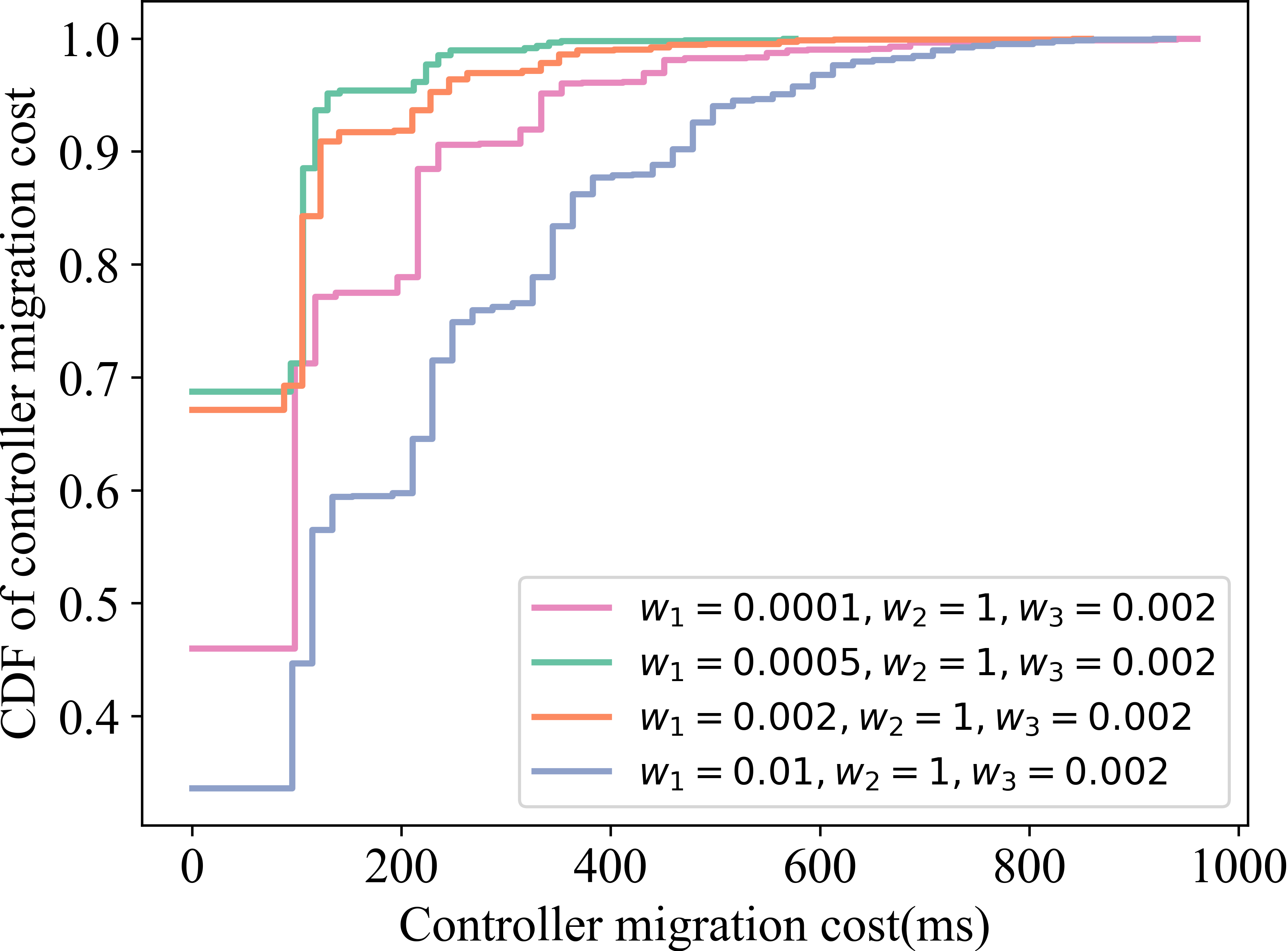}%
  \label{2}}
      \hfil
  \subfloat[]{\includegraphics[width=1.7in]{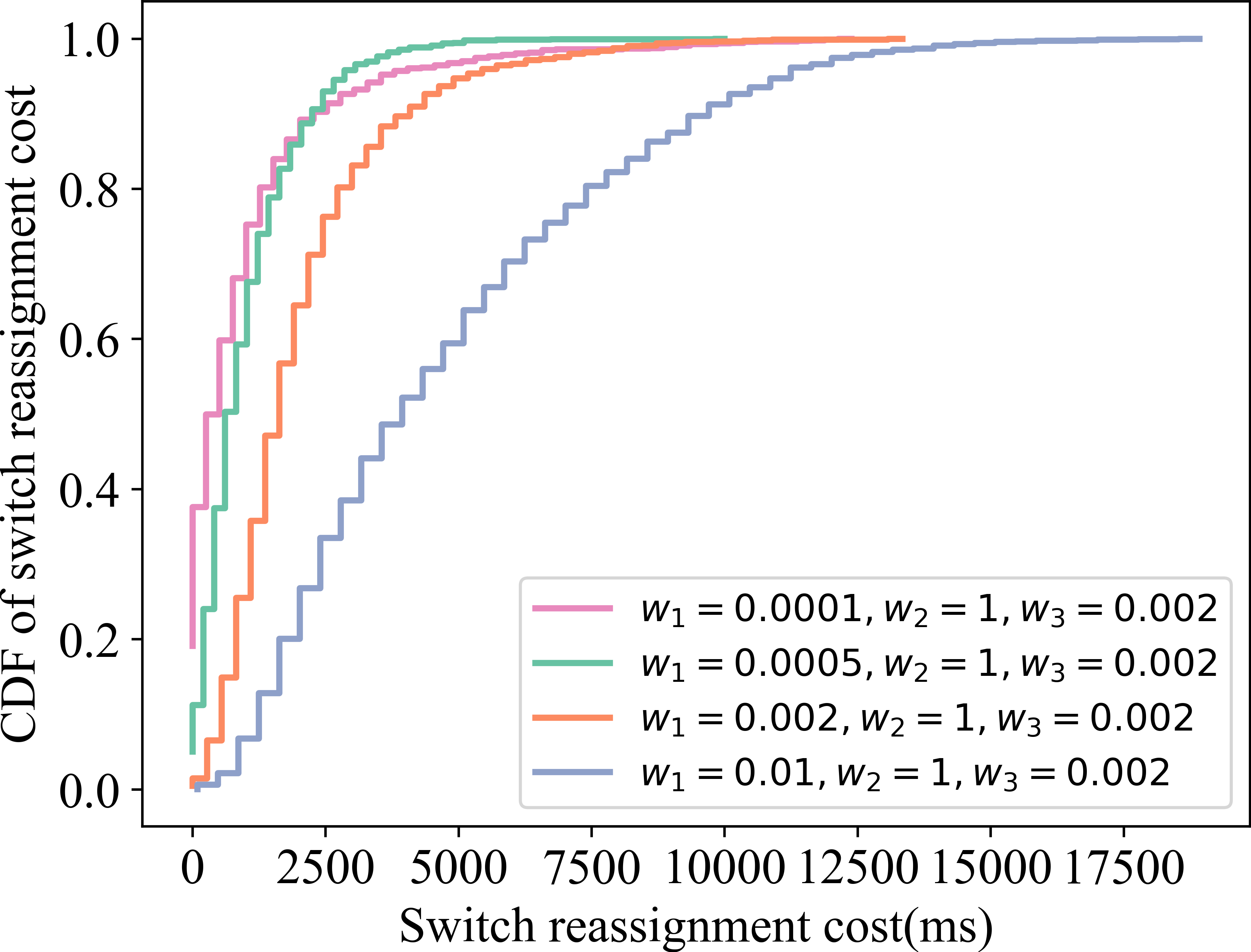}%
  \label{3}}
  \hfil
  \subfloat[]{\includegraphics[width=1.7in]{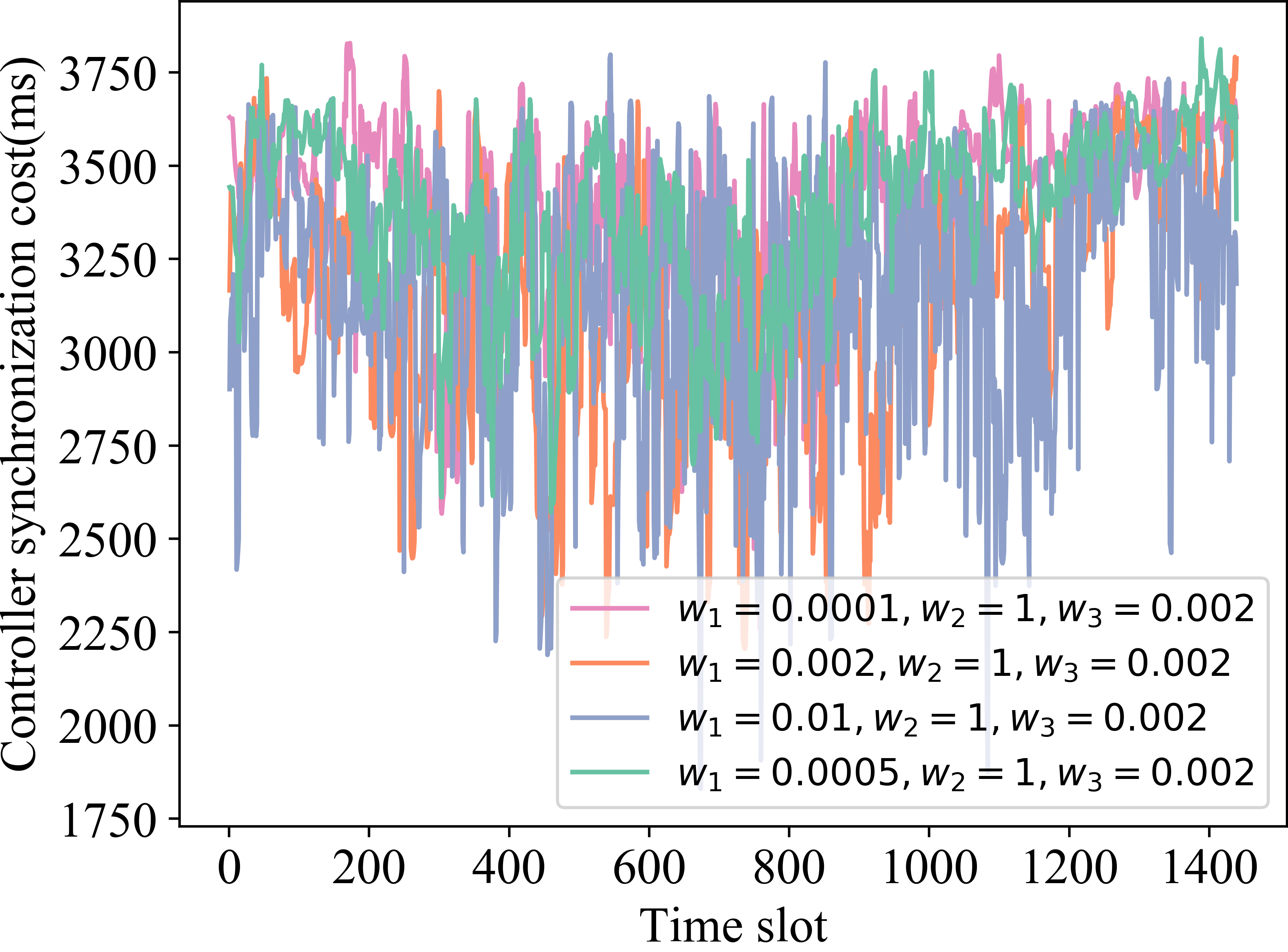}%
  \label{4}}
  \hfil
  \subfloat[]{\includegraphics[width=2in]{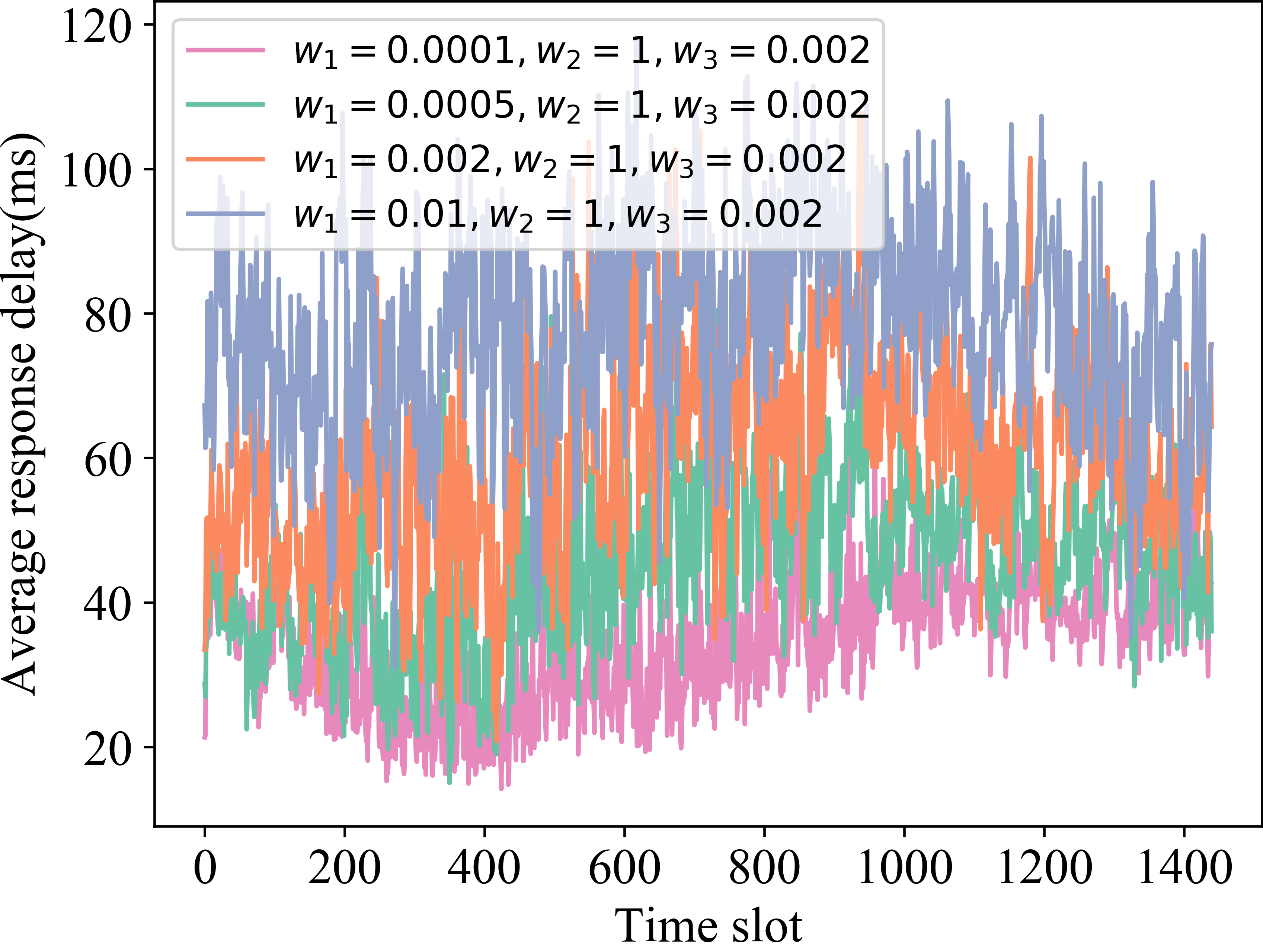}%
  \label{5}}
  \caption{Performance with different $w_1$. (a)Load balance factor. (b)Controller migration cost. (c)Switch reassignment cost. (d)Controller synchronization cost. (e)Average controller response delay.}
  \label{weight111}
  \end{figure}

\emph{1)Performance with different $w_1$:}
First, we set $w_2$ and $w_3$ to fixed values of 1 and 0.002, respectively, to explore the effect of $w_1$ on performance.
It can be observed from Fig.\ref{weight111}\subref{1} that the greater the value of $w_1$, the better the optimization effect of the load balance factor. 
Fig.\ref{weight111}\subref{3} and Fig.\ref{weight111}\subref{5} show that the average controller response delay and switch reassignment cost become worse with the increase of $w_1$.
It indicates that switch reassignment needs to occur more frequently, and the further average controller response delay needs to be sacrificed to improve load balance performance.
In addition, it can be seen from Fig.\ref{weight111}\subref{2} that the controller migration cost does not change regularly with the increase of $w_1$, because it does not conflict with the load balance factor.

\begin{figure}[h]
  \centering
  \subfloat[]{\includegraphics[width=1.7in]{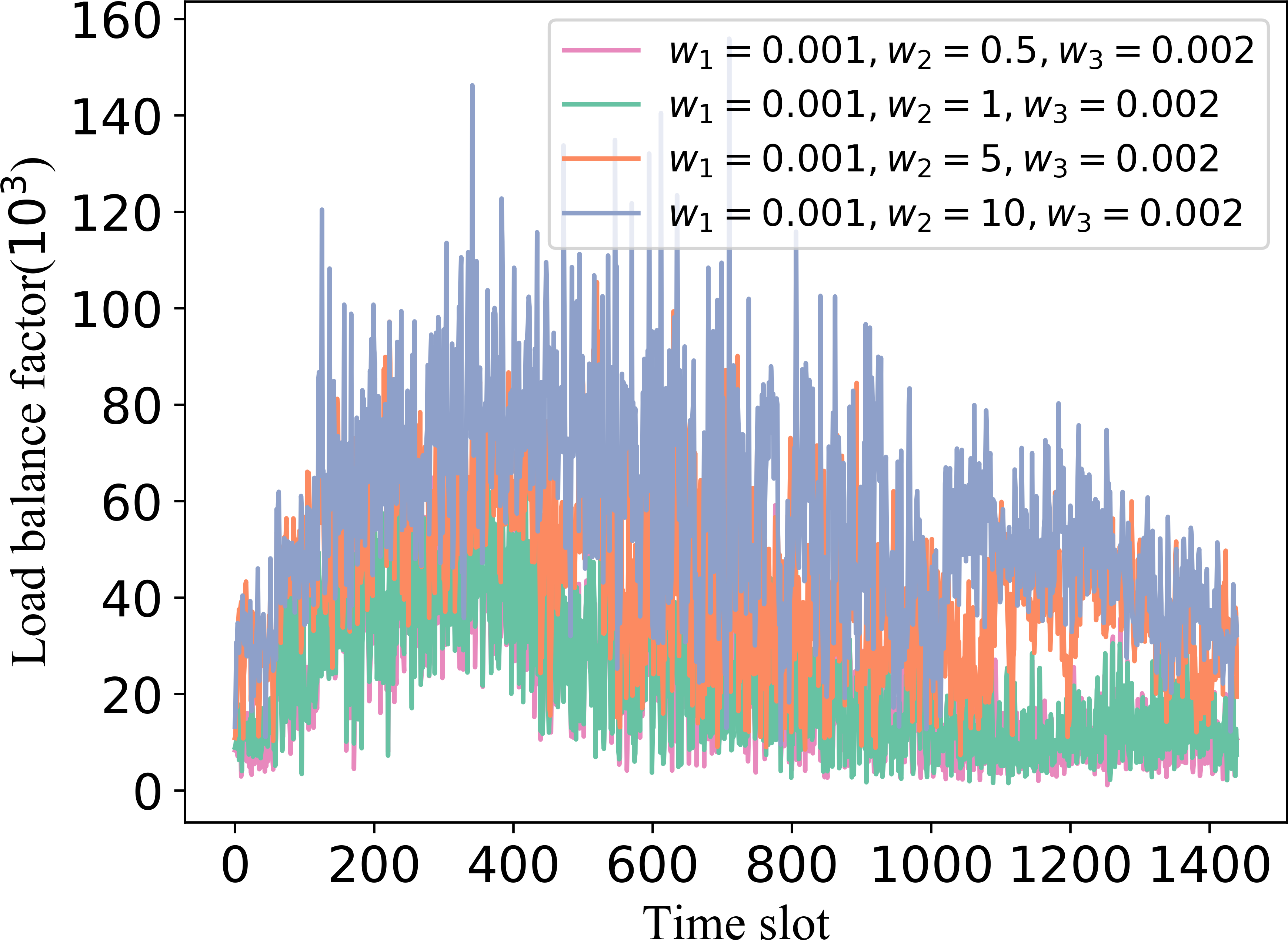}%
  \label{1}}
  \hfil
  \subfloat[]{\includegraphics[width=1.7in]{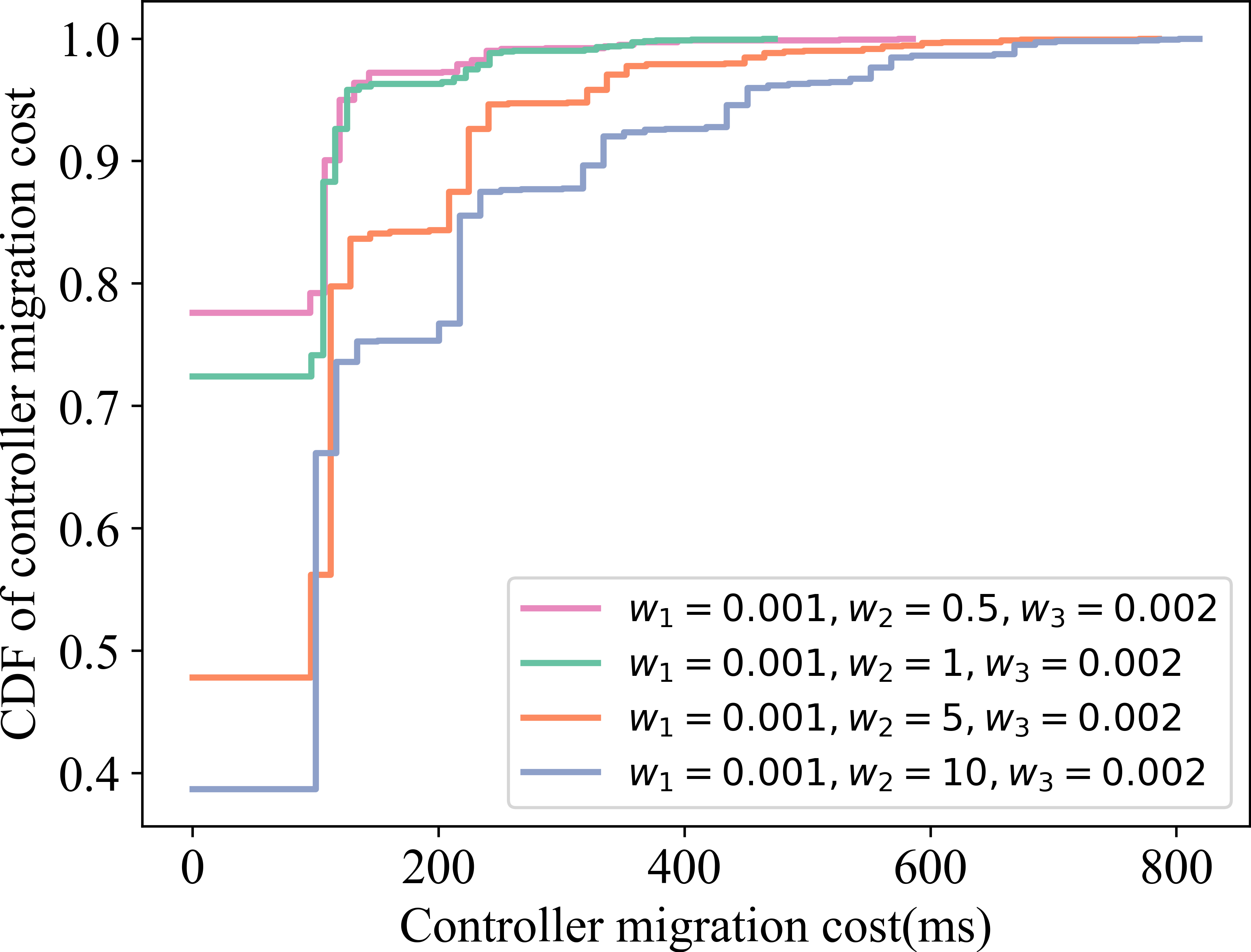}%
  \label{2}}
      \hfil
  \subfloat[]{\includegraphics[width=1.7in]{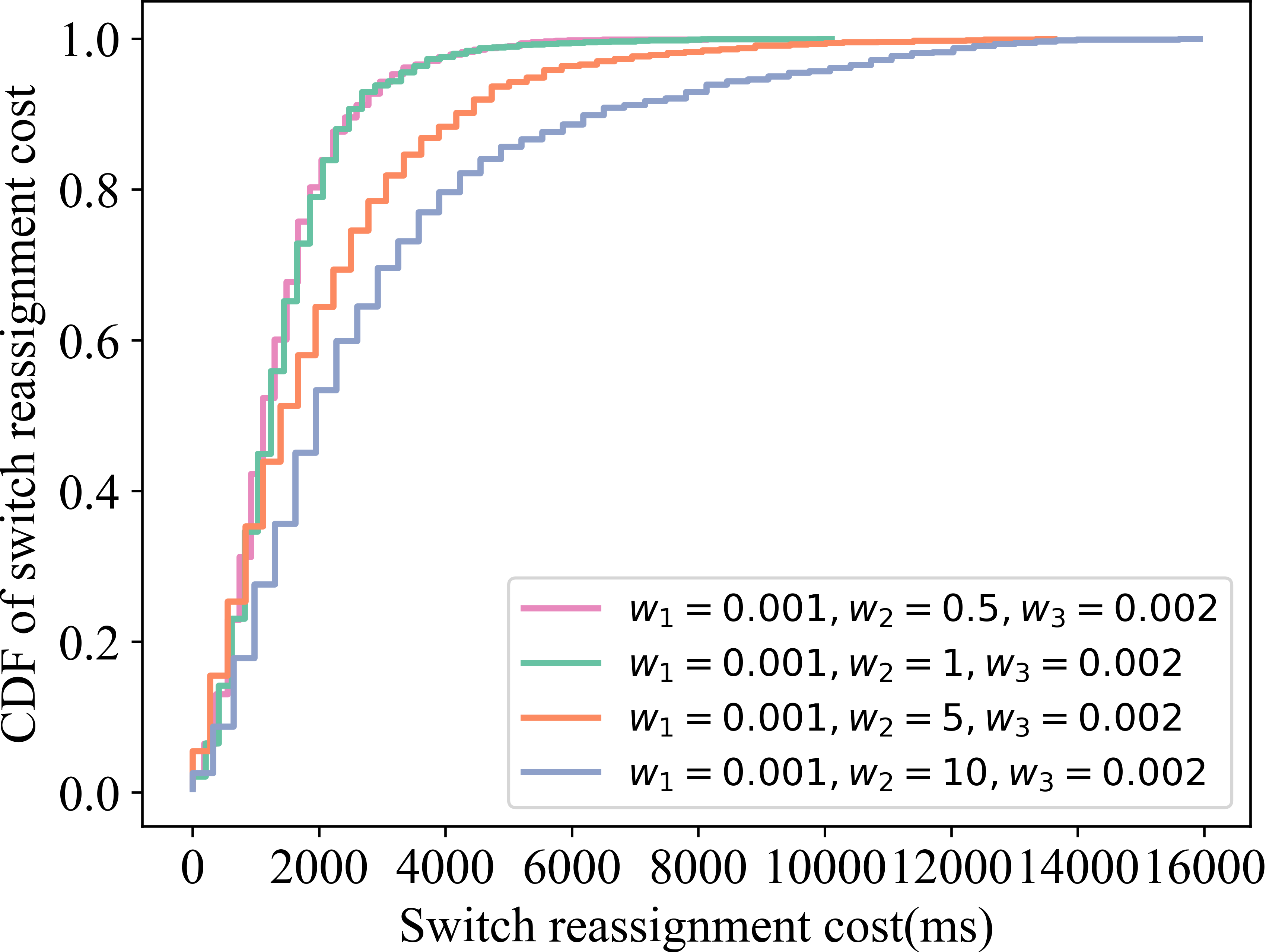}%
  \label{3}}
  \hfil
  \subfloat[]{\includegraphics[width=1.7in]{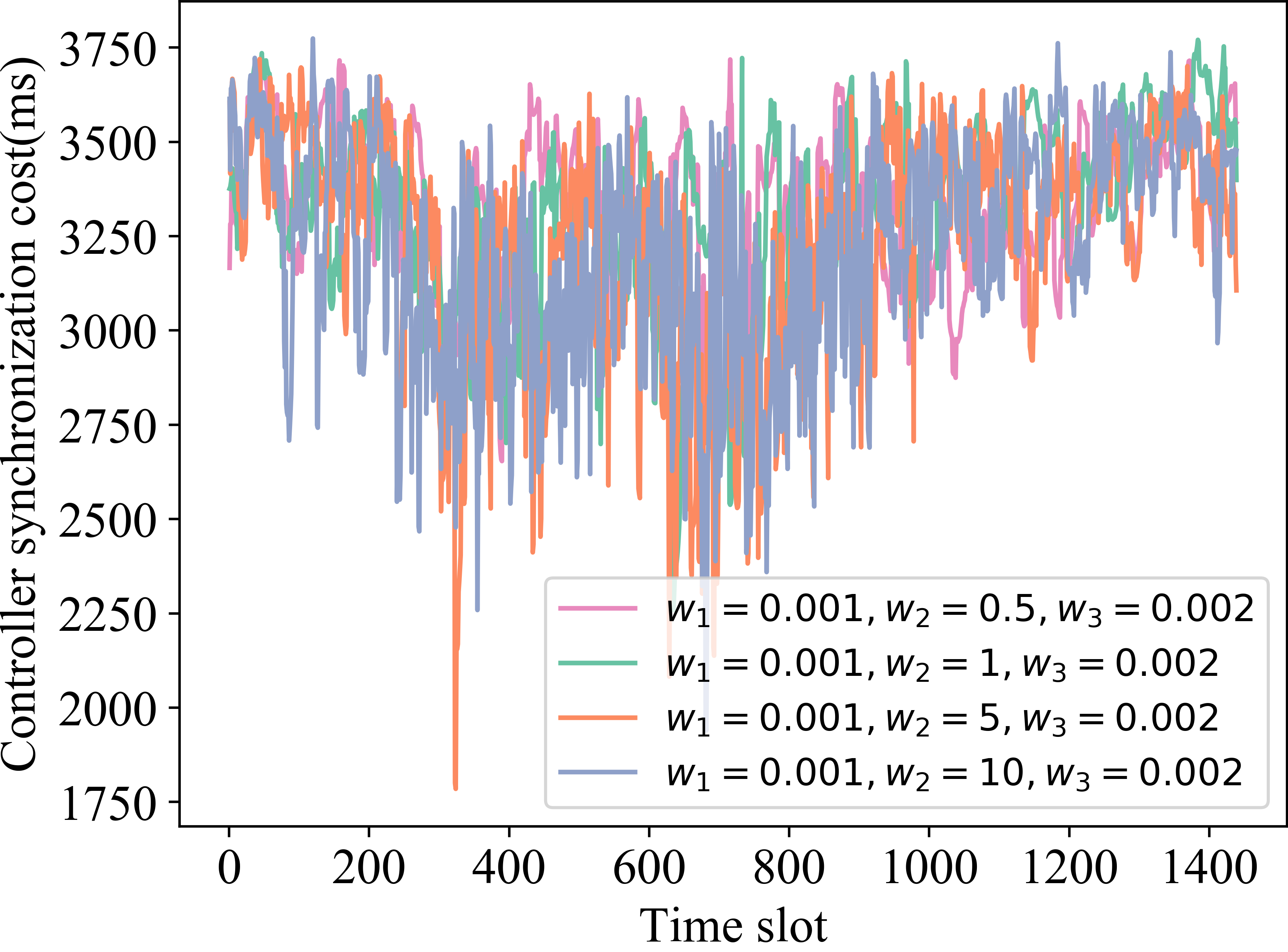}%
  \label{4}}
  \hfil
  \subfloat[]{\includegraphics[width=2in]{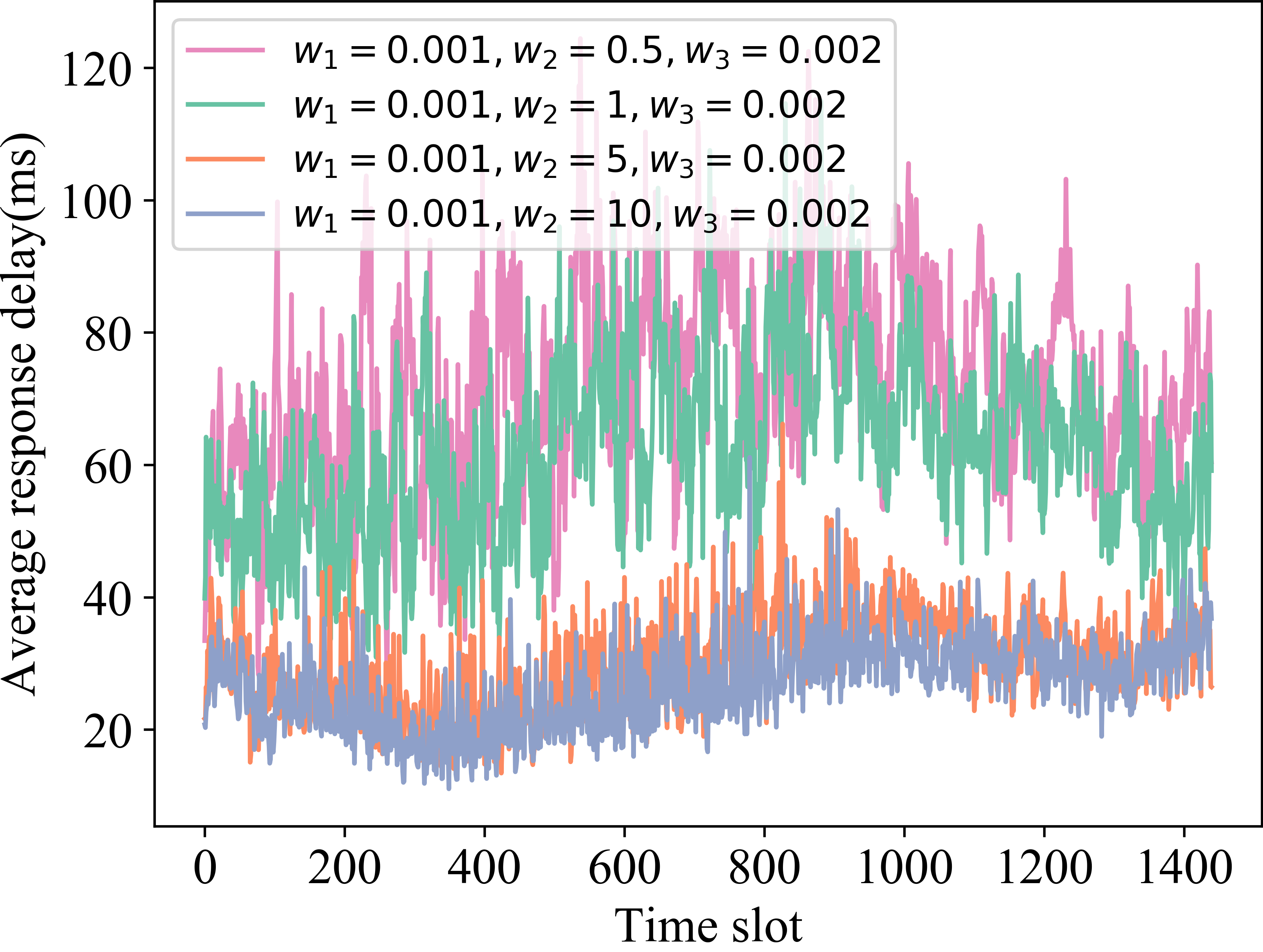}%
  \label{5}}
  \caption{Performance with different $w_2$. (a)Load balance factor. (b)Controller migration cost. (c)Switch reassignment cost. (d)Controller synchronization cost. (e)Average controller response delay.}
  \label{weight11}
  \end{figure}

\emph{2)Performance with different $w_2$:}
Next, we set $w_1$ and $w_3$ to fixed values of 0.001 and 0.002, respectively, to observe the effect of $w_2$ on performance.
From Fig.\ref{weight11}\subref{5}, we can observe that the growth of $w_2$ shortens the average controller response delay. 
Fig.\ref{weight11}\subref{1}, Fig.\ref{weight11}\subref{2} and Fig.\ref{weight11}\subref{3} show that the increase of $w_2$ leads to the increase of load balance factor, controller migration cost, and switch reassignment cost, respectively.
These results are in line with expectations. 
With the increase of $w_2$, the optimization of the average controller response delay becomes further remarkable.
While, for other performances, since $w_1$ and $w_3$ have been kept at fixed values, these performances are less optimized.

\begin{figure}[h]
    \centering
    \subfloat[]{\includegraphics[width=1.7in]{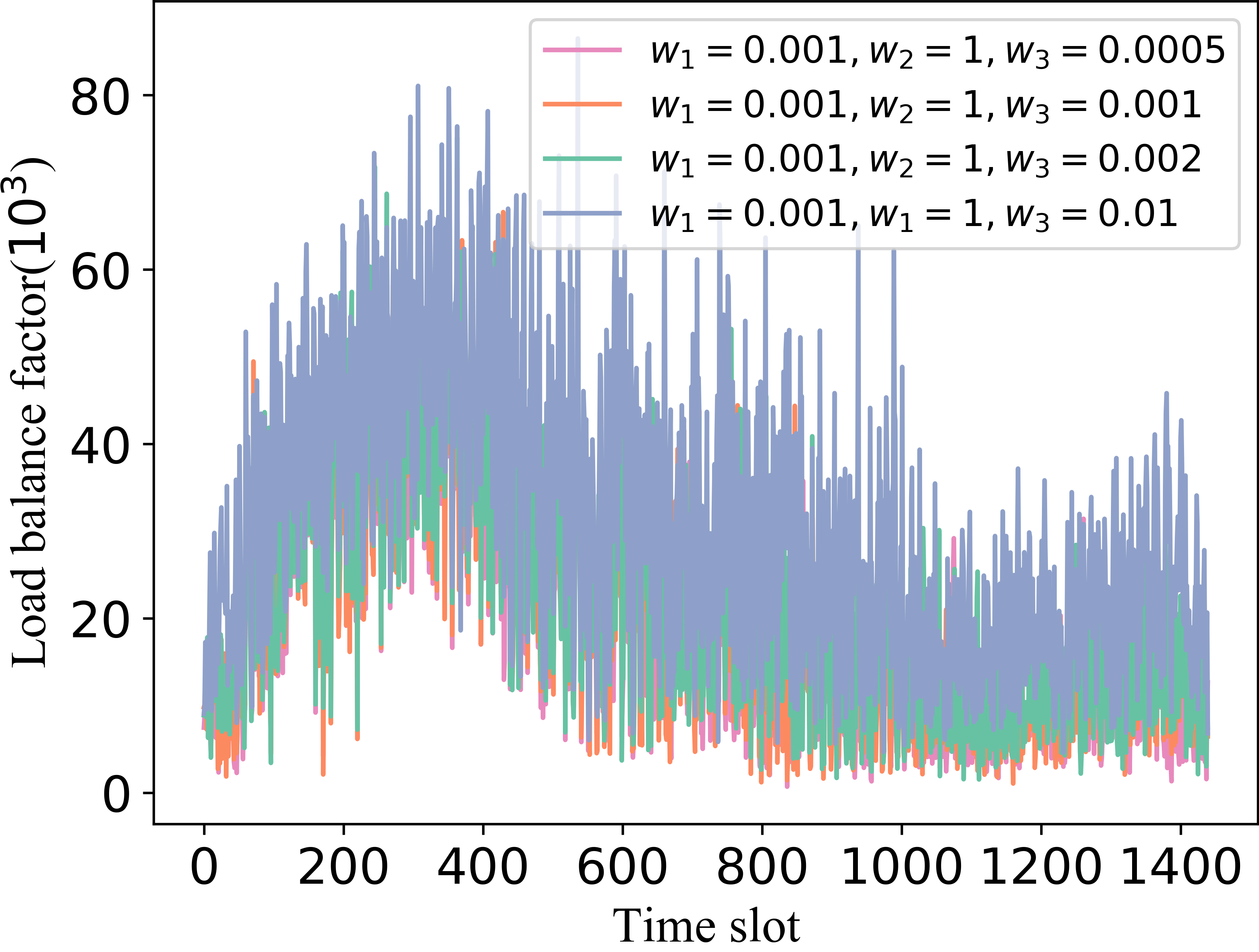}%
    \label{fig_first_case}}
    \hfil
    \subfloat[]{\includegraphics[width=1.7in]{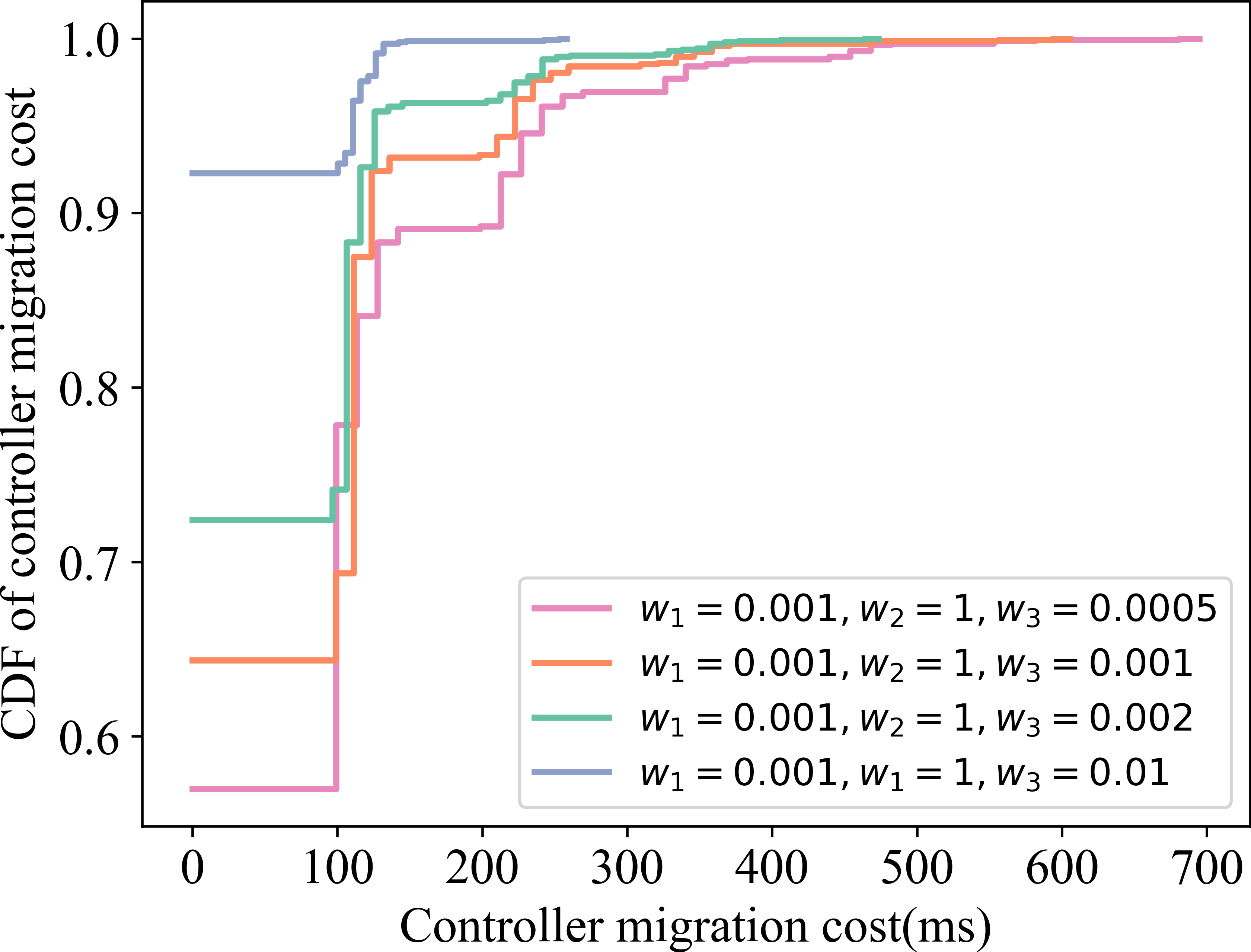}%
    \label{fig_second_case}}
        \hfil
    \subfloat[]{\includegraphics[width=1.7in]{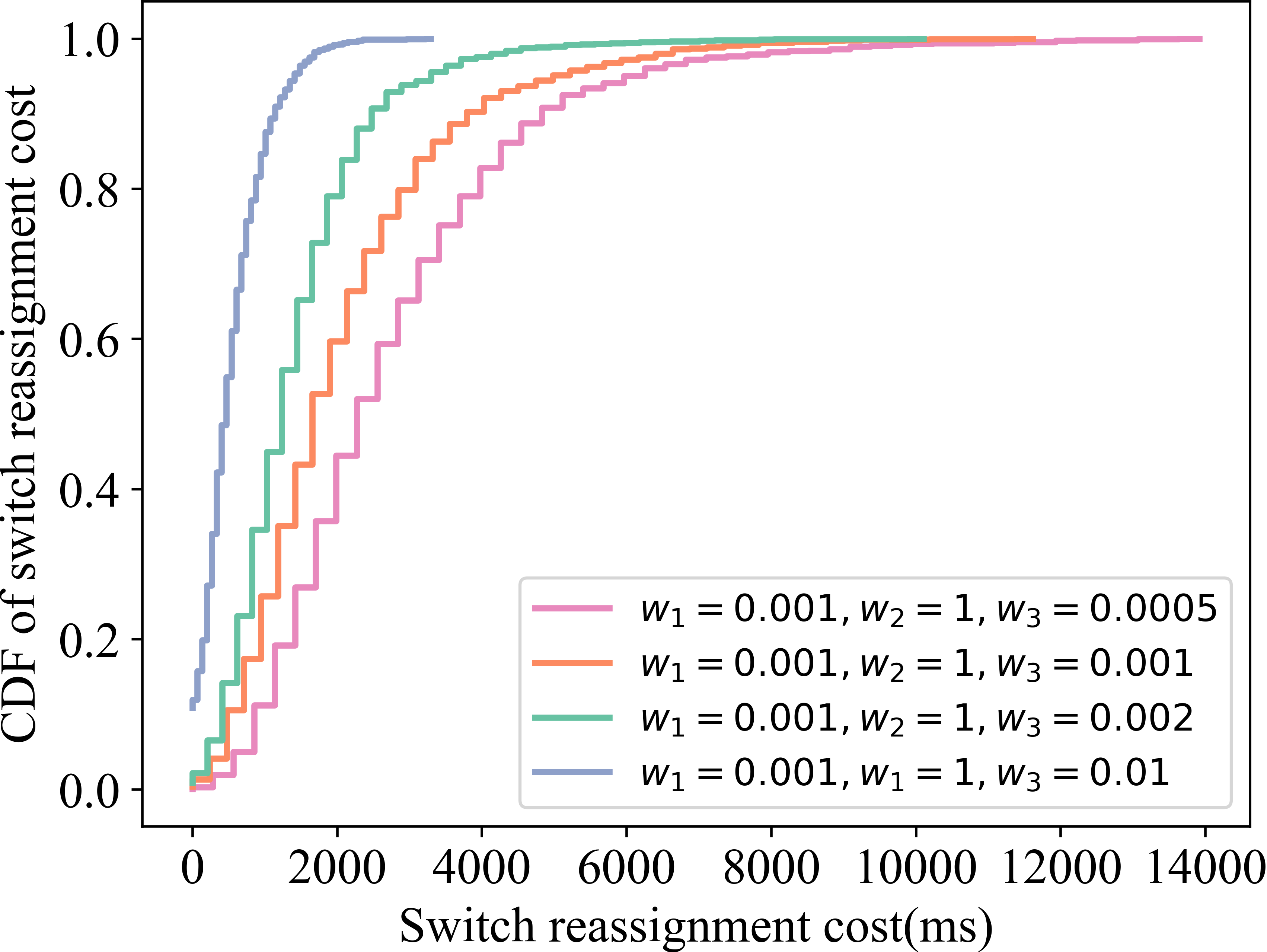}%
    \label{fig_third_case}}
    \hfil
    \subfloat[]{\includegraphics[width=1.7in]{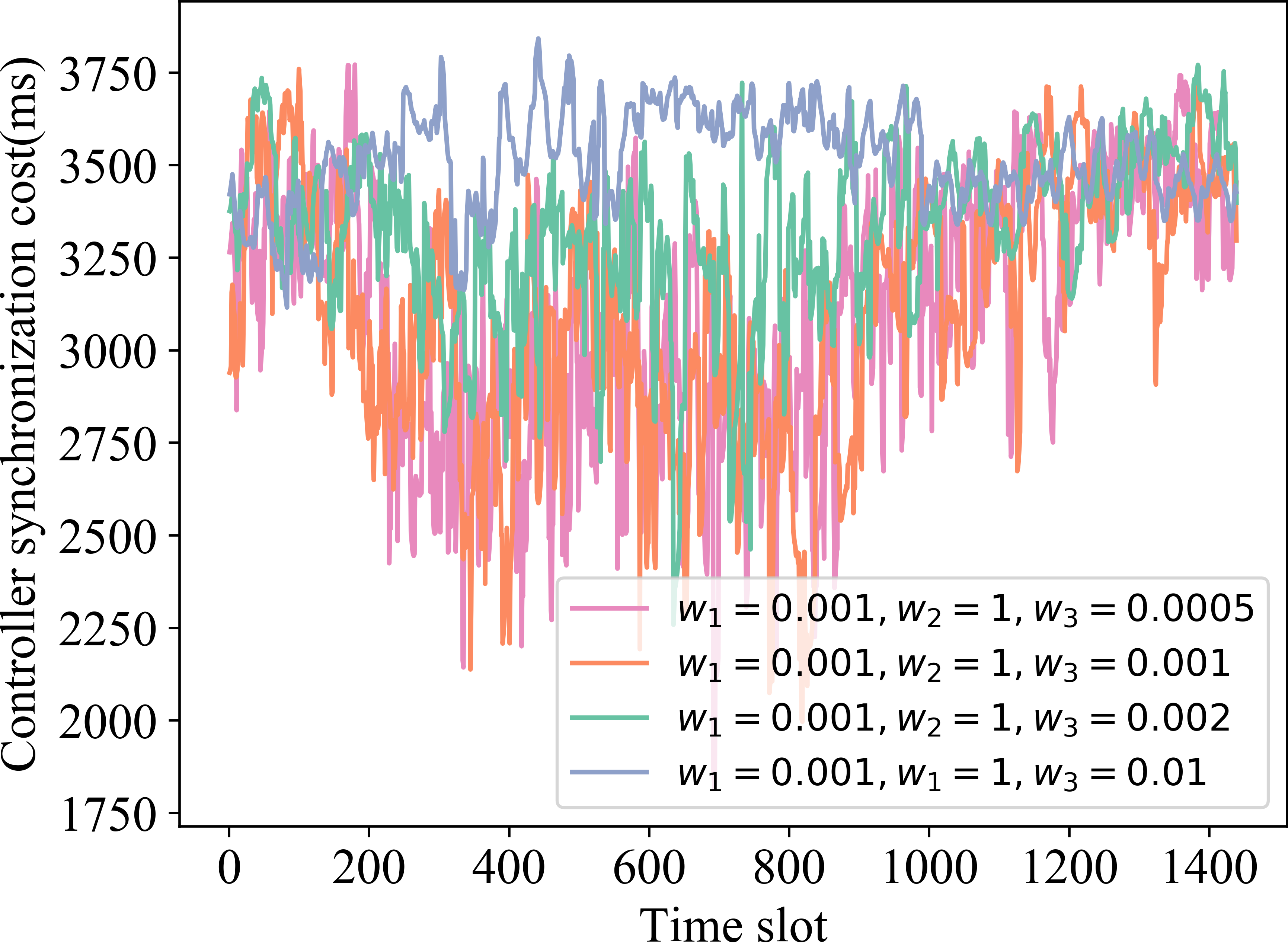}%
    \label{fig_forth_case}}
    \hfil
    \subfloat[]{\includegraphics[width=2in]{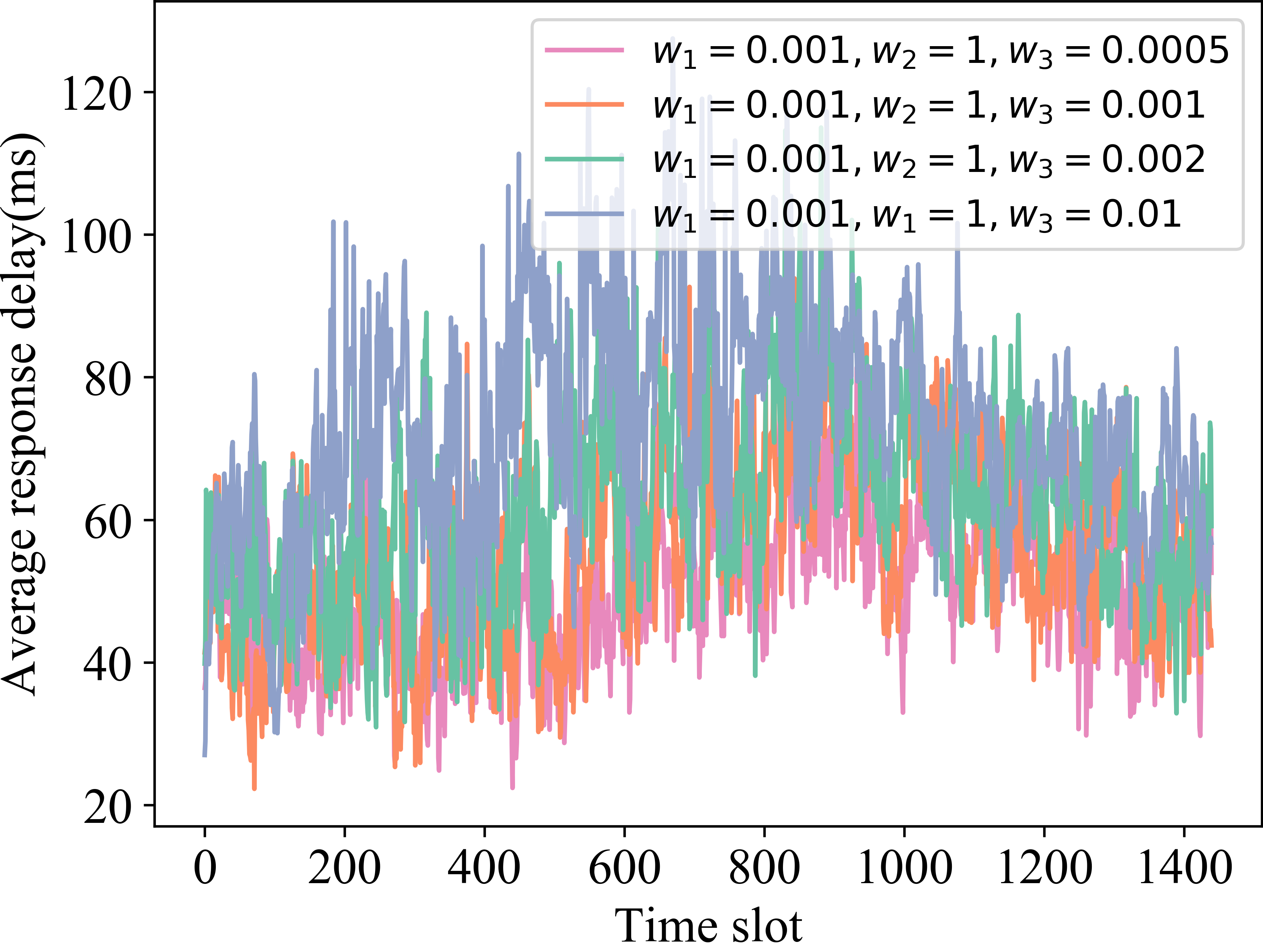}%
    \label{fig_fifth_case}}
    \caption{Performance with different $w_3$. (a)Load balance factor. (b)Controller migration cost. (c)Switch reassignment cost. (d)Controller synchronization cost. (e)Average controller response delay.}
    \label{weight1}
    \end{figure}

\emph{3)Performance with different $w_3$:}
Subsequently, we focus on the impact of $w_3$ on performance, with $w_1$ and $w_2$ set to fixed values of 0.001 and 1, respectively.
From Fig.\ref{weight1}\subref{fig_first_case} and Fig.\ref{weight1}\subref{fig_fifth_case}, we observe that as $w_3$ increases, the load balance factor and the average response delay become larger.
This means that these two performances need to rise to a bigger value to trigger migration and reassignment to prevent this rising trend.
It can be seen from Fig.\ref{weight1}\subref{fig_second_case} and Fig.\ref{weight1}\subref{fig_third_case} that as $w_3$ becomes larger, less controller migration and switch reassignment occurs.

\begin{figure}[h]
    \centering
    \subfloat[]{\includegraphics[width=1.7in]{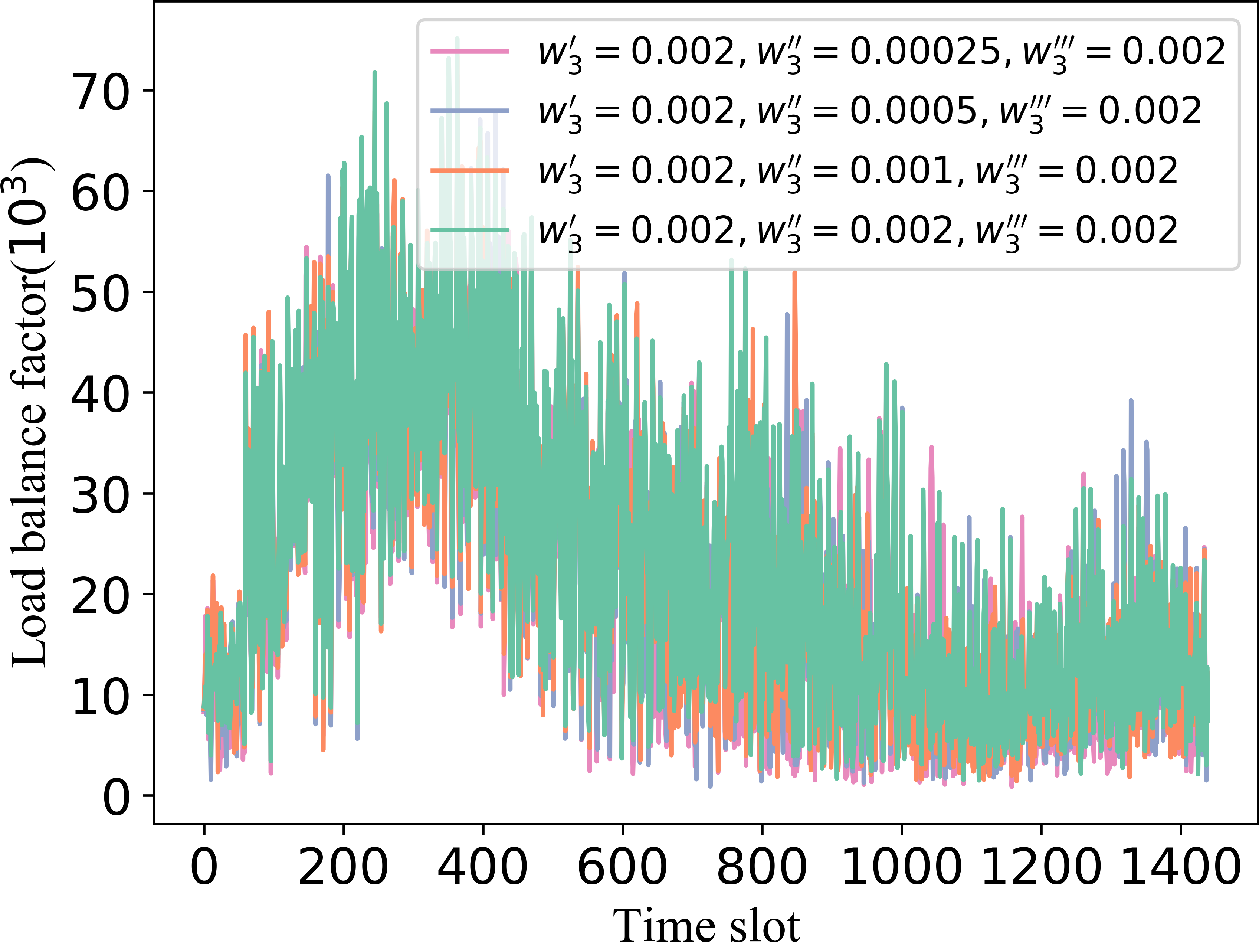}%
    \label{fig_first_case}}
    \hfil
    \subfloat[]{\includegraphics[width=1.7in]{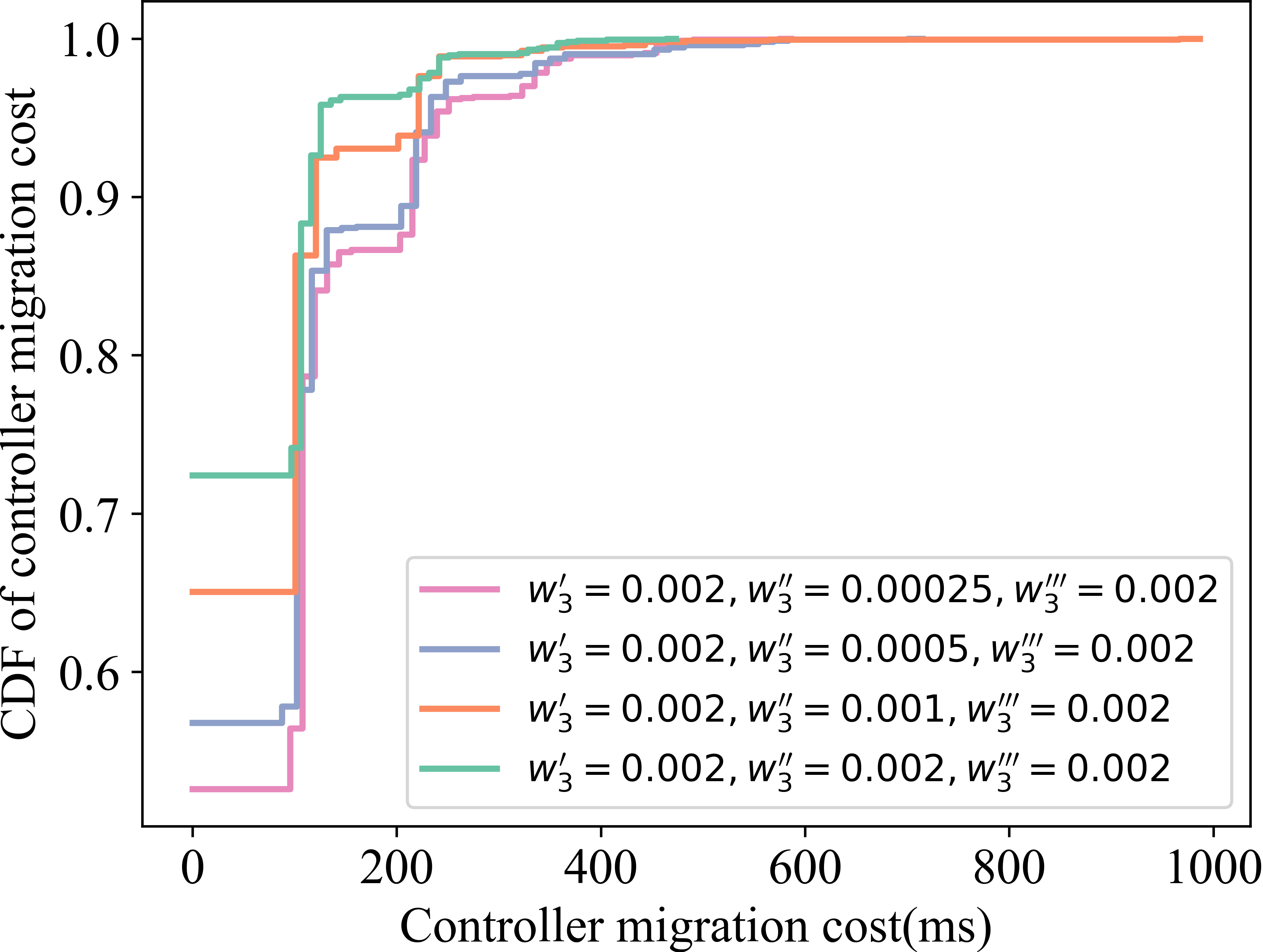}%
    \label{fig_second_case}}
        \hfil
    \subfloat[]{\includegraphics[width=1.7in]{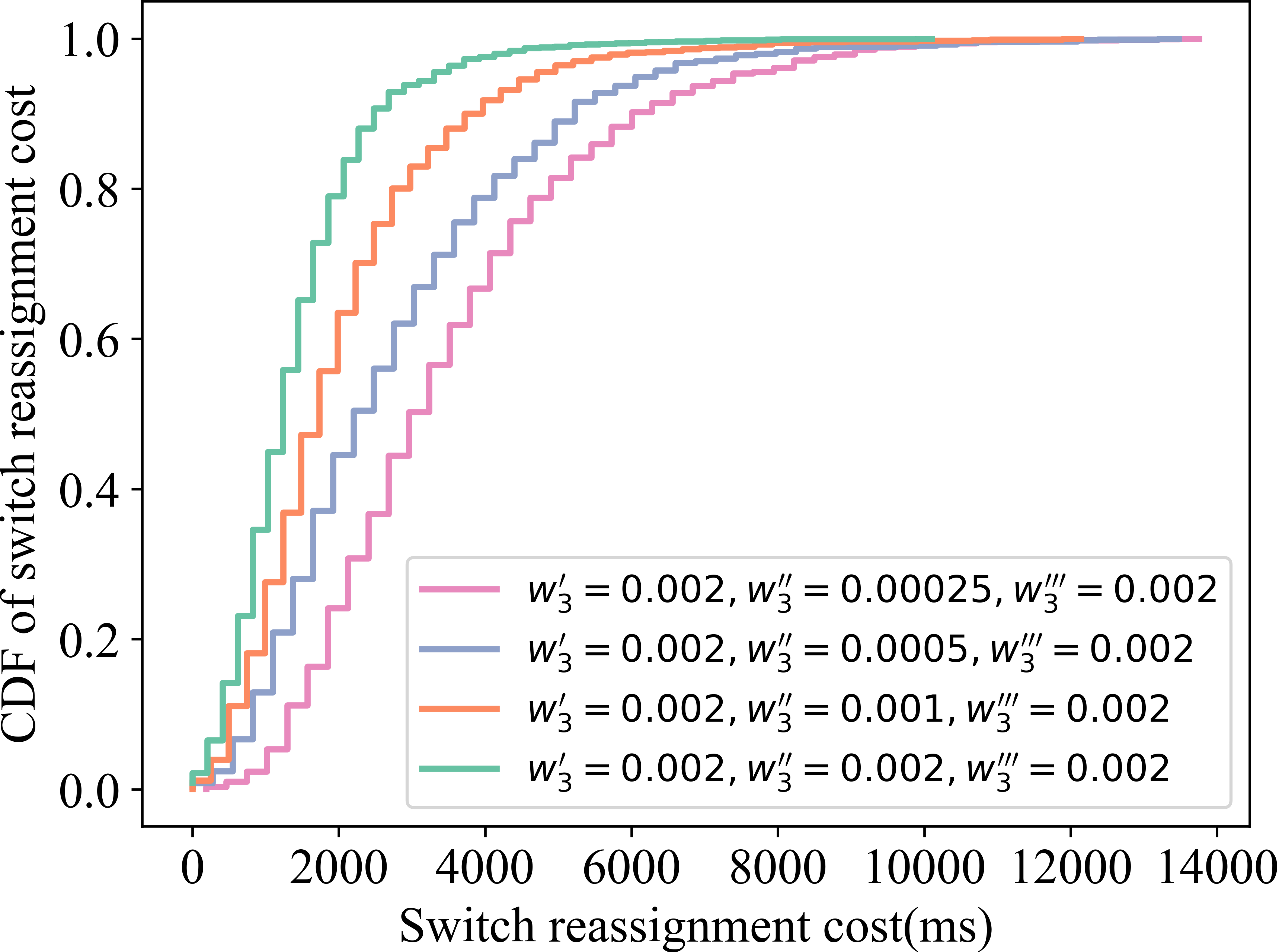}%
    \label{fig_third_case}}
    \hfil
    \subfloat[]{\includegraphics[width=1.7in]{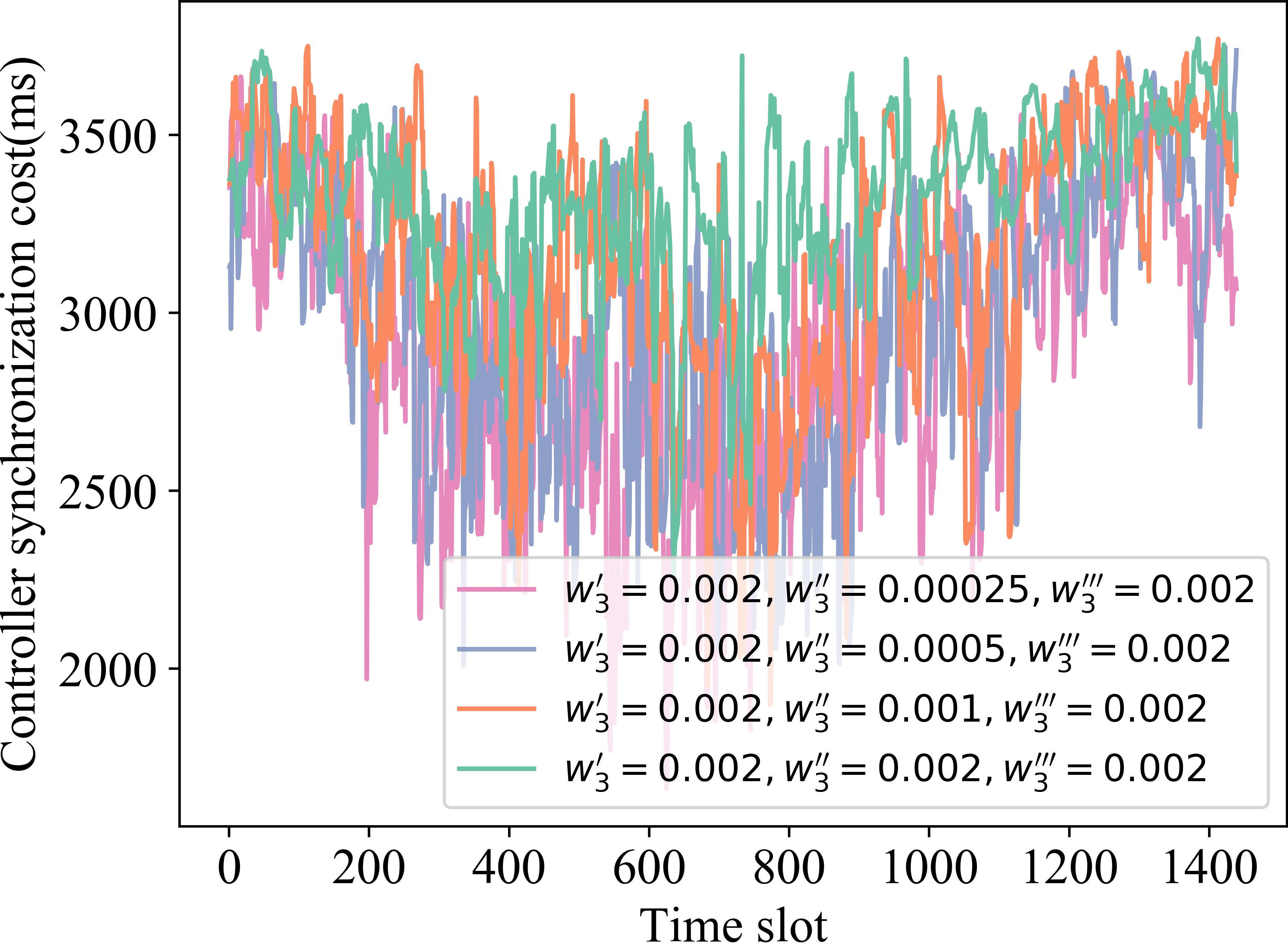}%
    \label{fig_forth_case}}
    \hfil
    \subfloat[]{\includegraphics[width=2in]{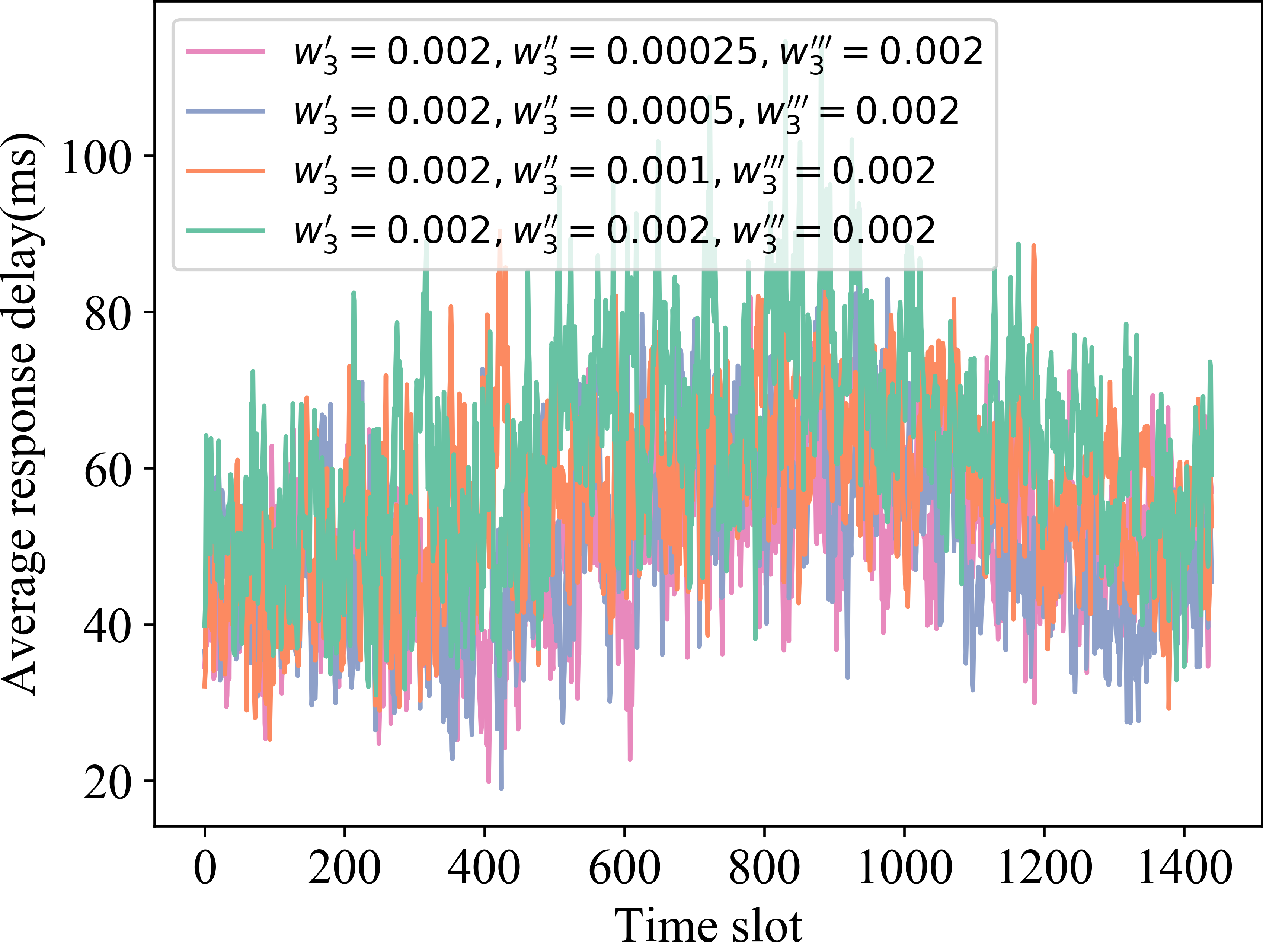}%
    \label{fig_fifth_case}}
    \caption{Performance with different $w_3^{\prime\prime}$. (a)Load balance factor. (b)Controller migration cost. (c)Switch reassignment cost. (d)Controller synchronization cost. (e)Average controller response delay.}
    \label{weight2}
    \end{figure}

\emph{4)Performance with different $w_3^{\prime\prime}$:}
In addition to the controller migration cost, the migration of a controller also results in the reassignment cost of switches within this controller's control domain.
This connection between controller migration cost and switch reassignment cost encourages us to divide $w_3$ into three sub weights, $w_3^{\prime}$, $w_3^{\prime\prime}$, $w_3^{\prime\prime\prime}$, corresponding to migration cost, reassignment cost and synchronization cost respectively, and then study the impact of $w_3^{\prime\prime}$ on performance.
In Fig.\ref{weight2}\subref{fig_second_case} and Fig.\ref{weight2}\subref{fig_third_case}, it is clear that as $w_3^{\prime\prime}$ increases, less controller migration cost and switch reassignment cost are incurred. 
This indicates that the smaller the $w_3^{\prime\prime}$, the more likely not only switch reassignment but also controller migration occurs.
Fig.\ref{weight2}\subref{fig_first_case} and Fig.\ref{weight2}\subref{fig_fifth_case} show the changes of load balance factor and average controller response delay with time, respectively. 
According to statistics, with the increase of $w_3^{\prime\prime}$, the total load balance factors are $3.0091\times 10^7$, $3.1391\times 10^7$, $3.1537\times 10^7$, and $3.2870\times 10^7$ in turn, and the total average controller response delay(ms) is $7.2252\times 10^4$, $7.2869\times 10^4$, $8.0884\times 10^4$, and $8.9790\times 10^4$ in turn. 
It indicates that the increase of $w_3^{\prime\prime}$ has a deterioration effect on these two performances.

\emph{5)Factors affecting the cost of controller synchronization:}
As shown in Fig.\ref{weight1}\subref{fig_forth_case}, the increase of $w_3$ leads to the increase of the controller synchronization cost, which urges us to analyze the impact of the weight setting on the controller synchronization cost.
Fig.\ref{weight111}\subref{4}, Fig.\ref{weight11}\subref{4}, Fig.\ref{weight1}\subref{fig_forth_case}, and Fig.\ref{weight2}\subref{fig_forth_case} show the change of controller synchronization cost with time under different weight settings. 
Since it is not easy to compare the controller synchronization cost under each weight setting in the figures, we list the value of the total controller synchronization cost in Table\ref{t3}.
We can clearly observe from the data in Table.\ref{t3} that with the increase of $w_1$, the controller synchronization cost changes irregularly.
The controller synchronization cost abates with the increase of $w_2$.
In contrast, it raises with the $w_3$'s increase.
In addition, when $w_3^{\prime}$ and $w_3^{\prime\prime\prime}$ are fixed values, the increase of $w_3^{\prime\prime}$ also causes an increment in the controller synchronization cost.
 
\begin{figure}[h]
    \centering
    \includegraphics[width=3.3in]{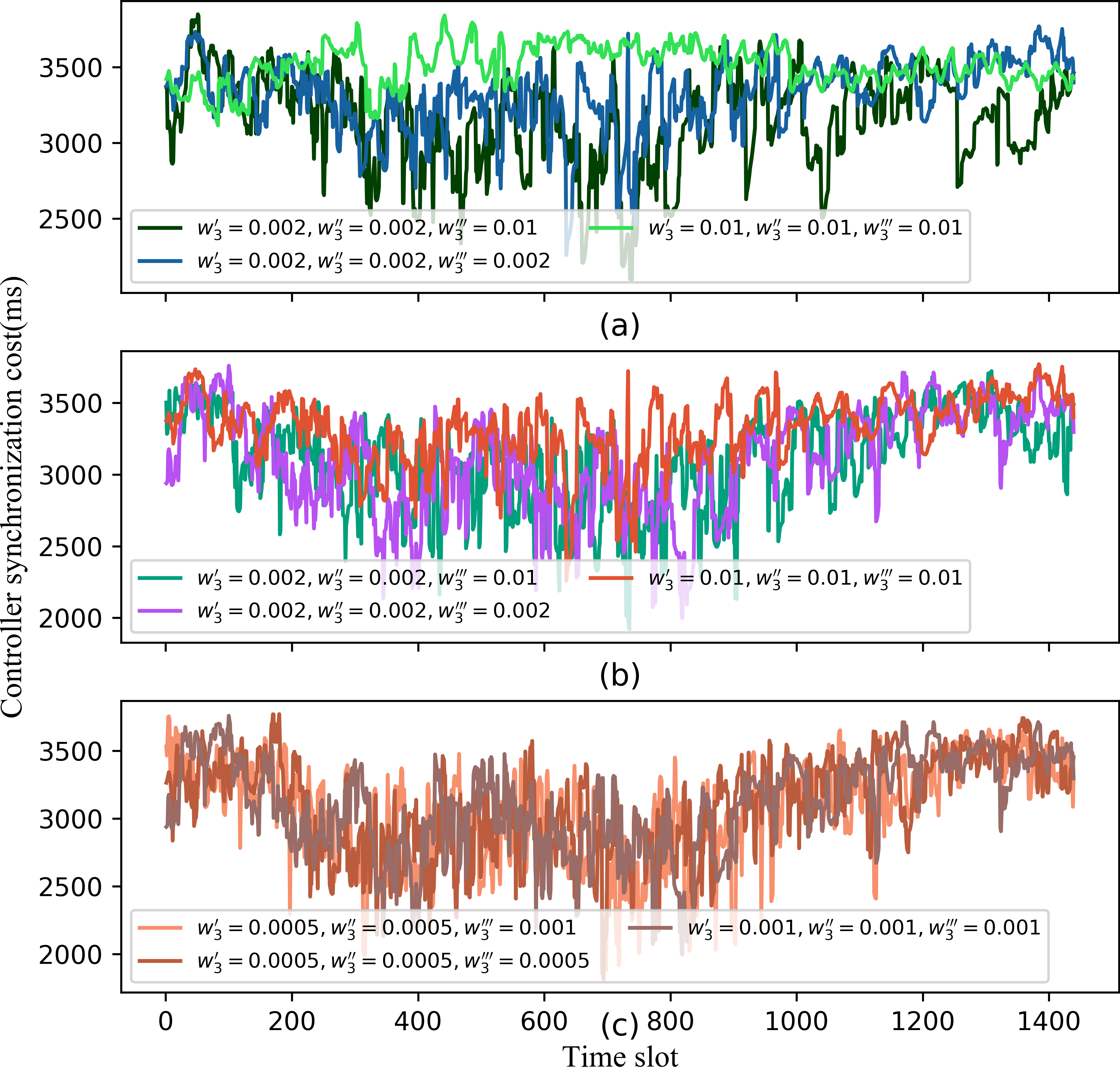}
    \caption{Controller synchronization cost with different weight settings.}
    \label{csc3}
\end{figure}

We also perform several experiments to demonstrate the effect of $w_3^{\prime\prime\prime}$ on the controller synchronization cost when $w_3^{\prime}$ and $w_3^{\prime\prime}$ are fixed values, and the effect of $w_3^{\prime}$ and $w_3^{\prime\prime}$ on the controller synchronization cost when $w_3^{\prime\prime\prime}$ is fixed, as shown in Fig.\ref{csc3}.
It is found that only increasing $w_3^{\prime\prime\prime}$, the controller synchronization cost can get better optimization results. 
While just increasing $w_3^{\prime}$ and $w_3^{\prime\prime}$, the synchronization cost of the controller gets worse results.

Based on the above observations, there is a preliminary indication that the optimization of controller migration cost and switch reassignment cost conflicts with the optimization of controller synchronization cost.
To fix the value of the controller migration cost and explore the impact of the optimization of the switch reassignment cost on the controller synchronization cost, we consider a scenario with an SPDA strategy. 
It is easy to draw the conclusion that whatever the weight of the switch reassignment cost is set to, it does not affect the controller synchronization cost. 
Therefore, in the dynamic placement dynamic assignment strategy, the real reason why the weight of switch reassignment affects the total controller synchronization cost is that the former affects the controller placement strategy and thus further affects the controller synchronization cost.
The reason for the increase in controller synchronization cost due to the increase in $w_3$ is that the optimization effect is less than the deterioration effect caused by the optimization of the controller migration cost.
The result shown in Fig.\ref{weight11}\subref{4} is caused by the fact that the growth of $w_2$ leads to the raise of controller migration cost, which further leads to the increase of controller synchronization cost.
As for the result of Fig.\ref{weight111}\subref{4}, combined with Fig.\ref{weight111}\subref{2}, the controller synchronization cost does not vary regularly with the change of the controller migration cost. 
This is because the change of $w_1$ breaks this regularity.

To sum up, the controller synchronization cost is in conflict with the controller migration cost and not with the switch reassignment cost.
Except $w_1$, the reduction of controller migration cost caused by other weights increases the controller synchronization cost.
Moreover, this increase is not linear.

\begin{table}[htbp]
\caption{Statistics of the total controller synchronization cost\label{t3}}
\centering
\begin{tabular}{|c|cc|}
\hline
Figure&Weight Settings&Values(ms)\\
\hline
\hline
\multirow{4}{*}{\shortstack{Fig.\ref{weight111}\subref{4}}}&$w_1=0.0001$&$4.8953\times 10^6$\\
&$w_1=0.0005$&$4.8721\times 10^6$\\
&$w_1=0.002$&$4.5452\times 10^6$\\
&$w_1=0.01$&$4.5721\times 10^6$\\
\hline
\hline
\multirow{4}{*}{\shortstack{Fig.\ref{weight11}\subref{4}}}&$w_2=0.5$&$4.7999\times 10^6$\\
&$w_2=1$&$4.7797\times 10^6$\\
&$w_2=5$&$4.6488\times 10^6$\\
&$w_2=10$&$4.6319\times 10^6$\\
\hline
\hline
\multirow{4}{*}{\shortstack{Fig.\ref{weight1}\subref{fig_forth_case}}}&$w_3=0.0005$&$4.4655\times 10^6$\\
&$w_3=0.001$&$4.4714\times 10^6$\\
&$w_3=0.002$&$4.7797\times 10^6$\\
&$w_3=0.01$&$5.0553\times 10^6$\\
\hline
\hline
\multirow{4}{*}{\shortstack{Fig.\ref{weight2}\subref{fig_forth_case}}}&$w_3^{\prime\prime}=0.00025$&$4.1992\times 10^6$\\
&$w_3^{\prime\prime}=0.0005$&$4.3093\times 10^6$\\
&$w_3^{\prime\prime}=0.001$&$4.5561\times 10^6$\\
&$w_3^{\prime\prime}=0.002$&$4.7797\times 10^6$\\
\hline
\hline
\multirow{3}{*}{\shortstack{Fig.\ref{csc3}(a)}}&$w_3^{\prime}=0.002,w_3^{\prime\prime}=0.002,w_3^{\prime\prime\prime}=0.002$&$4.7797\times 10^6$\\
&$w_3^{\prime}=0.002,w_3^{\prime\prime}=0.002,w_3^{\prime\prime\prime}=0.01$&$4.5285\times 10^6$\\
&$w_3^{\prime}=0.01,w_3^{\prime\prime}=0.01,w_3^{\prime\prime\prime}=0.01$&$5.0553\times 10^6$\\
\hline
\hline
\multirow{3}{*}{\shortstack{Fig.\ref{csc3}(b)}}&$w_3^{\prime}=0.001,w_3^{\prime\prime}=0.001,w_3^{\prime\prime\prime}=0.001$&$4.4714\times 10^6$\\
&$w_3^{\prime}=0.001,w_3^{\prime\prime}=0.001,w_3^{\prime\prime\prime}=0.002$&$4.4669\times 10^6$\\
&$w_3^{\prime}=0.002,w_3^{\prime\prime}=0.002,w_3^{\prime\prime\prime}=0.002$&$4.7797\times 10^6$\\
\hline
\hline
\multirow{3}{*}{\shortstack{Fig.\ref{csc3}(c)}}&$w_3^{\prime}=0.0005,w_3^{\prime\prime}=0.0005,w_3^{\prime\prime\prime}=0.0005$&$4.4655\times 10^6$\\
&$w_3^{\prime}=0.0005,w_3^{\prime\prime}=0.0005,w_3^{\prime\prime\prime}=0.001$&$4.4167\times 10^6$\\
&$w_3^{\prime}=0.001,w_3^{\prime\prime}=0.001,w_3^{\prime\prime\prime}=0.001$&$4.4714\times 10^6$\\
\hline
\end{tabular}
\end{table}

\section{Conclusion}
In this paper, we have investigated the joint optimization problem of controller placement and switch assignment for objective function value minimization in SDN-based LEO satellite networks.
Five metrics were considered in the objective function, including load balancing, average controller response delay, controller migration cost, switch reassignment cost, and controller synchronization cost.
To solve this problem, we proposed a prior population-based genetic algorithm, which uses the prior population to concatenate algorithms to achieve continuous optimization in the time dimension.
Besides, the prior population also accelerated the algorithm's convergence for each time slot.
Then, we built a LEO satellite network as our experimental scenario and developed a visualization tool. 
Based on this environment, we conducted numerous experiments, and the results show the proposed algorithm's effectiveness.
In addition, we analyzed the effect of the number of controllers and the weights on the performance of each aspect.

In the future, we will focus on studying the strategy in the case of uncertainty in the number of required controllers.
The number of controllers is considered as part of the objective function, and the number of controllers is determined by the algorithm.
Eventually, the number of controllers, the location of controllers, and the division of control domains can be dynamically adjusted according to the network state.
\bibliographystyle{IEEEtran}
\bibliography{main}

\end{document}